\documentclass{article}
\usepackage{graphicx}  % standard LaTeX graphics tool;
\usepackage{amsmath}   % For getting proper math eqns
\usepackage{adjustbox}
\usepackage[compress]{cite}
\usepackage{amssymb}   % Used for getting various symbols eg. gtrsim, lesssim etc.
\usepackage{bm} % For bold math (esp of lower greek letters)
\usepackage{dcolumn}% Align table columns on decimal point
\usepackage{color}
\usepackage{colortbl}
\usepackage{colordvi}
\usepackage{mathrsfs}
\usepackage{amsfonts}
\usepackage{varioref}
\usepackage{float}
\usepackage{graphicx}
\usepackage{multirow}
\usepackage{subcaption}
\usepackage[bottom=0.5in,top=0.5in,right=1.5in,left=1.5in]{geometry}
\usepackage{tabularx}

\RequirePackage[colorlinks,citecolor=blue,urlcolor=magenta,linkcolor=blue]{hyperref}
\input epsf
\usepackage{tablefootnote}
\labelformat{section}{Section #1} 
\labelformat{subsection}{Section #1} 
\labelformat{subsubsection}{Section #1}
\labelformat{subsubsubsection}{Section #1}
\labelformat{equation}{Eq.~(#1)} 
\labelformat{figure}{Fig.~#1} 
\labelformat{subfigure}{Fig.~\thefigure#1} 
\labelformat{table}{Table~#1} 
\labelformat{appendix}{Appendix #1}
\title{Simpson Visser}
\author{Anirban Dasgupta}
\date{}

\title{ Constraining the rotating Simpson-Visser spacetime from the observed quasi-periodic oscillations in black holes
}

\author{Anirban Dasgupta \footnote{522ph1007@nitrkl.ac.in}~$^{1}$ and Indrani Banerjee \footnote{banerjeein@nitrkl.ac.in}~$^{1}$\\
{\small{$^{1}$Department of Physics and Astronomy, National Institute of Technology, Rourkela, Odisha-769008, India}}}

\begin{document}

\maketitle

\begin{abstract}
Regular black holes (BHs) which are singularity-free alternatives to the standard black hole  paradigm in General Relativity (GR), offer effective models for probing the interface between classical and quantum gravity. They serve as promising candidates for exploring the nature of strong gravity and potential extensions of GR by providing testing grounds to understand how quantum corrections might manifest in astrophysical black holes. In the present work, we investigate the regular BH scenario described by the Simpson-Visser (SV) spacetime and explore its imprints on the high-frequency quasi-periodic oscillations (HFQPOs) observed in the black hole power spectrum. The Simpson-Visser spacetime represent the simplest, globally regular extensions of the Schwarzschild scenario, through the presence of a regularizing parameter. We explore the imprints of the regularizing parameter on the orbital and epicyclic frequencies associated with the motion of test particles in the rotating SV spacetime. Models aimed to explain the observed HFQPOs often invoke these fundamental frequencies and hence can potentially constrain the regularizing parameter from the available HFQPO data. We test eleven well-established HFQPO models against available observations from six black hole sources, obtaining spin constraints that, when compared with previous independent estimates, help identify the observationally favored models for each source. Based on the present data, we report that the observationally favored models cannot discriminate between the Kerr and the Simpson-Visser scenario. This when coupled with the large discrepancy in previous spin estimates of these sources, may plausibly indicate some deviation from GR in the strong gravity regime near BHs which requires further investigation.

\end{abstract}

\section{Introduction}
Classical black hole solutions in General Relativity contain central curvature singularities, where physical quantities diverge and the theory loses its predictive power \cite{Hawking:1973uf}. Since such singularities are generally considered unphysical, they motivate the development of alternative models that avoid these pathologies. Regular black holes offer a natural resolution by replacing the singular core with a non-singular one, typically sustained through nonlinear electrodynamics, quantum gravity-inspired effects, or modifications to Einstein’s theory of gravity. Such regular BHs not only address the long-standing singularity problem in gravitational physics, but also allow for testable astrophysical predictions, such as black hole shadows, quasi-normal modes, and accretion dynamics. This has led to the investigation of several regular BH scenarios in the literature \cite{PhysRevLett.86.5227,bardeen1968non,Dymnikova:1992ux,Ayon-Beato:2000mjt,Hayward:2005gi,Ansoldi:2008jw,Hazarika:2025axz,Banerjee:2022ffu,Chamseddine:2016ktu,de2020singularity,babichev2020regular,cano2020electromagnetic,Ashtekar:2006wn,Ashtekar:2006rx,Barcelo:2007yk,Roman:1983zza,Junior:2023qaq,Frolov:2014jva,Frolov:2016pav,Frolov:2017rjz,Carballo-Rubio:2018pmi,Carballo-Rubio:2018jzw,Vagnozzi:2022moj,Pedrotti:2024znu,Calza:2024fzo,Calza:2024xdh,Calza:2022ioe,Mitra:2024olo}. 
In this work we aim to study the astrophysical implications of the regular BH scenario proposed by Simpson and Visser (SV) \cite{simpson2019black} which represent the simplest globally regular 
extensions of classical black hole solutions in General Relativity, e.g.
the Schwarzschild BH. The Simpson-Visser spacetime can not only describe a regular BH but also wormhole geometries, such as the one-way wormholes or the traversable wormholes, depending on the values of the regularizing parameter, which marks the deviation from the standard Schwarzschild scenario \cite{simpson2019black}. Since wormholes require exotic matter and are not yet observationally confirmed (unlike black holes), we will concentrate on the regular BH sector of the SV spacetime in this work. Since astrophysical BHs are in general rotating we consider here the rotating SV spacetime which was first put forward by Mazza \cite{mazza2021novel}. The SV spacetime has been tested with several astrophysical observations, e.g. shadows, gravitational lensing, quasi-normal modes, black hole spectra etc.  \cite{Riaz:2022rlx,Stuchlik:2021tcn,Tsukamoto:2021caq,Bronnikov:2021liv,Tsukamoto:2020bjm,Shaikh:2021yux,Lobo:2020ffi,Guerrero:2021ues,Huang:2019arj,Xu:2021lff,Chatzifotis:2021hpg,Guo:2021wid,Islam:2021ful,Jiang:2021ajk,Lu:2024nev,Yang:2024mro}.
In this work, we will explore the signatures of the Simpson-Visser regular BH from the high-frequency quasi-periodic oscillations (HFQPOs) observed in the power spectrum of certain black holes. Although work has been done in this direction in the past \cite{Lu:2024nev,Jiang:2021ajk,Stuchlik:2021tcn,Yang:2024mro}, not all the available BH sources exhibiting HFQPOs were considered and the available data have not been tested with all the available HFQPO models. The present work is therefore a more comprehensive approach in this direction and aims to bridge this gap in the literature.

Quasi-periodic oscillations (QPOs) are primarily detected in the power spectra of certain Low-Mass X-ray Binaries (LMXRBs) hosting neutron stars (NSs) or black holes (BHs), and more infrequently in active galactic nuclei (AGN) \cite{2006csxs.book.....L,vanderKlis:2000ca,Torok:2004xs}. These features, manifesting as sharp peaks in the X-ray power spectrum, were first identified in BH and NS systems by NASA’s Rossi X-Ray Timing Explorer (RXTE) mission \cite{2006csxs.book.....L}. On the basis of their observed frequencies, QPOs are broadly divided into two categories: low-frequency QPOs (LFQPOs), with frequencies in the range of $\sim \rm mHz$, and high-frequency QPOs (HFQPOs), with characteristic frequencies of a few hundred Hz for stellar-mass BH sources \cite{vanderKlis:2000ca,Maselli:2014fca}.
The present study focuses on HFQPOs in BH systems \cite{Torok:2004xs,Abramowicz:2011xu,Aschenbach:2004kj,Kotrlova:2017wyq,Torok:2011qy}. HFQPOs in LMXRBs correspond to timescales of order $0.1–1\rm ms$, which are commensurate with the dynamical timescales of accreting matter in the innermost regions ($r<10 R_{\rm g}$) of compact objects \cite{1971SvA....15..377S,1973SvA....16..941S,PhysRevLett8217}. This connection arises from the scaling of the fluid velocity in the gravitational field of a compact object, $v \sim \sqrt{GM/r}$, leading to the dynamical timescale, $
t_{\rm d} \sim \sqrt{\frac{r^{3}}{GM}} 
$,
which for a $10 M_\odot$ BH at $r \sim 100\rm{km}$ becomes $t_{\rm d}\sim 1~\rm ms$ \cite{2006csxs.book.....L,vanderKlis:2000ca}.
These dynamical timescales, of the order of milliseconds for stellar-mass BHs \cite{1971SvA....15..377S,1973SvA....16..941S}, translate into frequencies of a few hundred Hz in the Fourier domain, in direct agreement with observed HFQPOs \cite{2006csxs.book.....L,vanderKlis:2000ca}. This remarkable consistency underscores the physical relevance of HFQPOs as probes of accretion dynamics and strong-field gravity in the vicinity of BHs \cite{vanderKlis:2000ca}. Moreover, since HFQPO frequencies are closely linked to the orbital and epicyclic frequencies of matter near the compact object, they exhibit a mass scaling relation inversely proportional to the central object’s mass. Consequently, HFQPOs appear at $\sim \rm kHz$ frequencies in NS systems, while for supermassive BHs such as Sgr A* they are shifted down to $\sim \rm mHz$.
Owing to their intrinsic association with the motion of accreting matter in the immediate vicinity of compact objects, HFQPOs are believed to encode valuable information about the nature of gravity in the strong-field, high-curvature regime \cite{PhysRevLett8217}. Their study therefore provides a unique observational window into near-horizon physics and the fundamental properties of BH spacetimes.

To account for the observed HFQPOs in black hole systems, a variety of theoretical models have been put forward \cite{Stella:1997tc,PhysRevLett8217,Cadez:2008iv,Kostic:2009hp,Germana:2009ce,Kluzniak:2002bb,Abramowicz:2003xy,Rebusco:2004ba,Nowak:1996hg,Torok:2010rk,Torok:2011qy,Kotrlova:2020pqy,1980PASJ...32..377K,Perez:1996ti,Silbergleit:2000ck,Rezzolla:2003zy,Rezzolla:2003zx}. These models are typically formulated in terms of the orbital and epicyclic frequencies of test particle motion in the gravitational field of compact objects. We compare the HFQPO data of six BH sources with eleven well established HFQPO models existing in the literature. We determine the best-fit model parameters for each source using both the grid-search and the MCMC approaches. 
This enables us to place constraints on the black hole spin as well as on additional parameters such as the regularizing parameter. 
A comparison of these constraints with independent spin measurements then offers insights into which of the proposed HFQPO models are most consistent with observations for a given source. This also enables us to comment on the viability of the Simpson-Visser scenario in explaining the HFQPOs associated with these sources.  

The paper is structured as follows: In the next section, we provide a brief overview of the regular BH scenario described by the Simpson-Visser spacetime while 
\ref{S3} summarizes the dynamics of massive test particles in this spacetime, which serve as the foundation for the various HFQPO models. In \ref{error-analysis}, we describe these models in detail and discuss their comparison with HFQPO observations of black holes using the grid-search method. \ref{S5} presents constraints on the model parameters obtained using MCMC simulations, and \ref{S6} concludes with a summary and implications of our findings. Throughout this work, we adopt the metric signature $(-,+,+,+)$ and employ geometrized units with $G=c=1$.

\section{Review of Simpson-Visser black hole}
\label{S2}
One of the main predictions of GR is the existence of a singularity where the space-time curvature becomes infinity and the laws of physics break down. But recent studies show that this singularity can be bypassed by quantum gravitational effects. Many regular black holes (BHs) have been proposed \cite{bardeen1968non,Dymnikova:1992ux,Ayon-Beato:2000mjt,Hayward:2005gi,Ansoldi:2008jw,Chamseddine:2016ktu,de2020singularity,babichev2020regular,cano2020electromagnetic} where the singularity is absent. Simpson-Visser metric \cite{simpson2019black} is one of those which suggests a BH with no singularity. 
%Apart from that, the topology of the metric also gives rise to a traversable wormhole structure.\\
The static non-rotating Simpson-Visser BH is given by the following metric:
\begin{eqnarray}
ds^2 = -\left(1 - \frac{2M}{\sqrt{r^2 + l^2}}\right) dt^2 
+ \left(1 - \frac{2M}{\sqrt{r^2 + l^2}}\right)^{-1} dr^2 
+ (r^2 + l^2)\left(d\theta^2 + \sin^2\theta \, d\phi^2\right),
\label{1}
\end{eqnarray}
where \(M\) is the ADM mass and \(l\) is the regularising length scale.\\
The Ricci scalar \(R\), the quadratic Ricci invariant \(R_{\mu\nu}R^{\mu\nu}\) and the Kretchmann scalar \(K = R_{\mu\nu\rho\sigma}R^{\mu\nu\rho\sigma}\) at \(r=0\) are given by the following:
\begin{eqnarray}
    R &=& -\frac{2}{l^2} + \frac{6M}{l^3}, \nonumber\\
    R_{\mu\nu}R^{\mu\nu} &=& \frac{4}{l^4} - \frac{12 M}{l^5} + \frac{18 M^2}{l^6}, \nonumber\\
    K &=& \frac{12}{l^4} - \frac{32 M}{l^5} + \frac{36 M^2}{l^6}.
\end{eqnarray}
Clearly, none of the curvature scalars blow up at \(r=0\), which implies the non-existence of a singularity at that point. Since the areal radius of the two-sphere in this spherically symmetric metric is \(R=\sqrt{r^2+l^2}\), the range of the coordinate function \(r\) can be extended to \(r \in (-\infty, +\infty)\), revealing the existence of a parallel universe at negative \(r\) values. The Simpson-Visser metric can be derived from an action principle comprising of a scalar field and non-linear electrodynamics minimally coupled to gravity \cite{Bronnikov:2021uta}.

In our present work, we want to test theoretical models with astrophysical observations. Since astrophysical BHs are generally rotating we need to consider rotating counterpart of the Simpson-Visser metric which was obtained by Mazza et al. \cite{mazza2021novel}. The rotating axisymmetric Simpson-Visser black hole with spin $a$ in Boyer-Lindquist coordinate is given by the following metric,
\begin{eqnarray}
ds^2 = -\left(1 - \frac{2M \sqrt{r^2 + l^2}}{\Sigma}\right) dt^2 + \frac{\Sigma}{\Delta} dr^2 + \Sigma d\theta^2 - \frac{4Ma \sin^2\theta \sqrt{r^2 + l^2}}{\Sigma} dtd\phi + \frac{A \sin^2\theta}{\Sigma} d\phi^2,
\label{SV}
\end{eqnarray}
where
\[
\Sigma = r^2 + l^2 + a^2 \cos^2\theta, \quad \Delta = r^2 + l^2 + a^2 - 2M\sqrt{r^2 + l^2}, \quad A = (r^2 + l^2 + a^2)^2 - \Delta a^2 \sin^2\theta.
\]
The event horizon of the black hole is given by $\Delta=0$ which is located at,
\begin{eqnarray}
r_\pm = \sqrt{\left(M \pm \sqrt{M^2 - a^2}\right)^2 - l^2}.
\label{S2-11}
\end{eqnarray}
where, 
$\rho_\pm = M \pm \sqrt{M^2 - a^2}.$ \nonumber\\
The topology of the metric gives rise to various configuration of space-time depending on the value of $l$ and $a$. From the expression of event horizon, it is clear that the spin $a$ can assume a maximum value of M which is the extremal case. We observe that \( r_+ \) is real only if \( l \leq \rho_+ \), and similarly, \( r_- \) is real only if $ l \leq \rho_- $. Consequently, depending on the parameter values, we get that  two horizons exist if \( a < M \) and \( l < \rho_- \) and a single horizon exists if \( a < M \) and \( \rho_- < l < \rho_+ \). When \( a < M \) and \( l > \rho_+ \), or \( a > M \) the above metric represents a traversable wormhole \cite{simpson2019black,mazza2021novel}. In cases where the equality holds, the solutions correspond to extremal or limiting configurations of the above scenarios. Since the maximal value of $\rho_+$ can be $2M$, when $l>2M$ the above metric represents a wormhole geometry. In this work we aim to understand the role of the regularizing parameter in explaining the HFQPOs in BHs. Since the objects exhibiting HFQPOs are believed to be BH sources (based on other independent observations) \cite{motta2014precise,Orosz:2011ki,Ghez:2008ms,Gillessen:2008qv,Reid:2014ywa}  hence we shall focus in the BH regime of the above spacetime (\ref{SV}) and confine ourselves between $0\leq l\leq 2M$.
In what follows we will scale the radial distance $r$ and the spin $a$ with the gravitational radius $R_g=M$ ($G=c=1)$, such that $r\equiv r/M$ and $a\equiv a/M$.

\section{Epicyclic frequencies of test particles orbiting the Simpson-Visser black hole}
\label{S3}
The stationary, axisymmetric spacetime  can generally be expressed as:
\begin{eqnarray}
ds^2 = g_{tt} dt^2 + 2 g_{t\phi} dt d\phi + g_{\phi \phi} d\phi^2 + g_{rr} dr^2 + g_{\theta \theta} d\theta^2
\end{eqnarray}
where the metric components \( g_{\mu \nu} = g_{\mu \nu} (r, \theta) \) satisfy the symmetry \( g_{\mu \nu}(r, \theta) = g_{\mu \nu}(r, -\theta) \). Here, we examine the motion of massive test particles in circular equatorial geodesic. In the Lagrangian approach, we identify two conserved quantities: the energy \( E \) per unit mass and the angular momentum \( L \) per unit mass. \nonumber\\

The Lagrangian for a test particle in this background is given by:
\begin{eqnarray}
2\mathcal{L} = g_{tt} \dot{t}^2 + 2 g_{t\phi} \dot{t} \dot{\phi} + g_{\phi \phi} \dot{\phi}^2 + g_{rr} \dot{r}^2 + g_{\theta \theta} \dot{\theta}^2
\end{eqnarray}
From the Euler-Lagrange equation, we have:
\begin{eqnarray}
\frac{d}{d\lambda} \left( \frac{\partial \mathcal{L}}{\partial \dot{t}} \right) - \frac{\partial \mathcal{L}}{\partial t} = 0
\end{eqnarray}
Since \( \mathcal{L} \) does not depend explicitly on \( t \), we get \( \frac{\partial \mathcal{L}}{\partial \dot{t}} = {\rm constant}= -2 E \), implying that:
\begin{eqnarray}
g_{tt} \dot{t} + g_{t\phi} \dot{\phi} = -E = p_t
\end{eqnarray}
Similarly, as \( \mathcal{L} \) is independent of \( \phi \), we have \( \frac{\partial \mathcal{L}}{\partial \phi} = 0 \), leading to:
\begin{eqnarray}
g_{\phi \phi} \dot{\phi} + g_{t\phi} \dot{t} = L = p_\phi
\end{eqnarray}
Hence, \( E \) and \( L \) are two conserved quantities of the motion.
With these conserved quantities, we define an effective potential that governs the particle’s motion:
\begin{eqnarray}
U(r, \theta) = g^{tt} - 2 \left( \frac{L}{E} \right) g^{t\phi} + \left( \frac{L}{E} \right)^2 g^{\phi \phi}
\label{16}
\end{eqnarray}
Using the normalization condition \( g_{\mu \nu} u^{\mu} u^{\nu} = -1 \), we get:
\begin{eqnarray}
g_{rr} \dot{r}^2 + g_{\theta \theta} \dot{\theta}^2 + E^2 U(r, \theta) = -1
\label{17}
\end{eqnarray}
Assuming the test particle’s motion is restricted to the equatorial plane, this reduces to \( \dot{r}^2 = V(r) \) where
\begin{eqnarray}
V(r) = -\frac{1 + E^2 U(r_c, \pi / 2)}{g_{rr}}
\end{eqnarray}
The marginally stable circular orbit ($r_{ms}$) can be obtained by solving \( V(r) = 0 \), \( V'(r) = 0 \), and \( V''(r) = 0 \). This marks the boundary beyond which a particle would no longer maintain a stable circular orbit and would instead move radially toward the black hole.

\begin{figure}[t!]
    %\centering
    \begin{subfigure}{0.5\textwidth}
        \includegraphics[width=\linewidth]{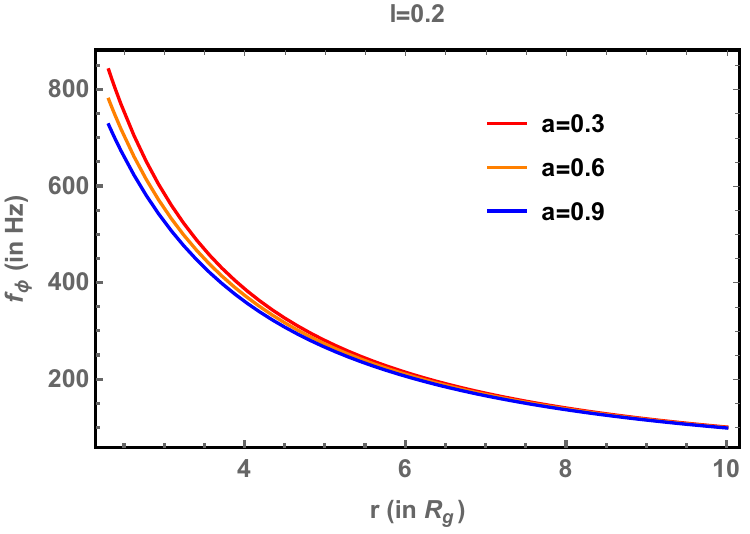}
        \caption{}
        \label{a1}
    \end{subfigure}
    % \hspace{0.02\textwidth} % Adjust the horizontal space between figures
    \begin{subfigure}{0.5\textwidth}
        \includegraphics[width=\linewidth]{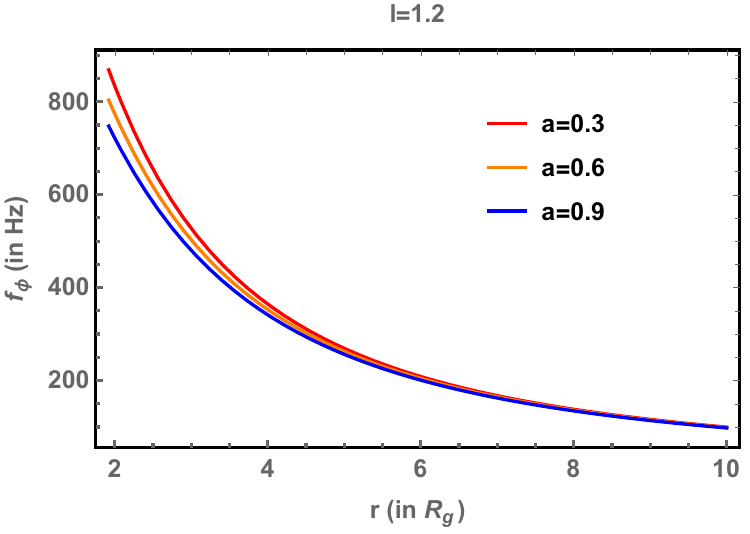}
        \caption{}
        \label{a3}
    \end{subfigure}
    \caption{The above figure shows radial variation of $f_{\phi}$ for (a) l=0.2 and  (b) l= 1.2 for different choices of spin, a=0.3(red), a=0.6(orange), a=0.9(blue) }
    \label{fphi}
\end{figure}

The angular velocity of a particle in circular orbit is given by \( \Omega = \frac{u^{\phi}}{u^t} \), where \( u^{\mu} = \dot{x}^{\mu} = \frac{dx^{\mu}}{d\tau} \). From the radial component of the Euler-Lagrange equation and for circular, equatorial geodesics, we obtain:
\begin{eqnarray}
\frac{\partial g_{tt}}{\partial r} + 2 \Omega \frac{\partial g_{t\phi}}{\partial r} + \Omega^2 \frac{\partial g_{\phi \phi}}{\partial r} = 0
\end{eqnarray}
Solving this quadratic equation yields \( \Omega \) as:
\begin{eqnarray}
\Omega = \frac{-\partial_r g_{t\phi} \pm \sqrt{(\partial_r g_{t\phi})^2 - (\partial_r g_{\phi \phi})(\partial_r g_{tt})}}{\partial_r g_{\phi \phi}}
\end{eqnarray}
Since \( \Omega = 2\pi f_\phi \), \( f_\phi \) represents the orbital frequency of the particle around the black hole. {\bf In \ref{fphi}, we have shown variation of orbital frequency $f_\phi$ with radius for (a) $l=0.2$ and (b) $l=1.2$ for different choices of spin a=0.3 (red), a=0.6 (orange), a=0.9 (blue) for a black hole having 10 $M_\odot$. From the figure we note that, as we move closer to the black hole, the orbital frequency increases. However, it decreases as the spin value of black hole increases. We also note that the orbital frequency is not much sensitive to the variation in $l$ for a fixed spin. }\\ 

\begin{figure}[t!]
   % \centering
    \begin{subfigure}{0.5\textwidth}
        \includegraphics[width=\linewidth]{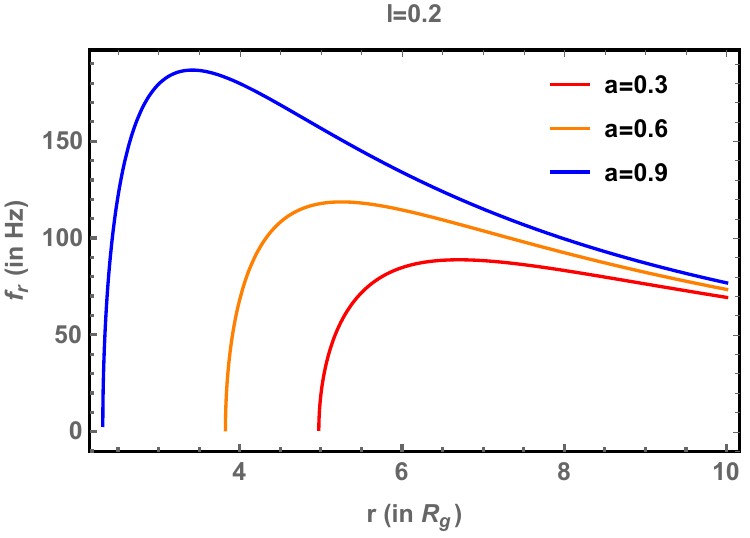}
        \caption{}
        \label{b1}
    \end{subfigure}
     % Adjust the horizontal space between figures
    \begin{subfigure}{0.5\textwidth}
        \includegraphics[width=\linewidth]{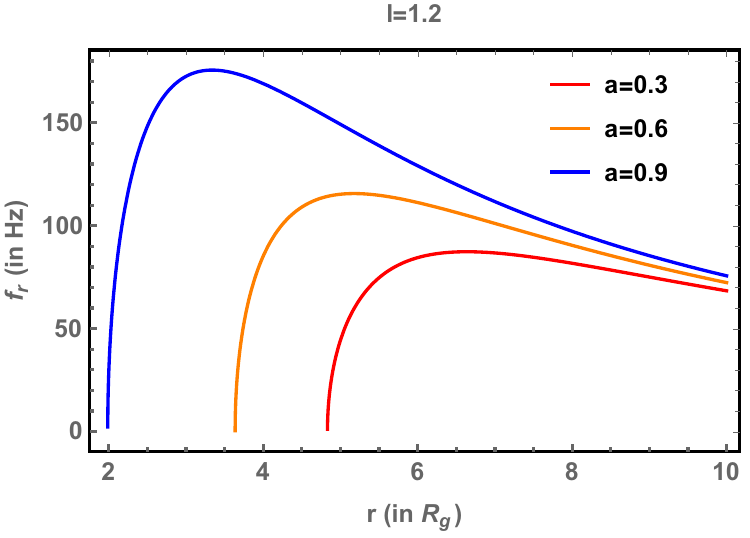}
        \caption{}
        \label{b3}
    \end{subfigure}
    \caption{The above figure shows the radial variation of $f_r$ for (a) l=0.2 and  (b) l= 1.2 for different choices of spin, a=0.3  (red), a=0.6(orange), a=0.9 (blue) }
    \label{fr}
\end{figure}

\begin{figure}[t!]
  %  \centering
    \begin{subfigure}{0.5\textwidth}
        \includegraphics[width=\linewidth]{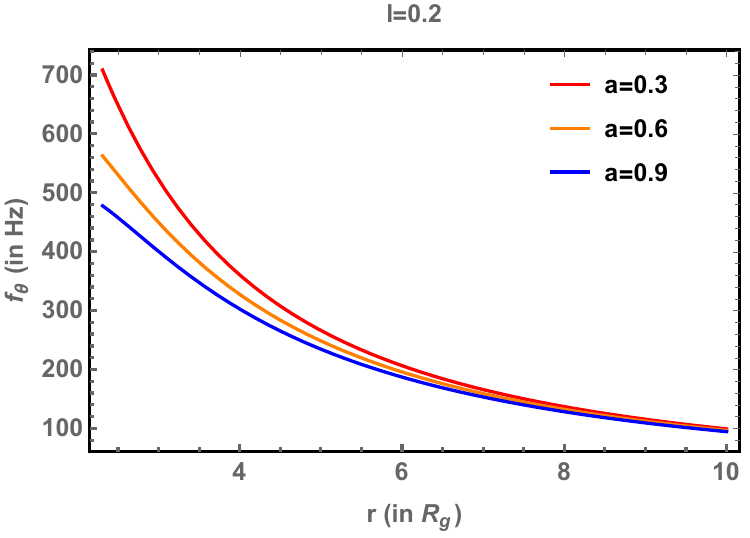}
        \caption{}
        \label{c1}
    \end{subfigure}
    \begin{subfigure}{0.5\textwidth}
        \includegraphics[width=\linewidth]{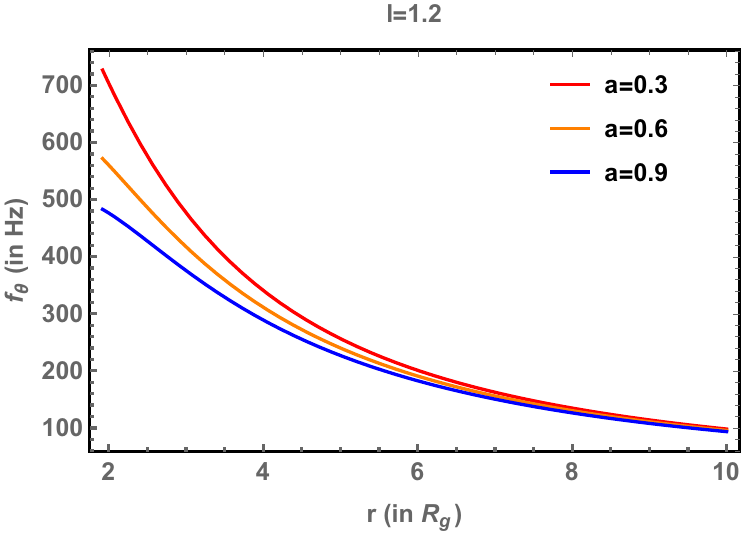}
        \caption{}
        \label{c3}
    \end{subfigure}
    \caption{The above figure shows the radial variation of $f_{\theta}$ for (a) l=0.2 and (b) l= 1.2 for different choices of spin, a=0.3(red), a=0.6(orange), a=0.9(blue)}
    \label{ftheta}
\end{figure}

To derive the epicyclic frequencies, we perturb the stable circular equatorial orbit radially and vertically, such that:
\begin{eqnarray}
r(t) = r_c + \delta r, \quad \delta r \sim e^{i \omega_r t}, \quad \delta r \ll r \nonumber \\
\theta(t) = \frac{\pi}{2} + \delta \theta, \quad \delta \theta \sim e^{i \omega_\theta t}, \quad \delta \theta \ll \frac{\pi}{2}
\end{eqnarray}
In the first approximation, these two oscillations are considered decoupled. From \ref{17}, we get:
\begin{eqnarray}
-g_{rr} (u^0 2 \pi f_r \delta r)^2 - g_{\theta \theta} (u^0 2 \pi f_{\theta})^2 + E^2 \left[ U(r_c, \frac{\pi}{2}) + \frac{1}{2} \frac{\partial^2 U}{\partial r^2} (r_c, \frac{\pi}{2}) \delta r^2 + \frac{1}{2} \frac{\partial^2 U}{\partial \theta^2} (r_c, \frac{\pi}{2}) \delta \theta^2 \right] = 0
\end{eqnarray}
By setting the coefficients of \( \delta r^2 \) and \( \delta \theta^2 \) to zero, we find the radial and vertical epicyclic frequencies as:
\begin{eqnarray}
f_r^2 = \frac{c^6}{G^2 M^2} \left[ \frac{(g_{tt} + g_{t\phi} \Omega)^2}{2 (2\pi)^2 g_{rr}} \left( \frac{\partial^2 U}{\partial r^2} \right)_{r_c, \frac{\pi}{2}} \right]
\label{22}
\end{eqnarray}
\begin{eqnarray}
f_{\theta}^2 = \frac{c^6}{G^2 M^2} \left[ \frac{(g_{tt} + g_{t\phi} \Omega)^2}{2 (2\pi)^2 g_{\theta \theta}} \left( \frac{\partial^2 U}{\partial \theta^2} \right)_{r_c, \frac{\pi}{2}} \right]
\label{23}
\end{eqnarray}
The factor $\frac{c^6}{G^2M^2}$ has been used for  matching the dimension of frequency squared. {\bf In \ref{fr} and \ref{ftheta}, we have respectively shown the variation of radial and vertical epicyclic frequencies with radius for a black hole of 10 solar mass for different choices of regularisation length scale $l$ and spin $a$. The radial epicyclic frequency $f_r$ attains a maxima and then falls to zero at the $r_{\rm ms}$ as one approaches the BH. For a given $l$, $f_r$ increases with $a$, but decreases marginally with an increase in $l$ when $a$ is fixed. 
The vertical epicyclic frequency also increases as one goes inward, decreases with an increase in spin, but does not change much with $l$ for a fixed spin. }

We now review the main theoretical models proposed to explain high-frequency QPOs (HFQPOs) observed in black hole and neutron star systems. These HFQPOs typically appear in pairs with a characteristic 3:2 frequency ratio. As listed in \ref{t2}, the model-dependent QPO frequencies are expressed as linear combinations of the fundamental orbital frequencies 
$f_\phi$, $f_r$, $f_\theta$. Broadly, the models fall into two categories: kinematic models and resonance models, both assuming that HFQPOs originate at the same circular radius $r_{em}$ \cite{stuchlik2016models,Yagi:2016jml,Kotrlova:2020pqy}. In addition, diskoseismic models suggest QPOs arise from oscillation modes at different disk radii; however, they generally fail to reproduce the 3:2 ratio in simulations \cite{Tsang:2008fz,Fu:2008iw,Fu:2010tf} and are therefore not considered further here. Notably, the Relativistic Precession Model (RPM) stands out, as it provides a unified framework for explaining both HFQPOs and low-frequency QPOs (LFQPOs).

In \ref{t1} we have shown the six BH sources which exhibit HFQPOs in their power spectrum. In our analysis, we include five stellar-mass BHs and one supermassive BH, as these are the only sources known to exhibit twin-peak HFQPOs in their power spectra. While many BHs show QPOs, only a few display the twin-peak HFQPOs, which are listed in \ref{t1}. Since our work aims to probe signatures of strong gravity, we focus exclusively on sources with observed HFQPOs \cite{torok2011confronting,Abramowicz:2011xu,kotrlova2017super,aschenbach2004measuring,torok2005possible}. For stellar-mass black holes, these frequencies typically lie in the hundreds of Hertz, whereas for supermassive black holes such as Sgr A*, they are of the order of milliHertz \cite{Torok:2004xs,Aschenbach:2004kj}. This shift in frequency scale directly reflects the vast difference in the masses of these objects (\ref{22} and \ref{23}). 
The observed HFQPOs are denoted by $f_{up1}$ and $f_{up2}$ with $\Delta f_{up1}$ and $\Delta f_{up2}$ the associated errors.  The LFQPO which incidentally is only observed in the source GRO J1655-40 along with the two HFQPOs is denoted by $f_{up3}$ and its error by $\Delta f_{up3}$.

\begin{table}[t!]
    \centering
\begin{tabular}[width=\textwidth]{|c|c|c|c|}
\hline
     Models & $f_1$ & $f_2$ & $f_3$  \\
\hline
\hline
Parametric Resonance Model \cite{kluzniak2002parametric,Abramowicz:2003xy,Rebusco:2004ba}&$f_\theta$&$f_r$&-\\
\hline
Forced Resonance Model 1 \cite{kluzniak2002parametric} &$f_\theta$&$f_{\theta}-f_r$&-\\
\hline
Forced Resonance Model 2 \cite{kluzniak2002parametric}&$f_\theta+f_r$&$f_\theta$&- \\
\hline
Keplerian Resonance Model 1 \cite{nowak199767}&$f_\phi$&$ f_r$&-\\
\hline
Keplerian Resonance Model 2 \cite{nowak199767}&$f_\phi$&$2f_r$&-\\
\hline
Keplerian Resonance Model 3 \cite{nowak199767}&$3f_r$&$f_\phi$&-\\
\hline
Warped Disk Oscillation Model\cite{kato2001basic, kato2004wave} & $2f_\phi-f_r$ & $2(f_\phi-f_r)$ & -\\
\hline
Non-axisymmetric Disk Oscillation Model 1\cite{bursa2004upper,Bursa:2005th} &$f_\theta$&$f_\phi-f_r$&-\\
\hline
Non-axisymmetric Disk Oscillation Model 2 \cite{Kotrlova:2020pqy,Torok:2011qy,Torok:2010rk}&$2f_\phi-f_\theta$&$f_\phi-f_r$&-\\
\hline
Relativistic Precession Model \cite{Stella:1998mq,stella1997lense,stella1999correlations}&$f_\phi$&$f_\phi-f_r$&$f_\phi-f_\theta$ \\
\hline  
Tidal Disruption Model \cite{Cadez:2008iv,Kostic:2009hp,Germana:2009ce}&$f_\phi+f_r$&$f_\theta$&- \\
\hline
\end{tabular}
    \caption{Theoretical models proposed to explain the HFPQOs in black holes. Note that the Relativistic Precession Model also explains low-frequency QPO.}
    \label{t2}
\end{table}
\begin{table}[t!]
    \centering
\begin{adjustbox}{width=0.9\textwidth}
    \begin{tabularx}{\textwidth}{|X|X|X|X|X|}
    \hline
    \textbf{Source} & Mass \boldmath$(M_{\boldsymbol{\odot}})$ & $f_{up1} \pm \Delta f_{up1}$ \textbf{(Hz)} & $f_{up2} \pm \Delta f_{up2}$ \textbf{(Hz)} & $f_{up3} \pm \Delta f_{up3}$ \textbf{(Hz)}\\
    \hline
    \hline
    GRO J1655-40 & $5.4 \pm 0.3$ \cite{beer2002quiescent} & $441 \pm 2$ \cite{motta2014precise} & $298 \pm 4$ \cite{motta2014precise} & $17.3 \pm 0.1$ \cite{motta2014precise} \\
    \hline
    XTE J1550-564 & $9.1 \pm 0.61$\cite{Orosz:2011ki} & $276 \pm 3$ & $184 \pm 5$ & - \\
    \hline
    GRS 1915+105 & $12.4^{+2.0}_{-1.8}$ \cite{Reid:2014ywa} & $168 \pm 3$ & $113 \pm 5$ & - \\
    \hline
    H 1743+322 & $8.0 - 14.07$ \cite{Pei:2016kka,Bhattacharjee:2017rbl,Petri:2008jc} & $242 \pm 3$ & $166 \pm 5$ & - \\
    \hline
    Sgr A* & $(3.5-4.9)\times 10^6$\cite{Ghez:2008ms,Gillessen:2008qv} & $(1.445 \pm 0.16)\times10^{-3}$\cite{Torok:2004xs,Stuchlik:2008fy} & $(0.886 \pm 0.04)\times10^{-3}$\cite{Torok:2004xs,Stuchlik:2008fy} & - \\
    \hline
    XTE J1859+226 & $7.85 \pm 0.46$\cite{motta2022black} & $227.5^{+2.1}_{-2.4}$\cite{motta2022black} & $128.6^{+1.6}_{-1.8}$\cite{motta2022black} & $3.65 \pm 0.01$ \\
    \hline
    \end{tabularx}
\end{adjustbox}
\caption{BH Sources where High-Frequency Quasi-Periodic Oscillations (HFQPOs) are observed.}
\label{t1}
\end{table}

\section{Comparison of theoretical models with the observed QPO data}
\label{error-analysis}
In this section, we have compared the model dependent HFQPO frequencies with the observed values and conducted an error analysis to better assess the performance of each model. The $\chi^2$ method is used to minimize error between the theoretical predictions and the observational data. Since the Simpson-Visser metric depends on $l^2$ and not on $l$, hence in the rest of the paper we will consider $l^2=\beta$ as the parameter which is responsible for deviation of the metric from the Kerr scenario. Henceforth, $\beta$ is considered to be the regularisation parameter.

The $\chi^2$ function is defined as:
\begin{eqnarray}
\chi_i^{2}(\beta) =\frac{[f_{up1,i} - f_{1}(\beta, a_{min}, M_{min}, r_{min})]^2}{\sigma_{f_{up1},i}^2} 
+  \frac{[f_{up2,i} - f_{2}(\beta, a_{min}, M_{min}, r_{min})]^2}{\sigma_{f_{up2},i}^2}
\label{chi}
\end{eqnarray}
This expression evaluates the difference between the theoretical and observed frequencies. Here, $f_{up1,i}$ and $f_{up2,i}$ represent the upper and lower HFPQOs, respectively, for the $i$-th source listed in \ref{t1}. The uncertainties $\sigma_{f_{up1},i}$ and $\sigma_{f_{up2},i}$ correspond to the observational errors in HFQPO frequencies, also reported in \ref{t1}. Theoretical frequencies $f_1$ and $f_2$ for each model are based on $f_{\phi}$, $f_{\theta}$, and $f_r$, as outlined in \ref{t2}.

The steps of our method are detailed below:

\begin{itemize}
\item First, we select one of the models listed in \ref{t2}, where $f_1$, $f_2$, and possibly $f_3$ depend on the regularisation parameter $\beta$, spin $a$, black hole mass $M$, and the emission radius $r_{cm}$ for QPO generation.
\item We then select a black hole source from \ref{t1} with an established mass range, allowing us to calculate $f_1$, $f_2$, and possibly $f_3$ over this range.
\item The regularisation parameter $\beta$ is fixed to a specific value that yields a real, positive event horizon.
\item For the fixed $\beta$ the spin parameter $a$ varies within $-a_{max}\leq a \leq a_{max}$ (where $a_{max}=1$ from \ref{S2-11}), ensuring a real, postive event horizon.
\item The black hole mass $M$ is varied within $(M_0 - \Delta M) \le M \le (M_0 + \Delta M)$, where $M_0$ and $\Delta M$ are the mean mass and uncertainty values provided in (\ref{t1})
\item The primary objective is to compute $f_1$ and $f_2$. For each mass $M$, we find the emission radius $r_{cm}$ within the range $r_{ms}(a, \beta) \le r_{cm} \le r_{ms}(a, \beta) + 20 R_g$, where $R_g=GM/c^2$ is the gravitational radius and $r_{ms}$ is the radius of the marginally stable orbit.
\item For each combination of $M$, $a$, and $r_{\mathrm{cm}}$ at the specified $\beta$, 
the corresponding $\chi^{2}_{i}$ is evaluated for the selected source. 
The parameter values of $M$, $a$, and $r_{\mathrm{cm}}$ that minimize $\chi^{2}_{i}$ 
represent the most probable estimates of the mass, spin, and emission radius for the given $\beta$. 
These optimal values are denoted as $M_{\min}$, $a_{\min}$, and $r_{\mathrm{em},\min}$ in \ref{chi}\cite{avni1976energy}.
\item We repeat the above steps for different values of $\beta$ which gives the variation of $\chi^2$ with $\beta$ for the chosen source. 
\item The above procedure is repeated for all the BH sources for the chosen model.
\item The above process is then repeated for all the models given in \ref{t2}. The minimum $\chi^2$ value identifies the most favorable $\beta$, along with the corresponding values of $M_{min}$ and $a_{min}$ for each black hole source within the domain of a given model.  
\end{itemize}

We now review the various theoretical models proposed to interpret the observed high frequency quasi-periodic oscillations (HFQPOs) in the power spectrum of black holes (BHs). By comparing the theoretical predictions with the observed QPO frequencies, we aim to evaluate the effectiveness of these models.
\ref{t1} presents several BH sources that exhibit high-frequency QPOs (HFQPOs) within their power spectra. The fundamental frequencies associated with HFQPOs scale inversely with the mass of the BH, as discussed in the previous section. This scaling implies that while HFQPOs for stellar-mass BHs are in the range of hundreds of Hz, those for supermassive BH like Sgr A* appear in the mHz range. The mass estimates of these BH sources, derived from independent observations, are also included in \ref{t1}. From the table, it is evident that the observed HFQPOs often appear in commensurate frequency pairs ($f_{up1}$ and $f_{up2}$), with a characteristic ratio of 3:2. Since these frequencies are associated with the three fundamental frequencies ($f_r$, $f_\theta$ and $f_\phi$), hence they are primarily dependent on the spacetime geometry surrounding the BH, and less sensitive to the complex accretion physics. Hence, HFQPOs can serve as a cleaner probe of the underlying metric compared to other observational probes such as the continuum spectrum or the iron-line.

\ref{t2} lists the various theoretical models proposed to explain the HFQPOs detected in the BH power density spectra. For these models, the predicted HFQPOs are designated as $f_1$ and $f_2$, which correspond to the observed frequencies $f_{up1}$ and $f_{up2}$ in \ref{t1}. Additionally, the low-frequency QPOs (LFQPOs) are denoted by $f_3$, which is comparable to $f_L$ in \ref{t1}. 
Given the wide range of theoretical models available to describe the HFQPOs, it is essential to gain an understanding of each model before proceeding further. Below, we discuss the main categories of HFQPO models, which can be broadly divided into two types: (a) Resonant models and (b) Kinetic models. At the same time, we will also discuss our results related to the constraints on the parameter $\beta$ to see whether it favors the Kerr scenario or the Simpson-Visser scenario.

\subsection{Resonant Models}
\label{subsec:resonant_models}
The resonant models are effective in explaining the 3:2 ratio observed in the twin peak HFQPOs, an outcome that is less satisfactorily addressed by other models \cite{torok2005orbital,Abramowicz:2003xy,Abramowicz:2001bi,kluzniak2001strong,Kluzniak:2002bb,torok2011confronting}. In these models, particles follow nearly circular geodesics in the equatorial plane, with small perturbations $\delta r = r - r_c$ and $\delta \theta = \theta - \pi/2$ around the circular radius $r_c$. These perturbations obey simple harmonic motion, characterized by the radial and vertical frequencies $\omega_r$ and $\omega_{\theta}$:
\begin{eqnarray}
\delta \ddot{r} + \omega_{r}^{2} \delta r = 0 \\
\delta \ddot{\theta} + \omega_{\theta}^{2} \delta \theta = 0.
\end{eqnarray}
Here, $\omega_{r} = 2\pi f_r$ and $\omega_{\theta} = 2\pi f_{\theta}$, where $f_r$ and $f_{\theta}$ represent the radial and vertical epicyclic frequencies, respectively. These equations describe two uncoupled harmonic oscillators, lacking any external forcing term. However, to model more realistic scenarios, we must introduce non-linear terms arising from pressure and dissipative effects in the accretion fluid. This leads to coupled equations with forcing terms, expressed as:
\begin{eqnarray}
\delta \ddot{r} + \omega_{r}^{2} \delta r = \omega_{r}^{2} F_r(\delta r, \delta \theta, \delta \dot{r}, \delta \dot{\theta})\\
\delta \ddot{\theta} + \omega_{\theta}^{2} \delta \theta = \omega_{\theta}^{2} F_{\theta}(\delta r, \delta \theta, \delta \dot{r}, \delta \dot{\theta}).
\end{eqnarray}
In this framework, $F_r$ and $F_{\theta}$ are the forcing terms, which are non-linear functions of their arguments. The specific forms of $F_r$ and $F_{\theta}$ are determined by the accretion flow model, although it is challenging to identify their exact forms. In practice, analytical approximations are often used based on relevant physical conditions \cite{Abramowicz:2003xy, Horak:2004hm}. Below, we discuss various resonant models, each assuming different plausible forms for the forcing terms.

\subsubsection{Parametric Resonance Model}
In the Parametric Resonance Model (PRM), it is assumed that radial epicyclic oscillations drive vertical epicyclic oscillations such that $\delta r \gg \delta \theta$ \cite{torok2005orbital, kluzniak2002parametric, Abramowicz:2003xy, Rebusco:2004ba, Abramowicz:2001bi}. Under this assumption, the equations simplify to:
\begin{eqnarray}
\delta \ddot{r} + \omega_r^2 \delta r = 0 \\
\delta \ddot{\theta} + \omega_{\theta}^{2} \delta \theta = -\omega_{\theta}^{2} \delta r \delta \theta.
\end{eqnarray}
In these equations, $F_r = 0$ and $F_{\theta} = \delta r \delta \theta$. Parametric resonance is achieved when the ratio of radial to vertical frequencies satisfies $\frac{f_r}{f_{\theta}} = \frac{2}{n}$, where $n$ is a positive integer. Resonance effects are the strongest for the smallest values of $n$. Since for Simpson-Visser black holes $f_{\theta} > f_r$, the smallest allowable value of $n$ is 3, naturally leading to the 3:2 frequency ratio observed in the HFQPOs. In this model, we take $f_1 = f_{\theta}$ and $f_2 = f_r$ to represent the HFQPO frequencies.
\begin{figure}[htbp!]
  \begin{subfigure}{0.5\textwidth}
    \centering
    \includegraphics[width=\linewidth]{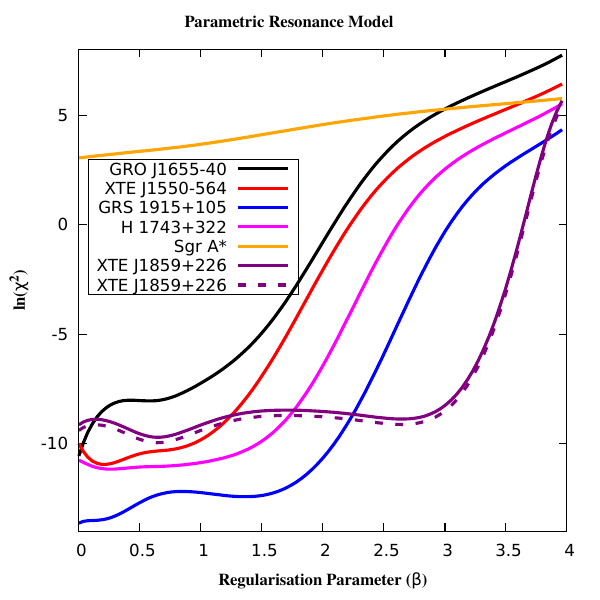}
    \caption{}
    \label{PRM1}
  \end{subfigure}
  \begin{subfigure}{0.5\textwidth}
    \centering
    \includegraphics[width=\linewidth]{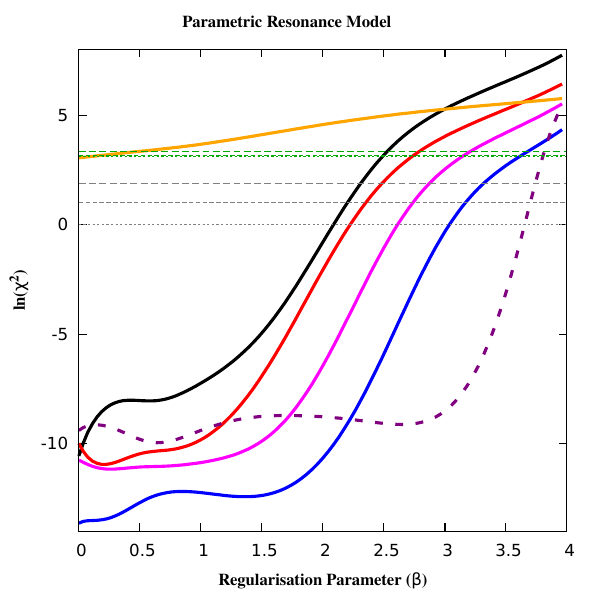}
    \caption{}
    \label{PRM2}
  \end{subfigure}
  \caption{The above figure (a) demonstrates the variation of $\ln \chi^{2}$ with the regularisation parameter $\beta$ for the six black holes sources shown in \ref{t1} assuming PRM. Figure (b) plots the variation of $\ln\chi^{2}$ with $\beta$ (assuming the same model) along with the confidence lines. The grey dotted line corresponds to the 1-$\sigma$ contour, the grey short dashed line corresponds to the 2-$\sigma$ contour and the grey long dashed line is associated with the 3-$\sigma$ contour for GRO J1655-40, XTE J1550-564, GRS 1915+105, H 1743+322, XTE J1859+226. The green dotted, short dashed and long dashed lines are the 1-$\sigma$, 2-$\sigma$ and 3-$\sigma$ contours respectively, corresponding to the source Sgr A*.}
  \label{PRM}
\end{figure}

Now, we discuss the results obtained when the HFQPOs observed in the six BH sources are compared with the PRM. 
In \hyperref[PRM1]{\ref*{PRM1}}, we present the plots of $\ln \chi^2$ (given in \ref{chi}) as a function of the regularisation parameter $\beta$ for each of the six black holes listed in \ref{t1}. 
Taking the logarithmic value of the error enables us to clearly display the variations in $\chi^2$ across all the black holes within a single frame, despite the wide range of values of $\chi^2$ each black hole exhibits. Note that, for the source XTE J1859+226, the positive and negative errors in $f_{up1}$ and $f_{up2}$ are not the same (\ref{t1}). Hence, we present two $\chi^2$ versus $\beta$ plots for this source in \ref{PRM1}, such that the solid purple curve gives the highest possible $\chi^2$ by considering the minimum of the positive and the negative errors (in our case, +$\Delta f_{up1}$ and +$\Delta f_{up2}$ in \ref{t1}) while the dashed purple curve gives the lowest possible $\chi^2$ by considering the maximum of the positive and the negative errors (in our case, -$\Delta f_{up1}$ and -$\Delta f_{up2}$ in \ref{t1}).

For GRO J1655-40 (the black curve in \hyperref[PRM]{\ref*{PRM}}), XTE J1550-564 (the red curve in  \hyperref[PRM]{\ref*{PRM}}), GRS 1915+105 (the blue curve in  \hyperref[PRM]{\ref*{PRM}}), H 1743+322 (the magenta curve in  \hyperref[PRM]{\ref*{PRM}}) and XTE J1859+226 (the purple curve in  \hyperref[PRM]{\ref*{PRM}}), the $\chi^2$ minimizes at $\beta\sim0$, $\beta\sim0.2$, $\beta\sim0$, $\beta\sim0.2$, $\beta\sim0.6$ respectively. However, the magnitude of $\chi^2_{min}\approx 0$ for all of them. Therefore, the $\Delta\chi^2$ values from $\chi^2_{{min}}$ corresponding to the 1-$\sigma$, 2-$\sigma$, and 3-$\sigma$ confidence levels are 1, 2.71, and 6.63, respectively for all the aforesaid five BHs \cite{avni1976energy}. These are represented by the grey lines in  \hyperref[PRM2]{\ref*{PRM2}}, where the dotted line indicates the 1-$\sigma$ contour, the short-dashed line denotes the 2-$\sigma$ contour, and the long-dashed line marks the 3-$\sigma$ contour. 
From \hyperref[PRM2]{\ref*{PRM2}}, we note that GRO J1655-40 , XTE J1550-564 , GRS 1915+105, H 1743+322 and XTE J859+226 respectively rule out $\beta>2.1$, $\beta>2.3$, $\beta>3$, $\beta>2.5$, $\beta>3.6$ outside $1-\sigma$. Considering the $3-\sigma$ interval GRO J1655-40 , XTE J1550-564 , GRS 1915+105, H 1743+322, and XTE J1859+226 respectively rule out $\beta>2.3$, $\beta>2.5$, $\beta>3.3$, $\beta>2.9$ and $\beta>3.7$. The strongest constrain on the regularisation parameter ($\beta$) is established by the HFQPO data of Sgr A* (the orange curve in  \hyperref[PRM]{\ref*{PRM}}). From \hyperref[PRM2]{\ref*{PRM2}} one may note that the minimum of $\chi^2$ occurs at $\beta\sim 0$ and $\beta>0.5$ is ruled out outside the $3-\sigma$ interval (denoted by the green long-dashed line). All these BHs include the Kerr scenario within 1-$\sigma$.

\subsubsection{Forced Resonance Model 1 (FRM1)}
\label{subsec:FRM1}
\begin{figure}[t!]
  \begin{subfigure}{0.5\textwidth}
    \centering
    \includegraphics[width=\linewidth]{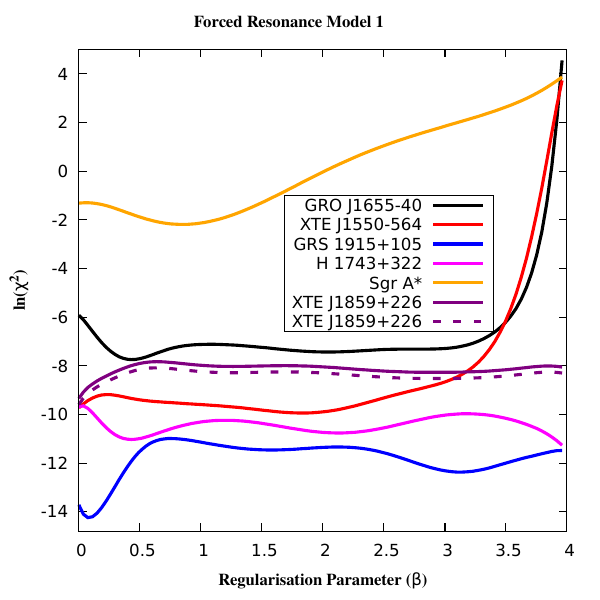}
    \caption{}
    \label{FRM11}
  \end{subfigure}
  \begin{subfigure}{0.5\textwidth}
    \centering
    \includegraphics[width=\linewidth]{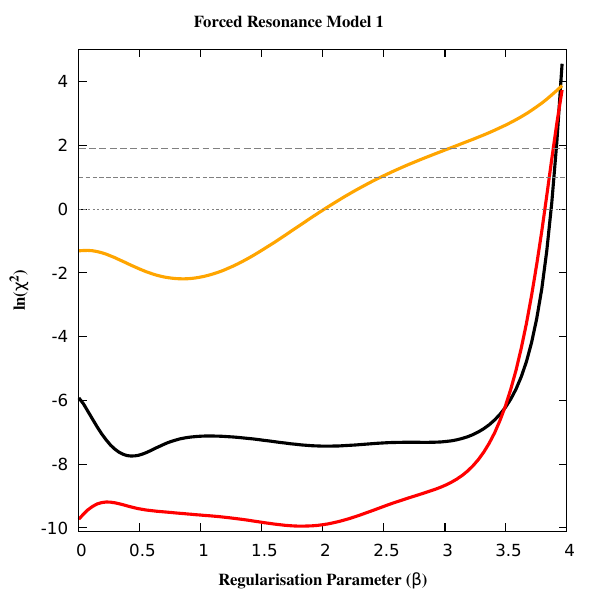}
    \caption{}
    \label{FRM12}
  \end{subfigure}
  \caption{The above figure (a) demonstrates change of $\ln\chi^2$ with regularisation parameter $\beta$ for individual black holes in \ref{t1} assuming FRM1, figure (b) plots the variation of $\ln\chi^{2}$ with regularisation parameter $\beta$ (assuming the same model) but for a subset of the six black holes where $\chi^2$  values are large and the variation with $\beta$ is also substantial such that the confidence lines can be drawn. The grey dotted line corresponds to the 1-$\sigma$ contour, the grey short-dashed line corresponds to the 2-$\sigma$ contour and the grey long-dashed line is associated with the 3-$\sigma$ contour. The confidence contour lines are the same for all the three BHs as $\chi^2_{min}\sim0$ for all of them.}
  \label{bbb}
\end{figure}

In the case where the accretion flow can be approximated by a nearly Keplerian disk, 
parametric resonance provides a natural explanation for the observed HFQPOs 
\cite{Kluzniak:2002bb,Abramowicz:2001bi,kluzniak2001strong}. However, when additional forces such as pressure, viscosity, 
and/or magnetic stresses are present, the non-linear couplings between 
$\delta r$ and $\delta \theta$ cannot be neglected alongside parametric resonance effects 
\cite{torok2005orbital}. Numerical simulations by Lee et al. \cite{lee2004resonance} demonstrated that pressure forces 
frequently induce resonant driving of vertical oscillations by radial oscillations 
\cite{Abramowicz:2001bi,lee2004resonance}, consistent with earlier results reported by Abramowicz \& Kluźniak \cite{Abramowicz:2001bi}. Under these circumstances, the perturbation equations are given by,
\begin{align}
\delta \ddot{r}&=-\omega_r^2 \delta r \\
\delta\ddot{\theta} + \omega_{\theta}^2 \delta \theta & = -\omega_{\theta}^2 \delta r \delta \theta + F_{\theta}(\delta \theta),
\end{align}
where $\omega_{\theta} = 2 \pi f_{\theta}$ represents the vertical epicyclic frequency, and $F_{\theta}$ is a forcing term that depends on the dynamics of the accretion disk. The solution to this equation can yield the following resonance condition:
\begin{eqnarray}
\frac{f_{\theta}}{f_r} = \frac{m}{n},
\end{eqnarray}
where $m$ and $n$ are natural numbers.
In FRM1, we take $m = 3$ and $n = 1$ which corresponds to a 3:1 frequency ratio. The upper HFQPO is then associated with the vertical epicyclic frequency $f_{\theta}$ and the lower HFQPO is given by $f_{\theta} - f_r$. Thus, we have:
\begin{eqnarray}
f_1 = f_{\theta} = 3 f_r \quad {and} \quad f_2 = f_{\theta} - f_r = 2 f_r
\end{eqnarray}
This model provides a natural explanation for the observed 3:2 frequency ratio in HFQPOs by setting $f_1 = f_{\theta}$ and $f_2 = f_{\theta} - f_r$. The resonance is driven by the forcing terms, which arise due to non-linear couplings in the accretion flow, as discussed in \cite{kluzniak2002parametric, abramowicz2001precise, kluzniak2001strong}.

Now we will discuss our result on Forced Resonance Model 1 (FRM1).
In \hyperref[FRM11]{\ref*{FRM11}}, we show the plots of ln$\chi^2$ as a function of the regularisation parameter $\beta$ for each of the six black holes listed in \ref{t1}. The figure shows that for GRS 1915+105, H1743+322, XTE J1859+226, $\chi^2$ minimizes at $\beta\sim0.1$, $\beta\sim0.4$, $\beta\sim0$. This however, cannot be considered as presence or absence of the regularisation parameter $\beta$ as  the $\chi^2$ for these sources is very small and also do not vary substantially such that
the $\Delta\chi^2$ intervals corresponding to 68\%, 90\% and 99\% confidence lines can rule out certain range of the allowed parameter space of $\beta$. For the other three black holes, the variation of $\chi^2$ is reasonable because GRO J1655-40, XTE J1550-564, Sgr A* rule out $\beta>3.8$, $\beta>3.7$, $\beta>2$ respectively outside $1-\sigma$ \hyperref[FRM12]{(\ref*{FRM12})}. If we consider $3-\sigma$ confidence interval, GRO J1655-40, XTE J1550-564, Sgr A* rule out $\beta>3.9$, $\beta>3.8$, $\beta>3$ respectively \hyperref[FRM12]{(\ref*{FRM12})}. 
The Kerr scenario is however allowed within 1-$\sigma$ for these three BHs.
However, for all these three BHs the $\chi^2$ minimizes at a non zero value of the regularisation parameter. For GRO J1655-40, XTE J1550-564 and Sgr A*, the $\chi^2$ minimizes at $\beta\sim0.4$, $\beta\sim1.8$ and $\beta\sim0.9$ respectively.

\subsubsection{Forced Resonance Model 2 (FRM2)}
\label{subsec:FRM2}
\begin{figure}[H]
  \begin{subfigure}{0.5\textwidth}
    \centering
    \includegraphics[width=\linewidth]{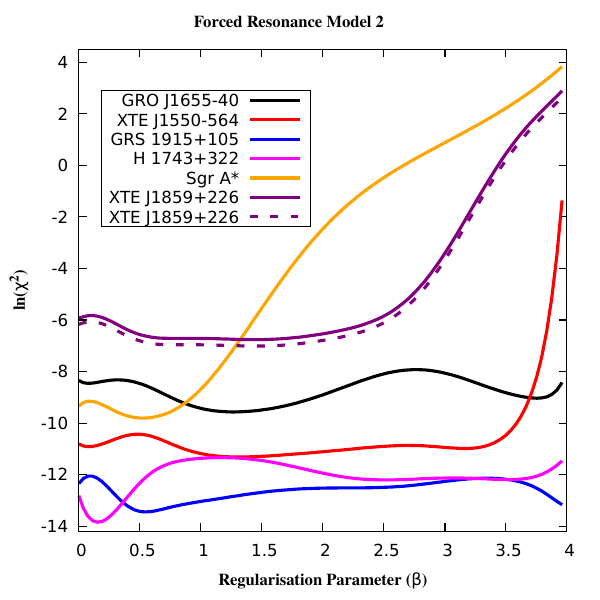}
    \caption{}
    \label{FRM21}
  \end{subfigure}
  \begin{subfigure}{0.5\textwidth}
    \centering
    \includegraphics[width=\linewidth]{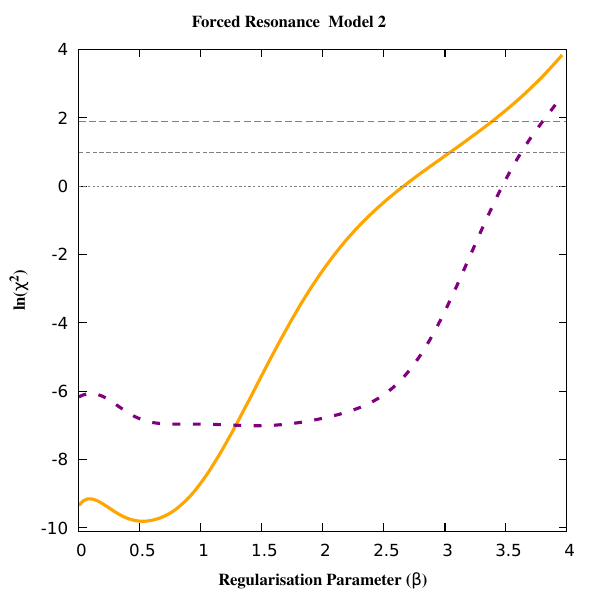}
    \caption{}
    \label{FRM22}
  \end{subfigure}
  \caption{The above figure (a) demonstrates change of $\ln\chi^2$ with regularisation parameter $\beta$ for individual black holes in \ref{t1} assuming FRM2, figure (b) plots the variation of $\ln\chi^{2}$ with regularisation parameter $\beta$ (assuming the same model) but for a subset of the  six black holes where $\chi^2$  values are large and the variation with $\beta$ is also substantial such that the confidence lines can be drawn. The grey dotted line corresponds to the 1-$\sigma$ contour, the grey short-dashed line corresponds to the 2-$\sigma$ contour and the grey long-dashed line is associated with the 3-$\sigma$ contour. The confidence contour lines are the same for all the two BHs as $\chi^2_{min}\sim0$ for all of them.}
  \label{ccc}
\end{figure}

In the Forced Resonance Model 2 (FRM2), a 2:1 frequency ratio is considered between $f_\theta$ and $f_r$ \cite{Kluzniak:2002bb,Abramowicz:2001bi,kluzniak2001strong,kluzniak2002parametric}. Here, the differential equation for $\delta\theta$ is still described by the coupled motion with the forcing terms as in FRM1:

\begin{eqnarray}
\delta\ddot{\theta} + \omega_{\theta}^2 \delta \theta = -\omega_{\theta}^2 \delta r \delta \theta + F_{\theta}(\delta \theta).
\end{eqnarray}
However, the resonance condition is modified so that the frequency ratio is 2:1, where $f_{\theta}$ and $f_r$ satisfy:
\begin{eqnarray}
\frac{f_{\theta}}{f_r} = \frac{m}{n} = 2.
\end{eqnarray}
In this model, we take the upper HFQPO $f_1$ as $f_{\theta} + f_r$ and the lower HFQPO $f_2$ as $f_{\theta}$, leading to:
\begin{eqnarray}
f_1 = f_{\theta} + f_r = 3 f_r \quad {and} \quad f_2 = f_{\theta} = 2 f_r.
\end{eqnarray}
Thus in FRM2 as well, the 3:2 frequency ratio observed in HFQPOs is achieved between  $f_1 = f_{\theta} + f_r$ and $f_2 = f_{\theta}$. Both FRM1 and FRM2 leverage the non-linear forcing terms, which allow for combinations of frequencies that explain the resonant behavior seen in HFQPO observations.\nonumber\\

Now, we will discuss our results for Forced Resonance Model 2.
In \hyperref[FRM21]{\ref*{FRM21}}, we show the plots of $ln \chi^2$ as a function of the regularisation parameter $\beta$ for each of the six black holes listed in \ref{t1}. The figure shows that for GRO J1655-40, XTE J1550-564, GRS 1915+105 and H1743+322,  $\chi^2$ minimizes at $\beta\sim1.25$, $\beta\sim1.25$, $\beta\sim0.5$, $\beta\sim0.2$. This however, cannot be considered as a confirmatory deviation from the Kerr scenario as  the $\chi^2$ for these sources do not vary substantially such that the $\Delta\chi^2$ intervals corresponding to 68\%, 90\% and 99\% confidence lines can rule out certain range of the allowed parameter space of $\beta$. For the other two black holes, the variation of $\chi^2$ is reasonable because XTE J1859+226 and Sgr A* rule out $\beta>3.4$ and $\beta>2.5$ respectively outside $1-\sigma$ \hyperref[FRM22]{(\ref*{FRM22})}. If we consider $3-\sigma$ confidence interval, XTE J1859+226 and Sgr A* rule out $\beta>3.8$ and $\beta>3.4$ respectively \hyperref[FRM22]{(\ref*{FRM22})}. For all these two black holes, $\chi^2$ minimizes at a non-zero value of the regularisation parameter which turns out to be $\beta\sim0.6$ for both XTE J1859+226 and Sgr A*. The Kerr scenario is however allowed within 1-$\sigma$.

\subsubsection{Keplerian Resonance Model 1 (KRM1)}
\label{subsec:KRM1}
\begin{figure}[htbp]
    \centering
    \begin{subfigure}{0.3\textwidth}
        \includegraphics[width=\linewidth]{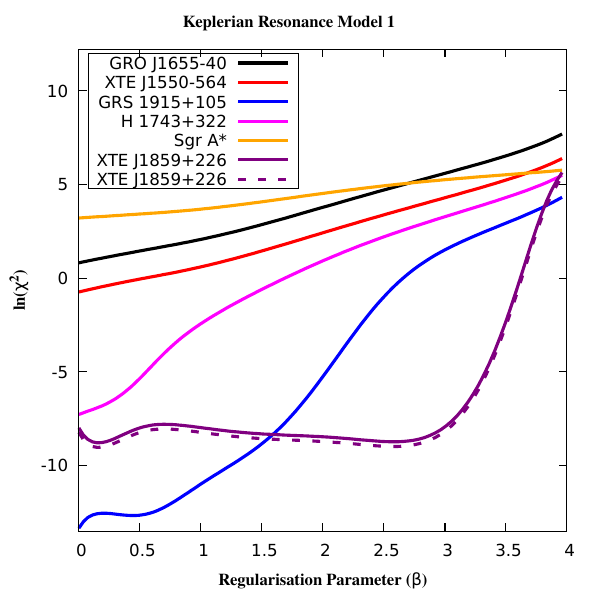}
        \caption{}
        \label{KRM11}
    \end{subfigure}
    \hspace{0.02\textwidth} % Adjust the horizontal space between figures
    \begin{subfigure}{0.3\textwidth}
        \includegraphics[width=\linewidth]{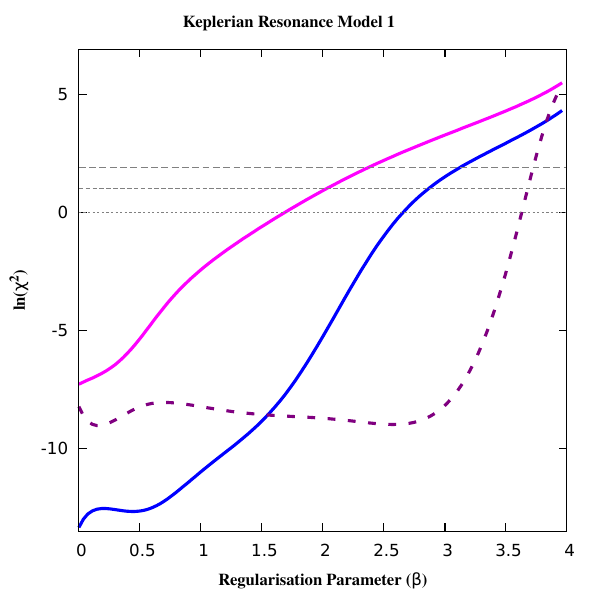}
        \caption{}
        \label{KRM12}
    \end{subfigure}
    \hspace{0.02\textwidth} % Adjust the horizontal space between figures
    \begin{subfigure}{0.3\textwidth}
        \includegraphics[width=\linewidth]{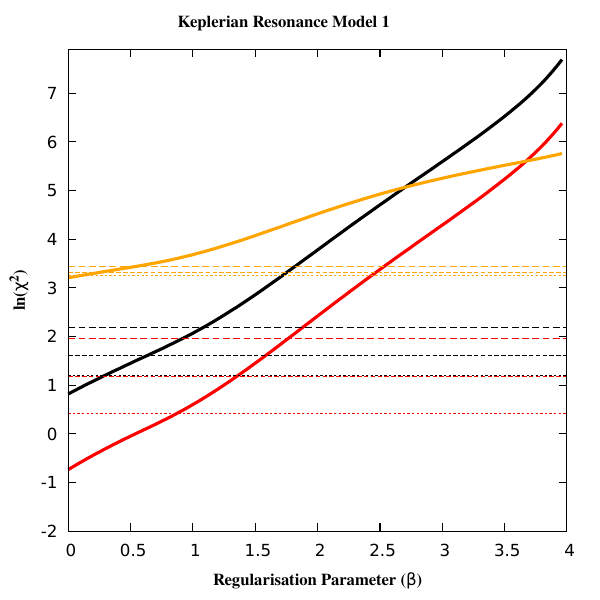}
        \caption{}
        \label{KRM13}
    \end{subfigure}
    \caption{The above figure (a) demonstrates the variation of $\ln\chi^{2}$ with the regularisation parameter $\beta$ for the six black holes sources shown in \ref{t1} assuming KRM1. Figure (b) plots the variation of $\ln\chi^{2}$ with $\beta$ for the sources GRS 1915+105, H 1743+322 and XTE J1859+226, along with the confidence lines. The grey dotted line corresponds to the 1-$\sigma$ contour, the grey short-dashed line corresponds to the 2-$\sigma$ contour and the grey long-dashed line is associated with the 3-$\sigma$ contour for  GRS 1915+105, H 1743+322 and XTE J1859+226. Figure (c) also plots the variation of $\ln\chi^{2}$ with $\beta$ for the sources GRO J1655-40, XTE J1550-564 and Sgr A* along with the confidence lines. The black dotted, short-dashed and long-dashed lines are the 1-$\sigma$, 2-$\sigma$ and 3-$\sigma$ contour respectively corresponding to the source GRO J1655-40. The red dotted, short-dashed and long-dashed lines are the 1-$\sigma$, 2-$\sigma$ and 3-$\sigma$ contours respectively corresponding to the source XTE J1550-564. The orange dotted, short-dashed and long-dashed lines are the 1-$\sigma$, 2-$\sigma$ and 3-$\sigma$ contours respectively corresponding to the source Sgr A*.}
    \label{fig:overall}
\end{figure}
In the Keplerian Resonance Models, we explore non-linear resonances involving the radial epicyclic frequency and the orbital angular frequency\cite{nowak199767,Abramowicz:2001bi,torok2005orbital,kato2001trapping}. Unlike the Parametric and Forced Resonance Models, where radial and vertical epicyclic frequencies are coupled, the Keplerian Resonance Models involve interactions between the radial epicyclic modes and the orbital angular motion. These resonances may occur in the inner region of relativistic thin accretion disks, potentially due to the trapping of non-axisymmetric g-mode oscillations induced by a corotation resonance.

However, it was observed that corotation resonance tends to dampen g-mode oscillations rather than excite them\cite{2001PASJ...53L..37K,Li:2002yi,kato2003damping}, which may limit the effectiveness of Keplerian resonance models in explaining HFQPOs in microquasars. Another scenario where Keplerian resonance might occur involves pairs of spatially separated, coherent vortices with opposite vorticities. These vortices oscillate with radial epicyclic frequencies and couple with the spatially varying orbital angular frequency \cite{torok2005orbital,abramowicz1998theory}. 

In the Keplerian Resonance Model 1 (KRM1)\cite{nowak199767,torok2005orbital,abramowicz1998theory}, the resonance occurs between the orbital angular frequency $f_{\phi}$ and the radial epicyclic frequency $f_r$. Here, we identify the upper and lower HFQPO frequencies $f_1$ and $f_2$ as follows:
\begin{eqnarray}
f_1 = f_{\phi}, \quad f_2 = f_r.
\end{eqnarray}
Such a resonance which may be significant in the inner regions of the accretion disk can potentially explain the HFQPOs, but it is sensitive to the detailed structure of the accretion flow and the specific disk parameters.

Now, we discuss our results on the Keplerian Resonance Model 1.
 We show the plots of $\ln \chi^2$ as a function of the regularisation parameter $\beta$ for each of the six black holes listed in  \ref{t1} in  \hyperref[KRM11]{\ref*{KRM11}}. Interestingly, this is the only model which is able to constrain $\beta$ from the HFQPO data of all the six black hole sources. In \hyperref[KRM12]{\ref*{KRM12}}, illustrated for GRS 1915+105, H 1743+322 and XTE J1859+226, $\chi^2$ minimizes at $\beta\sim0$, $\beta\sim0$ and $\beta\sim0.2$ respectively, but the value of minimum $\chi^2$ is nearly zero. Therefore, $\Delta\chi^2$ values from minimum $\chi^2$ corresponding to $1-\sigma$, $2-\sigma$ and $3-\sigma$ contours are 1, 2.71 and 6.63 respectively. These $1-\sigma$, $2-\sigma$ and $3-\sigma$ contours are shown by grey dotted, grey short-dashed and grey long-dashed lines. GRS 1915+105, H 1743+322 and XTE J1859+226 rule out $\beta>2.6$, $\beta>1.5$ and $\beta>3.6$ respectively outside $1-\sigma$ \hyperref[KRM12]{(\ref*{KRM12})}. If we consider the $3-\sigma$ confidence interval, these three sources rule out $\beta>3.1$ (for GRS 1915+105), $\beta>2.5$ (for H 1743+322) and $\beta>3.7$ (for XTE J1859+226) respectively \hyperref[KRM12]{(\ref*{KRM12})}.\\
 In \hyperref[KRM13]{\ref*{KRM13}}, we present the constraints on $\beta$ from the QPO data of GRO J1655-40, XTE J1550-564 and Sgr A*. For all these three black holes $\chi^2$ minimizes at $\beta\sim0$. The $1-\sigma$, $2-\sigma$ and $3-\sigma$ intervals for GRO J1655-40 are shown by black dotted, black short-dashed and black long-dashed lines. For $1-\sigma$ interval, it rules out $\beta>0.3$ while for $3-\sigma$ interval, it rules out $\beta>1.1$.  The $1-\sigma$, $2-\sigma$ and $3-\sigma$ contours for XTE J1550-564 are shown by red dotted, red short-dashed and red long-dashed lines. Within $1-\sigma$ interval, it rules out $\beta>0.9$ while for $3-\sigma$ interval, it rules out $\beta>1.7$.  The $1-\sigma$, $2-\sigma$ and $3-\sigma$ contours for Sgr A* are shown by orange dotted, orange short-dashed and orange long-dashed lines. For Sgr A*, $\beta>0.2$ is ruled out outside $1-\sigma$ interval while $\beta>0.6$ 
 is ruled out outside $3-\sigma$. Except for XTE J1859+226, the remaining five BHs favour the Kerr scenario.  XTE J1859+226 shows a mild preference towards $\beta\sim 0.2$, although GR is allowed within 1-$\sigma$.

\subsubsection{Keplerian Resonance Model 2 (KRM2)}
\label{subsec:KRM2}
Another variant of the Keplerian Resonance Model (KRM2)\cite{nowak199767,torok2005orbital,abramowicz1998theory} considers a resonance between the orbital angular frequency $f_{\phi}$ and twice the radial epicyclic frequency $f_r$ such that $f_1 = f_{\phi}, \quad f_2 = 2f_r$.

The orbital angular frequency and the radial epicyclic frequency may lead to resonant interactions that amplify oscillations in the accretion flow, potentially aligning with the HFQPO frequency ratios observed in certain black hole systems. This configuration depends on the strength of the non-linear coupling and the dynamical conditions in the accretion disk\cite{torok2005orbital,abramowicz1998theory}.\nonumber\\
\begin{figure}[htbp!]
  \begin{subfigure}{0.5\textwidth}
    \centering
    \includegraphics[width=\linewidth]{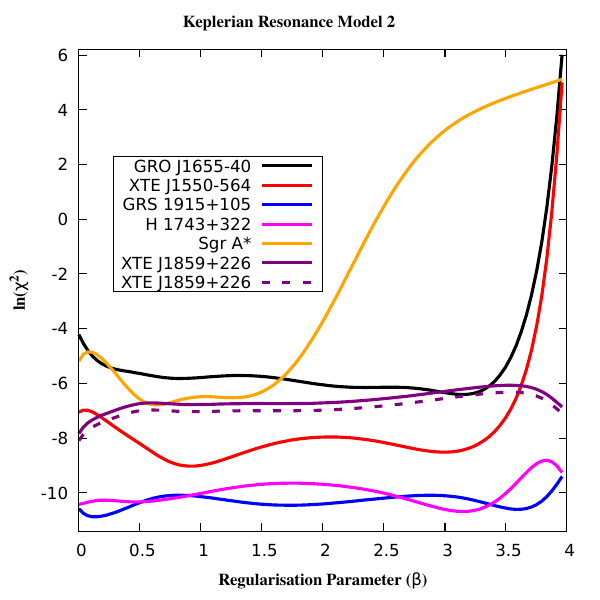}
    \caption{}
    \label{KRM21}
  \end{subfigure}
  \begin{subfigure}{0.5\textwidth}
    \centering
    \includegraphics[width=\linewidth]{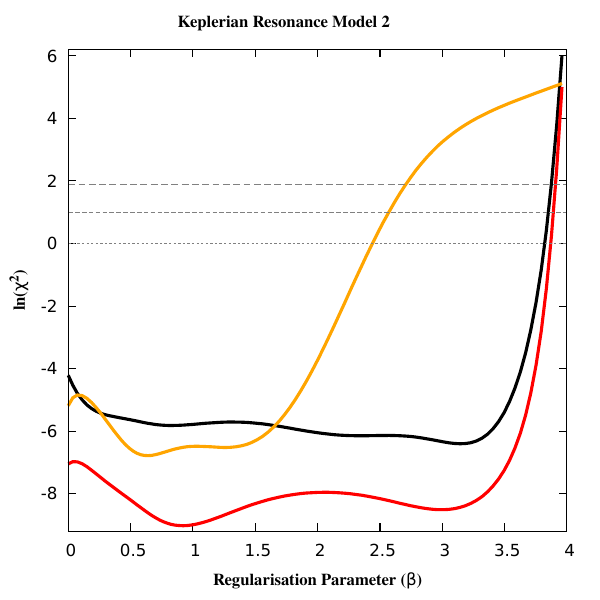}
    \caption{}
    \label{KRM22}
  \end{subfigure}
  \caption{The above figure (a) demonstrates the change of $\ln\chi^2$ with the regularisation parameter $\beta$ for individual black holes in \ref{t1} assuming KRM2, figure (b) plots the variation of $\ln\chi^{2}$ with the regularisation parameter $\beta$ (assuming the same model) but for a subset of the six black holes where $\chi^2$  values are large and the variation with $\beta$ is also substantial such that the confidence lines can be drawn. The grey dotted line corresponds to the 1-$\sigma$ contour, the grey short-dashed line corresponds to the 2-$\sigma$ contour and the grey long-dashed line is associated with the 3-$\sigma$ contour. The confidence contour lines are the same for all the three BHs as $\chi^2_{min}\sim0$ for all of them.}
  \label{eee}
\end{figure}
%Now, we will discuss our result on Keplerian Resonance Model 2\nonumber\\FRM1
In \hyperref[KRM21]{\ref*{KRM21}}, we show the plots of $\ln \chi^2$ as a function of the regularisation parameter $\beta$ for each of the six black holes listed in \ref{t1}. The figure shows that for GRS 1915+105, H1743+322 and XTE J1859+226, $\chi^2$ minimizes at $\beta\sim0.1$, $\beta\sim3.3$ and $\beta\sim0$. This however, cannot be considered as presence or absence of regularisation parameter $\beta$ as  the $\chi^2$ for these sources do not vary substantially such that the $\Delta\chi^2$ intervals corresponding to 68\%, 90\% and 99\% confidence lines can rule out certain range of the allowed parameter space of $\beta$. For the other three black holes, the variation of $\chi^2$ is reasonable because GRO J1655-40, XTE J1550-564 and Sgr A* rule out $\beta>3.7$, $\beta>3.8$ and $\beta>2.5$ respectively outside $1-\sigma$ \hyperref[KRM22]{(\ref*{KRM22})}. If we consider the $3-\sigma$ confidence interval, GRO J1655-40, XTE J1550-564 and Sgr A* rule out $\beta>3.8$, $\beta>3.9$, $\beta>2.7$ respectively \hyperref[KRM22]{(\ref*{KRM22})}. For all these three black holes $\chi^2$ minimizes at a non-zero value of the regularisation parameter although GR is allowed within 1-$\sigma$.

\subsubsection{Keplerian Resonance Model 3 (KRM3)}
\label{subsec:KRM3}
A third variant of the Keplerian Resonance Model 3 (KRM3)\cite{nowak199767} also exists in the literature where the resonance occurs between three times the radial epicyclic frequency $3f_r$ and the orbital angular frequency $f_{\phi}$ such that $f_1 = 3f_r, \quad f_2 = f_{\phi}.$
The feasibility of this model, however, depends on the specific physical conditions of the disk, as well as the effective coupling between the radial and the orbital modes\cite{torok2005orbital,abramowicz1998theory}.\nonumber\\
\begin{figure}[htbp!]
  \begin{subfigure}{0.5\textwidth}
    \centering
    \includegraphics[width=\linewidth]{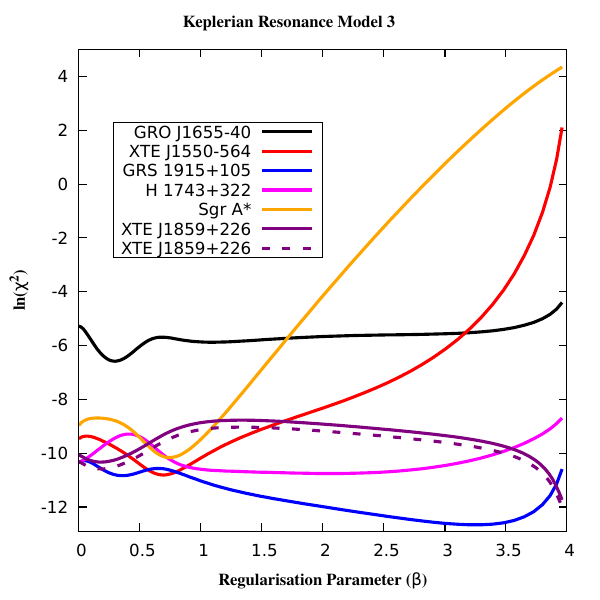}
    \caption{}
    \label{KRM31}
  \end{subfigure}
  \begin{subfigure}{0.5\textwidth}
    \centering
    \includegraphics[width=\linewidth]{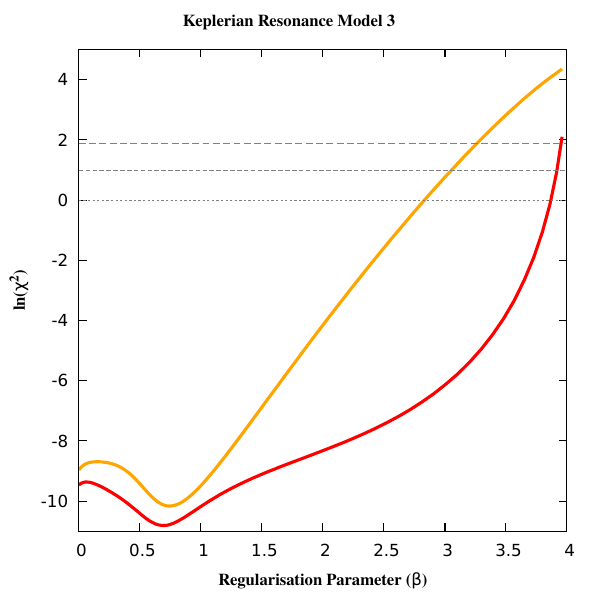}
    \caption{}
    \label{KRM32}
  \end{subfigure}
  \caption{The above figure (a) demonstrates the change of $\ln\chi^2$ with the regularisation parameter $\beta$ for individual black holes in \ref{t1} assuming KRM3, figure (b) plots the variation of $\ln\chi^{2}$ with the regularisation parameter $\beta$ (assuming the same model) but for a subset of the six black holes where $\chi^2$  values are large and the variation with $\beta$ is also substantial such that the confidence lines can be drawn. The grey dotted line corresponds to the 1-$\sigma$ contour, the grey short-dashed line corresponds to the 2-$\sigma$ contour and the grey long-dashed line is associated with the 3-$\sigma$ contour. The confidence contour lines are the same for all the three BHs as $\chi^2_{min}\sim0$ for all of them.}
  \label{fff}
\end{figure}
In \ref{KRM31} we show the plots of $ln\chi^2$ as a function of the regularisation parameter $\beta$ for each of the  six BHs listed in \ref{t1}. The figure shows that for GRO J1655-40, GRS 1915+105, H1743+322 and  XTE J1859+226,  $\chi^2$ minimizes at $\beta\sim 0.25$, $\beta\sim3.3$, $\beta\sim1$, $\beta\sim4$. This however, cannot be considered as presence or absence of regularisation parameter $\beta$ as  the $\chi^2$ for these sources do not vary substantially such that the $\Delta\chi^2$ intervals corresponding to 68\%, 90\% and 99\% confidence lines can rule out certain range of the allowed parameter space of $\beta$. For the other two black holes, the variation of $\chi^2$ is reasonable because XTE J1550-564 and Sgr A* rule out $\beta>3.8$ and $\beta>2.6$ respectively outside $1-\sigma$ \hyperref[KRM32]{(\ref*{KRM32})}. If we consider the $3-\sigma$ confidence interval, Sgr A* rules out $\beta>3.3$ while XTE J1550-564 allows all values of $\beta$ \hyperref[KRM32]{(\ref*{KRM32})}. For all these two black holes, $\chi^2$ minimizes at a non-zero value of the regularisation parameter although GR is allowed within 1-$\sigma$. For XTE J1550-564 and Sgr A*, the $\chi^2$ value minimizes at $\beta\sim 0.7$ and $\beta\sim 0.8$ respectively \hyperref[KRM32]{(\ref*{KRM32})}.

\subsubsection{Warped Disk Oscillation Model}
\label{section:WarpedDisk}
The Warped Disk Oscillation Model explores non-linear resonances between various disk oscillation modes within a relativistic, warped accretion disk \cite{kato2001basic, kato2004wave, kato2004resonant, kato2005vertical, Kato:2007ab}. This model posits that a disk deformed by a warp can exhibit resonant oscillations, driven by interactions between the warp and the oscillatory modes. Due to the relativistic nature of the disk, these resonances arise because the radial epicyclic frequency $f_r$ does not vary monotonically with the radial distance $r$, allowing for resonance at specific locations in the disk .

The model identifies two types of resonances that may play a role in exciting particular oscillation modes: (a) {The Horizontal Resonances} which can stimulate both p-mode and g-mode oscillations in the disk. Horizontal resonances occur when the frequencies of the oscillations align with the radial and azimuthal modes in a way that couples with the warp \cite{kato2004wave} and (b) {The Vertical Resonances} which are primarily responsible for exciting the g-mode oscillations in the disk. Vertical resonances are limited to oscillations perpendicular to the disk plane, leading to distinct behaviors compared to horizontal resonances \cite{kato2004wave}. In this model, the upper HFQPO is denoted by $f_1=2f_\phi-f_r$ and the lower HFQPO is denoted by $f_1=2(f_\phi-f_r)$. 
%These resonant oscillations are formulated by examining the epicyclic frequencies in the warped disk structure. In this setup, the resonant conditions are influenced by the warped geometry and lead to the following relationships for horizontal and vertical resonances:
\iffalse
\begin{eqnarray}
f_{{hor}} &=& f_{\phi} \pm f_r, \\
f_{{vert}} &=& f_{\theta} \pm f_r,
\end{eqnarray}
where $f_{\phi}$, $f_r$, and $f_{\theta}$ represent the azimuthal, radial, and vertical epicyclic frequencies, respectively. Here, $f_{{hor}}$ and $f_{{vert}}$ are the frequencies associated with horizontal and vertical resonances, which vary depending on the nature of the coupling within the warped disk.
\fi
Despite its promising approach to explaining HFQPOs, the Warped Disk Oscillation Model has limitations. One of the main drawback is its reliance on an unusual disk geometry, which may not be typical in astrophysical settings \cite{torok2011confronting, Yagi:2016jml}. Consequently, while the warped disk configuration provides a framework for understanding HFQPOs through resonant interactions, the physical applicability of such a configuration remains under investigation.\nonumber\\
\begin{figure}[htbp!]
  \begin{subfigure}{0.5\textwidth}
    \centering
    \includegraphics[width=\linewidth]{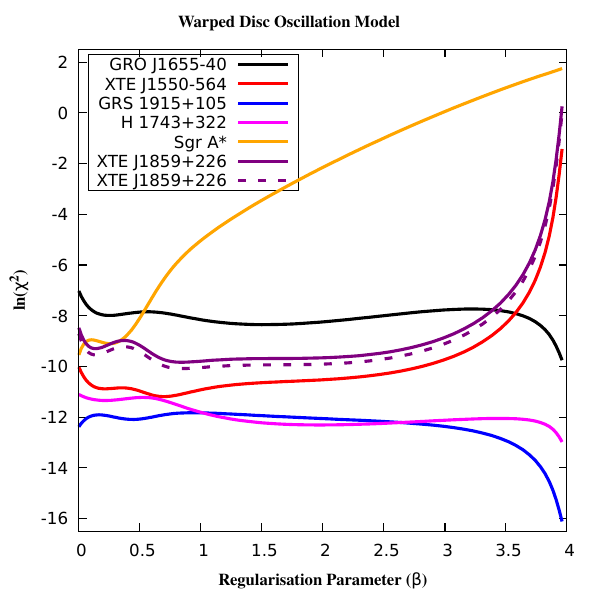}
    \caption{}
    \label{WDOM1}
  \end{subfigure}
  \begin{subfigure}{0.5\textwidth}
    \centering
    \includegraphics[width=\linewidth]{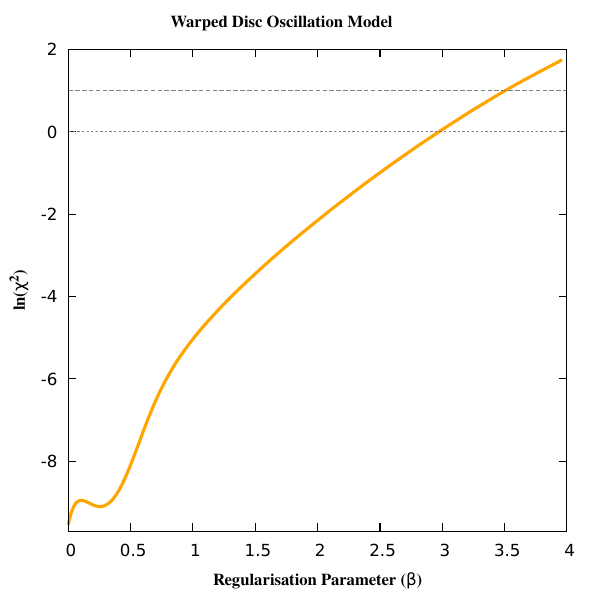}
    \caption{}
    \label{WDOM2}
  \end{subfigure}
  \caption{The above figure (a) demonstrates the variation of $\ln \chi^2$ with the regularisation parameter $\beta$ for individual black holes in \ref{t1} assuming WDOM, figure (b) plots the variation of $\ln \chi^{2}$ with the regularisation parameter $\beta$ assuming the same model but for a subset of the six black holes where $\chi^2$  values are large and the variation with $\beta$ is also substantial such that the confidence lines can be drawn. The grey dotted line corresponds to the 1-$\sigma$ contour, the grey short-dashed line corresponds to the 2-$\sigma$ contour.}
  \label{ggg}
\end{figure}
%Now, we will discuss our result Warped Disc Oscillation Model\nonumber\\
In \hyperref[WDOM1]{\ref*{WDOM1}} we show the plots of $\ln \chi^2$ as a function of the regularisation parameter $\beta$ for each of the six black holes listed in \ref{t1}. The figure shows that for GRO J1655-40, XTE J1550-564, GRS 1915+105, H1743+322,  XTE J1859+226,  $\chi^2$ minimizes at $\beta\sim3.99$, $\beta\sim0.7$, $\beta\sim3.99$, $\beta\sim1.5$,  $\beta\sim0.7$. This however, cannot be considered as a confirmatory signature of the Simpson-Visser BH as  the $\chi^2$ for these sources do not vary substantially such that the $\Delta\chi^2$ intervals corresponding to 68\%, 90\% and 99\% confidence lines can rule out certain range of the allowed parameter space of $\beta$. This model can constrain $\beta$ only for the supermassive black hole Sgr A*. For Sgr A*, the variation of $\chi^2$ is reasonable such that $\beta>3$ is ruled out outside $1-\sigma$ and $\beta>3.5$ is ruled out outside $2-\sigma$ \hyperref[WDOM2]{(\ref*{WDOM2})} and this BH favors the Kerr scenario compared to the SV scenario.

\subsubsection{Non-Axisymmetric Disk-Oscillation Model (NADO1)}
\label{subsec:NADO1}
Non-axisymmetric disk-oscillation models propose that HFQPOs may originate from various combinations of non-axisymmetric oscillation modes in the accretion disk \cite{Kotrlova:2020pqy,Torok:2015tpu,Sramkova:2015bha}. These models are closely related to the Relativistic Precession Model (RPM), as they consider oscillation frequencies that align with RPM predictions for slowly rotating black holes \cite{Torok:2011qy}. In these models, the accretion flow is approximated as a slightly non-slender, pressure-supported perfect fluid torus, allowing the incorporation of non-geodesic effects arising from the pressure forces in the disk.

The first variant, referred to as NADO1 or the Vertical Precession Resonance Model \cite{bursa2004upper,Bursa:2005th}, assumes a resonance between the $m = -1$ non-axisymmetric radial epicyclic frequency $f_{{per}} = f_{\phi} - f_{r}$ and the vertical epicyclic frequency $f_{\theta}$, where $m$ is the azimuthal wave number of the non-axisymmetric perturbation . The resonant frequencies in this model are defined as:
\begin{eqnarray}
f_1 = f_{\theta}, \quad f_2 = f_{{per}} = f_{\phi} - f_{r}.
\end{eqnarray}
This model yields results consistent with the Continuum Fitting method for the spin of the black hole GRO J1655-40, providing an alternative to the predictions of the Relativistic Precession Model in Kerr geometry. Using RPM, the spin of GRO J1655-40 turns out to be lower than that predicted by the Continuum Fitting or the Fe-line method \cite{Motta:2013wga,shafee2005estimating,Miller:2009cw}
\nonumber\\
\begin{figure}[htbp!]
  \begin{subfigure}{0.5\textwidth}
    \centering
    \includegraphics[width=\linewidth]{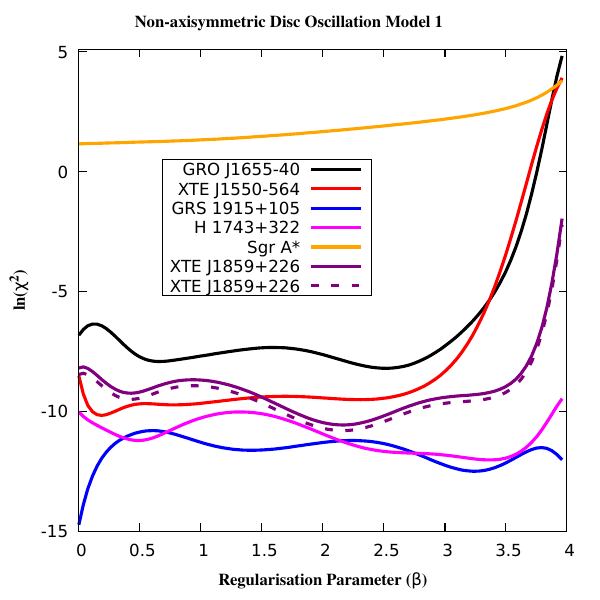}
    \caption{}
    \label{NADO11}
  \end{subfigure}
  \begin{subfigure}{0.5\textwidth}
    \centering
    \includegraphics[width=\linewidth]{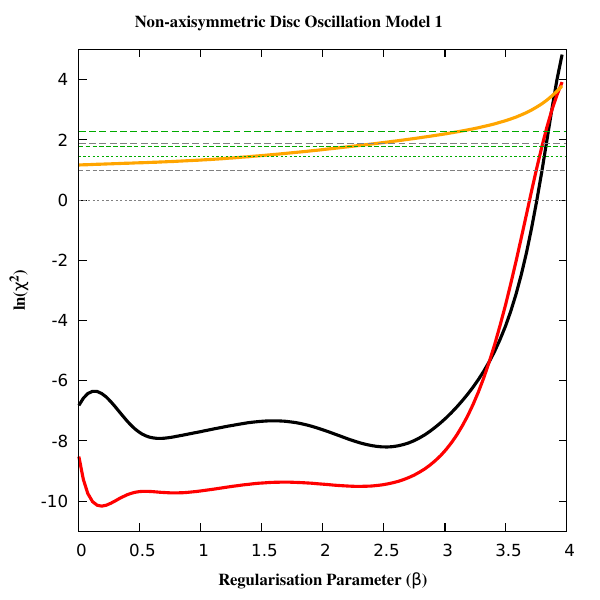}
    \caption{}
    \label{NADO12}
  \end{subfigure}
  \caption{The above figure (a) demonstrates the variation of $\ln \chi^{2}$ with the regularisation parameter $\beta$ for the six black holes sources shown in \ref{t1} assuming NADO1. Figure (b) plots the variation of $\ln \chi^{2}$ with $\beta$ (assuming the same model) along with the confidence lines. The grey dotted line corresponds to the 1-$\sigma$ contour, the grey short-dashed line corresponds to the 2-$\sigma$ contour and the grey long-dashed line is associated with the 3-$\sigma$ contour for GRO J1655-40 and XTE J1550-564. The green dotted, short dashed and long dashed lines are the 1-$\sigma$, 2-$\sigma$ and 3-$\sigma$ contours respectively corresponding to the source Sgr A*.}
  \label{hhh}
\end{figure}
%Now, we will discuss our result on Non-Axisymmetric Disc Oscillation Model

In  \hyperref[NADO11]{\ref*{NADO11}} we show the plots of $ln\chi^2$ as a function of the regularisation parameter $\beta$ for each of the six black holes listed in \ref{t1} . The figure shows that for GRS 1915+105, H1743+322 and XTE J1859+226,  $\chi^2$ minimizes at $\beta\sim0$, $\beta\sim3.5$, $\beta\sim2$ respectively. This however, cannot be considered as presence or absence of regularisation parameter $\beta$ as  the $\chi^2$ for these sources do not vary substantially such that the $\Delta\chi^2$ intervals corresponding to 68\%, 90\% and 99\% confidence lines can rule out certain range of the allowed parameter space of $\beta$. For GRO J1655-40 and XTE J1550-564, the magnitude of minimum $\chi^2$ is nearly 0, so their $1-\sigma$, $2-\sigma$ and $3-\sigma$ confidence intervals are shown by grey dotted, grey short-dashed, grey long-dashed lines. GRO J1655-40 rules out $\beta>3.7$ outside  $1-\sigma$ interval and $\beta>3.8$ outside $3-\sigma$ interval. For this black hole, $\chi^2$ minimizes at $\beta\sim2.6$ but GR is allowed within 1-$\sigma$.  XTE J1550-564 rules out $\beta>3.5$ outside  $1-\sigma$ interval and $\beta>3.7$ outside $3-\sigma$ interval. For this black hole, $\chi^2$ minimizes at $\beta\sim0.2$ \hyperref[NADO12]{(\ref*{NADO12})}. Both these BHs include the Kerr scenario within the 1-$\sigma$ interval.
 For Sgr A*,  $1-\sigma$, $2-\sigma$ and $3-\sigma$ confidence interval are shown by green dotted, green short-dashed and green long-dashed lines. Sgr A* rules out  $\beta>1.5$ outside  $1-\sigma$ interval and $\beta>3.2$ outside $3-\sigma$ interval. For this black hole, the $\chi^2$ minimizes at $\beta\sim0$ \hyperref[NADO12]{(\ref*{NADO12})}.\\

\subsubsection{Non-Axisymmetric Disk-Oscillation Model (NADO2)}
\label{subsec:NADO2}

The second variant, referred to as NADO2 involves a resonance between the $m = -1$ non-axisymmetric radial epicyclic frequency $f_2 = f_{\phi} - f_{r}$ and the $m = -2$ non-axisymmetric vertical epicyclic frequency $f_1 = 2f_{\phi} - f_{\theta}$ \cite{Torok:2010rk, Torok:2011qy, Kotrlova:2020pqy}. In this case, the resonant frequencies are defined as:
\begin{eqnarray}
f_1 = 2f_{\phi} - f_{\theta}, \quad f_2 = f_{{per}} = f_{\phi} - f_{r}.
\end{eqnarray}
The coupling between oscillation modes in NADO2 is theoretically allowed, though the precise mechanism inducing this coupling remains unclear. Unlike NADO1, which involves coupling between an axisymmetric and a non-axisymmetric mode, the pairs of oscillation modes in NADO2 involve non-axisymmetric modes only, which may make this variant more physically plausible .

\begin{figure}[htbp!]
  \begin{subfigure}{0.5\textwidth}
    \centering
    \includegraphics[width=\linewidth]{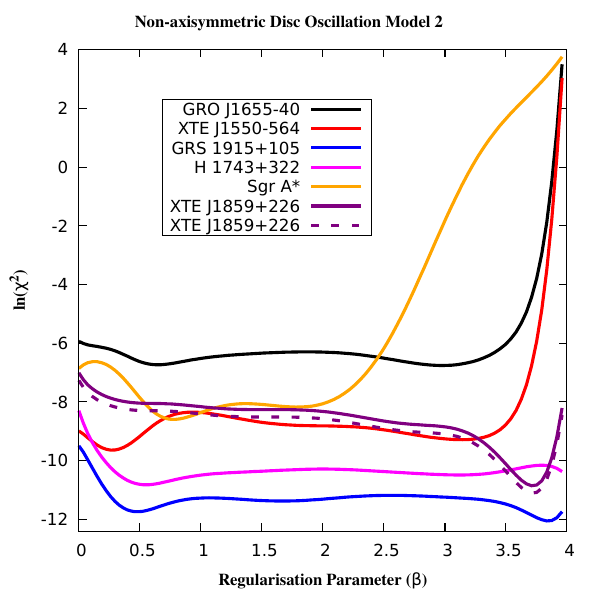}
    \caption{}
    \label{NADO21}
  \end{subfigure}
  \begin{subfigure}{0.5\textwidth}
    \centering
    \includegraphics[width=\linewidth]{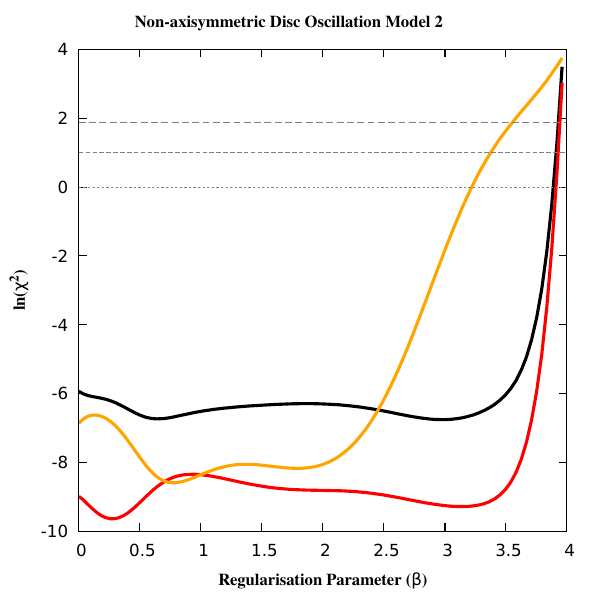}
    \caption{}
    \label{NADO22}
  \end{subfigure}
  \caption{The above figure (a) demonstrates variation of $\ln \chi^2$ with regularisation parameter $\beta$ for individual black holes in \ref{t1} assuming NADO2, figure (b) plots the variation of $\ln \chi^{2}$ with regularisation parameter $\beta$ (assuming the same model) but for a subset of the six black holes where $\chi^2$  values are large and the variation with $\beta$ is also substantial such that the confidence lines can be drawn. The grey dotted line corresponds to the 1-$\sigma$ contour, the grey short-dashed line corresponds to the 2-$\sigma$ contour and the grey long-dashed line is associated with the 3-$\sigma$ contour. The confidence contour lines are the same for all the three BHs as $\chi^2_{min}\sim0$ for all of them.}
  \label{iii}
\end{figure}
%Next we discuss our results related to Non-Axisymmetric Disc  Oscillation Model 2.\nonumber\\
In \hyperref[NADO21]{\ref*{NADO21}}, we show the plots of $\ln \chi^2$ as a function of the regularisation parameter $\beta$ for each of the six black holes listed in \ref{t1}. The figure shows that for GRS 1915+105, H1743+322 and XTE J1859+226, $\chi^2$ minimizes at $\beta\sim3.9$, $\beta\sim0.5$, $\beta\sim3.8$. This however, cannot be considered as presence or absence of regularisation parameter $\beta$ as  the $\chi^2$ for these sources do not vary substantially such that the $\Delta\chi^2$ intervals corresponding to 68\%, 90\% and 99\% confidence lines can rule out certain range of the allowed parameter space of $\beta$. For the other three black holes, the variation of $\chi^2$ is reasonable because GRO J1655-40, XTE J1550-564 and Sgr A* rule out $\beta>3.7$, $\beta>3.7$ and $\beta>3.2$ respectively outside $1-\sigma$ \hyperref[NADO22]{(\ref*{NADO22})}. If we consider the $3-\sigma$ confidence interval, GRO J1655-40, XTE J1550-564 and Sgr A* rule out $\beta>3.9$, $\beta>3.9$ and $\beta>3.5$ respectively (\hyperref[NADO22]{\ref*{NADO22}}). For all these three black holes, $\chi^2$ minimizes at a non-zero value of regularisation parameter although GR is allowed within 1-$\sigma$.

\subsection{Kinematic Models}
\label{section:KinematicModels}

Kinematic models propose that quasi-periodic oscillations (QPOs) originate from the local motion of matter within the accretion disk surrounding compact objects such as neutron stars and black holes. These models attribute QPOs to specific orbital and precessional motions of the accreting material, without requiring complex interactions within the disk. Among the various kinematic models, the Relativistic Precession Model (RPM) is particularly prominent, providing significant insights into QPO generation by linking QPO frequencies to relativistic effects near compact objects \cite{Stella:1998mq}. 

\subsubsection{Relativistic Precession Model (RPM)}
\label{section:RPM}
The Relativistic Precession Model (RPM)\cite{Stella:1998mq,stella1997lense,stella1999correlations} is a widely applied framework for analyzing high-frequency quasi-periodic oscillations (HFQPOs) in compact object systems, particularly neutron star and black hole sources. Initially developed to explain the twin HFQPOs in neutron star systems, RPM was later adapted to study similar oscillations in black hole sources \cite{stella1997lense}. According to this model, HFQPOs are associated with distinct fundamental frequencies that arise from the relativistic motion of matter in the accretion disk. Specifically, the twin HFQPOs are linked to the azimuthal (orbital) frequency $f_{\phi}$ and the periastron precession frequency $(f_{\phi} - f_r)$. Additionally, RPM provides an explanation for low-frequency QPOs, such as the $f_L$ observed in the black hole source GRO J1655-40, by associating them with the nodal precession frequency $(f_{\phi} - f_{\theta})$.

An essential assumption of RPM is that all QPOs originate from the same radial distance from the compact object, which is typically close to the innermost stable circular orbit (ISCO). The main frequencies predicted by RPM are:
\begin{itemize}
    \item $f_1 = f_{\phi}$: the azimuthal or orbital frequency, representing the fundamental orbital motion of matter around the compact object,
    \item $f_2 = f_{\phi} - f_r$: the periastron precession frequency, which corresponds to the advance of the periastron (closest approach in the orbit) due to spacetime curvature,
    \item $f_3 = f_{\phi} - f_{\theta}$: the nodal precession frequency, linked to the Lense-Thirring precession effect\cite{stella1997lense}.
\end{itemize}

The periastron precession frequency, $f_{\phi} - f_r$, arises due to the curved spacetime around a massive object, causing a shift in the orbit's closest approach point with each revolution. This precession effect is unique to relativistic orbits and is one of the key factors RPM uses to explain the twin-peak HFQPOs. The nodal precession frequency, $f_{\phi} - f_{\theta}$, is a result of the Lense-Thirring effect, a phenomenon where the rotation of a massive object, such as a black hole or neutron star, drags the surrounding spacetime. This dragging effect leads to the precession of the orbital plane around the spin axis of the central object. The nodal precession frequency has been instrumental in modeling low-frequency QPOs, making RPM one of the few models capable of explaining both the high and low-frequency QPOs.
Through these relativistic frequencies, RPM provides a powerful tool for probing the dynamics of matter close to compact objects and the strong gravitational fields surrounding them.
\begin{figure}[htbp!]
  \begin{subfigure}{0.5\textwidth}
    \centering
    \includegraphics[width=\linewidth]{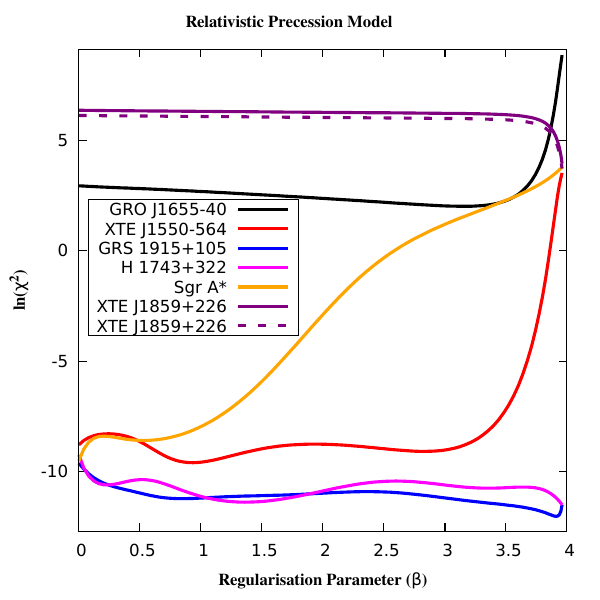}
    \caption{}
    \label{RPM1}
  \end{subfigure}
  \begin{subfigure}{0.5\textwidth}
    \centering
    \includegraphics[width=\linewidth]{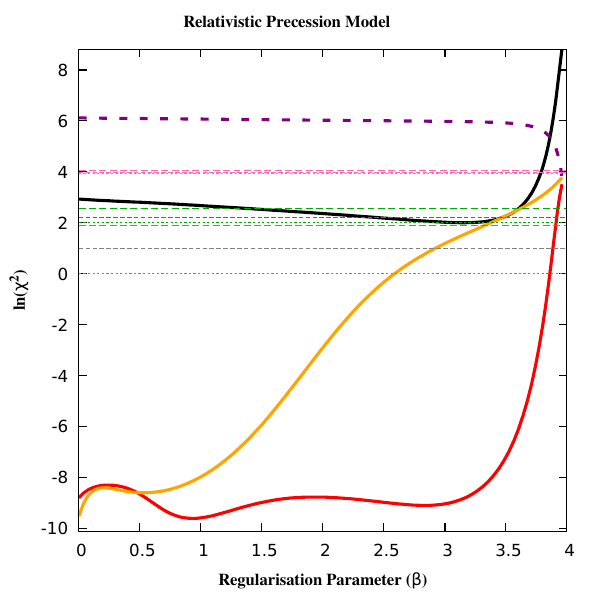}
    \caption{}
    \label{RPM2}
  \end{subfigure}
  \caption{The above figure (a) demonstrates the variation of ln$\chi^{2}$ with the regularisation parameter $\beta$ for the six black holes sources shown in \ref{t1} assuming RPM. Figure (b) plots the variation of ln$\chi^{2}$ with $\beta$ (assuming the same model) for a subset of the six BHs along with the confidence lines. The grey dotted line corresponds to the 1-$\sigma$ contour, the grey short-dashed line corresponds to the 2-$\sigma$ contour and the grey long-dashed line is associated with the 3-$\sigma$ contour for  XTE J1550-564 and SgrA*. The green dotted, short-dashed and long-dashed lines are the 1-$\sigma$, 2-$\sigma$ and 3-$\sigma$ contours respectively corresponding to the source GRO J1655-40. The magenta dotted, short-dashed and long-dashed lines are the 1-$\sigma$, 2-$\sigma$ and 3-$\sigma$ contours respectively corresponding to the source XTE J1859+226.}
  \label{jjj}
\end{figure}
Now, we discuss our results related to the Relativistic Precession Model.

 We show the plots of $ln\chi^2$ as a function of the regularisation parameter $\beta$ for each of the six black holes listed in \ref{t1} in  \hyperref[RPM1]{ \ref*{RPM1}}. The figure shows that for GRS 1915+105 and H1743+322,  $\chi^2$ minimizes at $\beta\sim3.99$ and $\beta\sim1.4$ respectively. This however, cannot be considered as a confirmatory signature of the Simpson-Visser BH as  the $\chi^2$ for these sources do not vary substantially such that the $\Delta\chi^2$ intervals corresponding to 68\%, 90\% and 99\% confidence lines can rule out certain range of the allowed parameter space of $\beta$. For XTE J1550-564 and Sgr A*, the value of minimum $\chi^2$ is nearly 0, so their $1-\sigma$, $2-\sigma$ and $3-\sigma$ confidence intervals are shown by grey dotted, grey short-dashed and grey long-dashed lines. XTE J1550-564 rules out $\beta>3.8$ outside  $1-\sigma$ interval and $\beta>3.9$ outside $3-\sigma$ interval. For this black hole, $\chi^2$ minimizes at $\beta\sim0.9$ but GR is included within 1-$\sigma$.  Sgr A* rules out $\beta>2.5$ outside  $1-\sigma$ interval and $\beta>3.3$ outside $3-\sigma$ interval. For this black hole, the $\chi^2$ minimizes at $\beta\sim0$ \hyperref[RPM2]{\ref*{RPM2}}.\\
 For GRO J1655-40,  the $1-\sigma$, $2-\sigma$ and $3-\sigma$ confidence interval are shown by the green dotted, the green short-dashed and the green long-dashed lines. GRO J1655-40 allows $2.8<\beta<3.4$ within the $1-\sigma$ interval and $1.5<\beta<3.7$ within the $3-\sigma$ interval. For this black hole, the $\chi^2$ minimizes at $\beta\sim3.3$ (\hyperref[RPM2]{\ref*{RPM2}}). Interestingly, using this model, the source GRO J1655-40 excludes GR outside 3-$\sigma$ and clearly exhibits a preference towards the regular BH scenario manifested by the SV model.
For XTE J1859+226, $\chi^2$ minimizes at $\beta\sim3.99$. This is a rare case,
where the extremal SV scenario is preferred and the Kerr scenario is completely ruled out (in the context of the model RPM) since we get very tight constrains on $\beta$ based on the $1-\sigma$, $2-\sigma$ and $3-\sigma$ confidence intervals denoted by the magenta lines in \hyperref[RPM2]{\ref*{RPM2}}. This might indicate non-existence of singularity inside this particular black hole. However, this result is model dependent and we will comment on this in the next section.

\subsubsection{Tidal Disruption Model}
\label{section:TDM}
\begin{figure}[t!]
  \begin{subfigure}{0.5\textwidth}
    \centering
    \includegraphics[width=\linewidth]{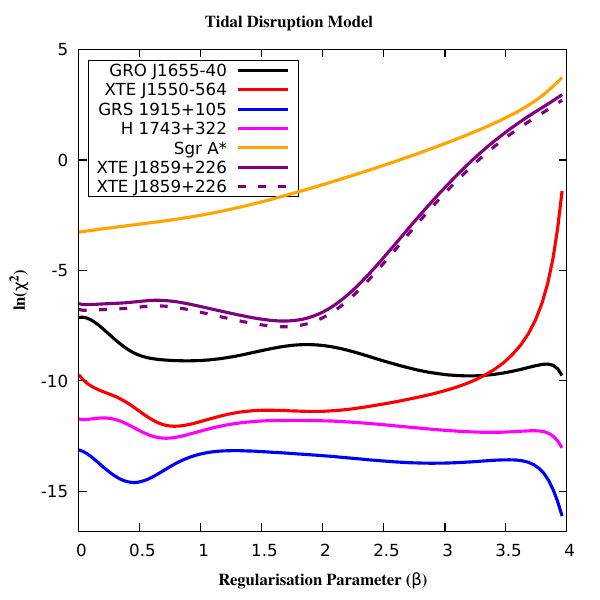}
    \caption{}
    \label{TDM1}
  \end{subfigure}
  \begin{subfigure}{0.5\textwidth}
    \centering
    \includegraphics[width=\linewidth]{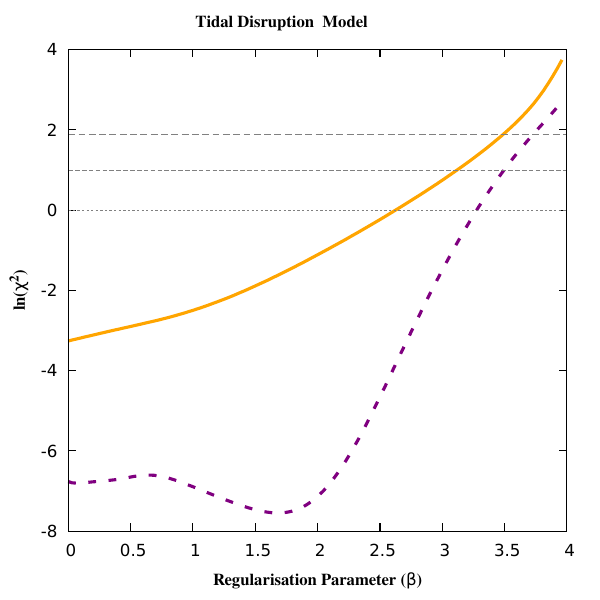}
    \caption{}
    \label{TDM2}
  \end{subfigure}
  \caption{The above figure (a) demonstrates the variation of $\ln\chi^2$ with the regularisation parameter $\beta$ for individual black holes in \ref{t1} assuming TDM, figure (b) plots the variation of $\ln \chi^{2}$ with the regularisation parameter $\beta$ (assuming the same model) but for a subset of the six black holes where $\chi^2$  values are large and the variation with $\beta$ is also substantial such that the confidence lines can be drawn. The grey dotted line corresponds to the 1-$\sigma$ contour, the grey short-dashed line corresponds to the 2-$\sigma$ contour and the grey long-dashed line is associated with the 3-$\sigma$ contour. The confidence contour lines are the same for all the two BHs as $\chi^2_{min}\sim0$ for all of them..}
  \label{kkk}
\end{figure}
The Tidal Disruption Model posits that quasi-periodic oscillations (QPOs) arise from the effects of tidal forces exerted by a black hole on clumps of plasma within the accretion disk \cite{Cadez:2008iv,Kostic:2009hp,Germana:2009ce}. These dense plasma clumps or blobs form due to inhomogeneities within the disk and are subject to strong tidal forces as they approach the black hole. This differential gravitational pull stretches and distorts the clumps, creating a ring-like structure in their orbit. As a result, distinct oscillatory motions are generated within these rings due to both the black hole’s gravitational field and the dynamics of accretion.

The primary high-frequency QPO (HFQPO) in this model is associated with the orbital motion of these tidally stretched structures. Specifically, the upper HFQPO frequency is given by $f_1 = f_{\phi} + f_r$, while the lower HFQPO, $f_2$, corresponds to $f_{\phi}$, the intrinsic orbital frequency of the ring. These frequencies are inherently dependent on the black hole’s mass, charge, spin, and the emission radius $r_{cm}$, thus providing a potential means to probe the black hole’s properties.
As the clumps undergo tidal disruption, the infalling material heats up due to viscous dissipation and the release of gravitational energy results in significant X-ray emissions. 
%The periodic nature of this infall causes the observed HFQPOs, which reflect the oscillatory behavior induced by the tidal forces on the accreting material. 
In this way, the Tidal Disruption Model links the observed HFQPOs to the fundamental orbital frequencies within the accretion disk and the strong tidal interactions near the black hole.
In \hyperref[TDM1]{\ref*{TDM1}}, we show the plots of $\ln \chi^2$ as a function of the regularisation parameter $\beta$ for each of the six black holes listed in \ref{t1}. The figure shows that for GRO J1655-40, XTE J1550-564, GRS 1915+105 and H1743+322, although the $\chi^2$ minimizes at a non-zero $\beta$, the variation of $\chi^2$ is not substantial such that the $\Delta\chi^2$ intervals corresponding to 68\%, 90\% and 99\% confidence lines can rule out certain range of the allowed parameter space of $\beta$. For these sources the Kerr and the SV scenario are equally favored.

For the other two black holes, the variation of $\chi^2$ is reasonable because Sgr A* and XTE J1859+226 rule out $\beta>2.6$ and $\beta>3.3$ respectively outside $1-\sigma$ (\hyperref[TDM2]{\ref*{TDM2}}). If we consider the $3-\sigma$ confidence interval, Sgr A* and XTE J1859+226 rule out $\beta>3.5$ and $\beta>3.7$ respectively (\hyperref[TDM2]{\ref*{TDM2}}). For Sgr A* and XTE J1859+226 the $\chi^2$ value minimizes at $\beta\sim 0$ and $\beta\sim 1.8$ respectively. Although the source XTE J1859+226 favors a non-zero $\beta$, GR is allowed within 1-$\sigma$.

\section{Constraining the model parameters using the MCMC method}
\label{S5}
Earlier, for parameter estimation we have used the grid search technique where we have varied the model parameters (regularisation parameter $\beta$, the spin a, the mass of the black hole M and the emission radius $r_{cm}$) and evaluated the $\chi^2$ for different combinations of the same. For every combination, we have calculated the $\chi^2$ and our best parameter estimation corresponds to the one where $\chi^2$ value is minimum. The necessary condition for this grid search method is that the gridding must be sufficiently fine so that we do not miss the global minima. To check whether our work meets this necessary condition, we verify it with the Bayesian approach, which employs the Markov chain Monte Carlo (MCMC) simulation. \\
The Bayesian posterior distribution for the model parameters is given by,
\begin{eqnarray}
\mathcal{P}(\theta|D)=\frac{\mathcal{P}(D|\theta)\mathcal{P}(\theta)}{\mathcal{P}(D)}
\label{3}
\end{eqnarray}
where, $\mathcal{P}(\theta)$ represents the prior distribution for the model parameters $\theta\equiv \lbrace \beta, a, M, r_{cm}\rbrace$, and $\mathcal{P}(D|\theta)$ is associated with the likelihood function and $\mathcal{P}(D)=1$ since the data is known \cite{Verde:2009tu}. We assume flat priors for $\beta$, $a$, $r_{cm}$ for all six black holes in the allowed range. For mass M, we choose flat priors for  GRS 1915 + 105, H 1743 + 322 and Sgr A* in the allowed range and Gaussian priors for  GRO J1655-40, XTE J1550-564 and XTE J1859+226 (see \ref{t1}), based on previous estimates.
The likelihood function considers the contribution from both the upper and lower HFQPO frequencies and also the LFQPO if exhibited by the source and addressed by the model (e.g., GRO J1655-40 exhibits twin-peak HFQPOs and a LFQPO, all of which can be explained by the RPM model). This is given by,
\begin{eqnarray}
\log \mathcal{L}=\log \mathcal{L}_{U_1} + \log \mathcal{L}_{U_2} + \log \mathcal{L}_{L}
\label{4}
\end{eqnarray}
where, 
\begin{eqnarray}
\log \mathcal{L}_{U_1} =-\frac{1}{2} \frac{\lbrace f_{\textrm{up1}{,i}}-f_1(\beta,a,M,r_{\rm em}) \rbrace ^2}{\sigma_{up_{\rm 1},i}^2}
\label{5}
\end{eqnarray}
\begin{eqnarray}
\log \mathcal{L}_{U_2} =-\frac{1}{2} \frac{\lbrace f_{\textrm{up2}{,i}}-f_2(\beta,a,M,r_{\rm em}) \rbrace ^2}{\sigma_{up_{\rm 2},i}^2}
\label{6}
\end{eqnarray}
\begin{eqnarray}
\log \mathcal{L}_{L} =-\frac{1}{2} \frac{\lbrace f_{\textrm{up3}{,i}}-f_3(\beta,a,M,r_{\rm em}) \rbrace ^2}{\sigma_{up_{\rm 3},i}^2}
\label{6}
\end{eqnarray}
Note that, except for RPM with source GRO J1655-40, $\log \mathcal{L}_{L}=0$.
We use the publicly available code \emph{emcee}\cite{Foreman-Mackey:2012any,foreman2013emcee} for our work where we consider 64 chains and in each chain we draw 20,000 samples for each of the four model parameters based on the prior information, thereby ensuring a thorough investigation of the multidimensional parameter space. We have also tested the convergence of MCMC using the Gelman-Rubin convergence criterion $R$. We have calculated $R$ corresponding to each model parameter, when a given model was tested with the HFQPO data of a particular source, and noted that $0.99\lesssim R\lesssim 1.1$, for every model parameter corresponding to each BH. 

\begin{figure}[H]
\vspace{-0.1cm}
\centering
\textbf{\underline{GRO J1655-40}}

% First row
\begin{subfigure}[b]{0.48\textwidth}
    \centering
    \includegraphics[width=\linewidth]{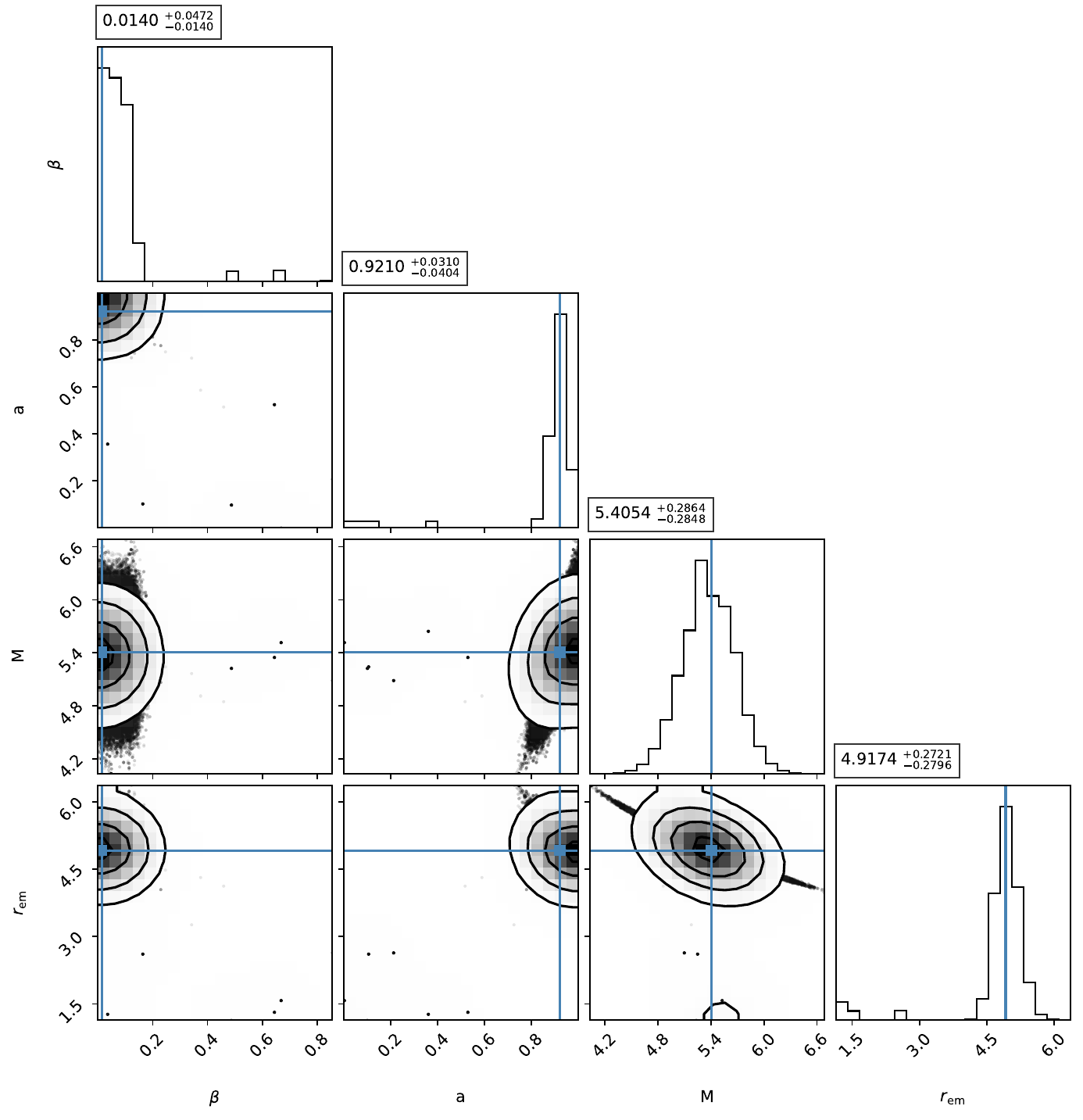}
    \caption*{(a) Parametric Resonance Model}
\end{subfigure}
\hfill
\begin{subfigure}[b]{0.48\textwidth}
    \centering
    \includegraphics[width=\linewidth]{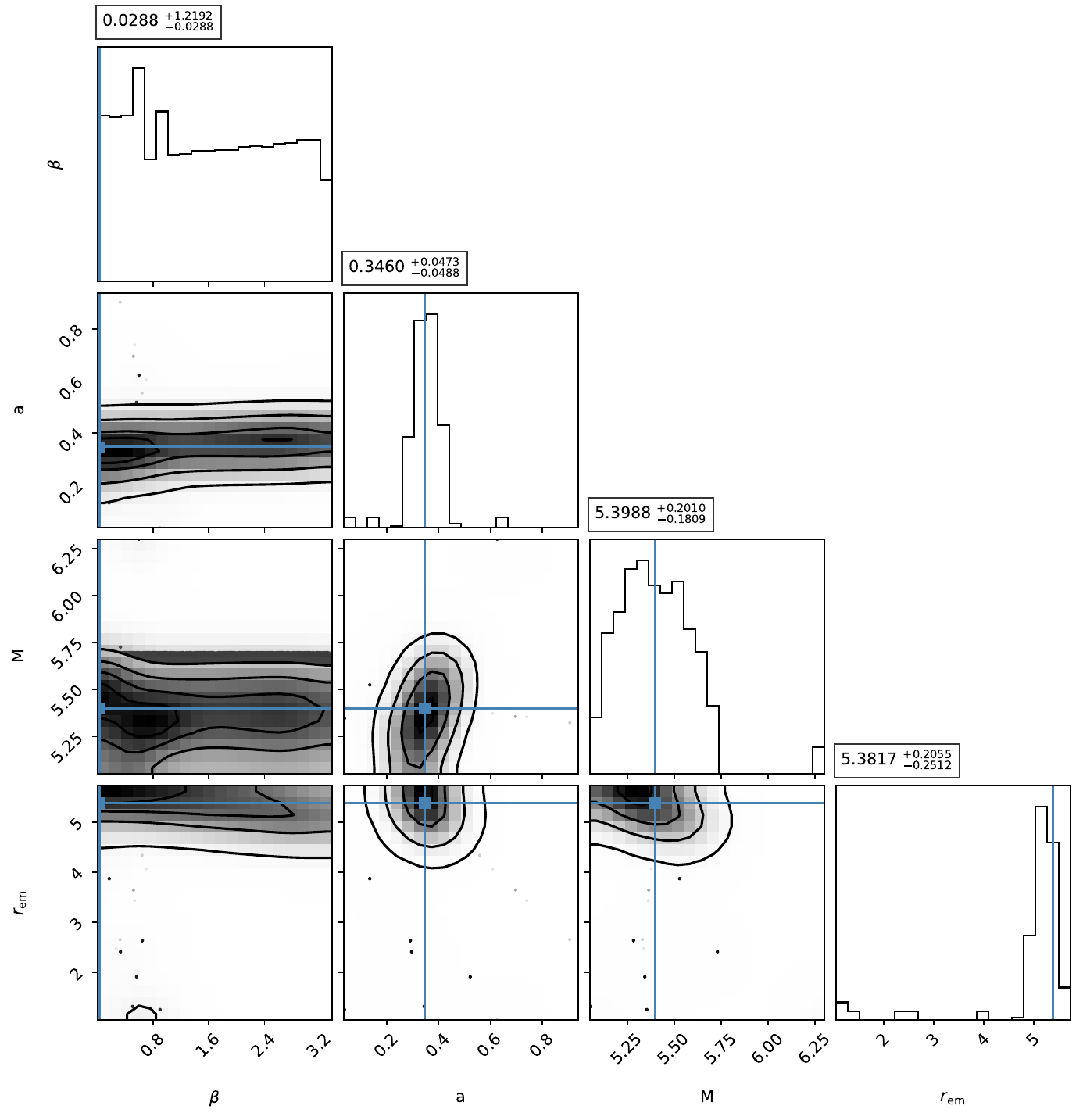}
    \caption*{(b) Forced Resonance Model 1}
\end{subfigure}

\vspace{0.3cm} % space between rows

% Second row
\begin{subfigure}[b]{0.48\textwidth}
    \centering
    \includegraphics[width=\linewidth]{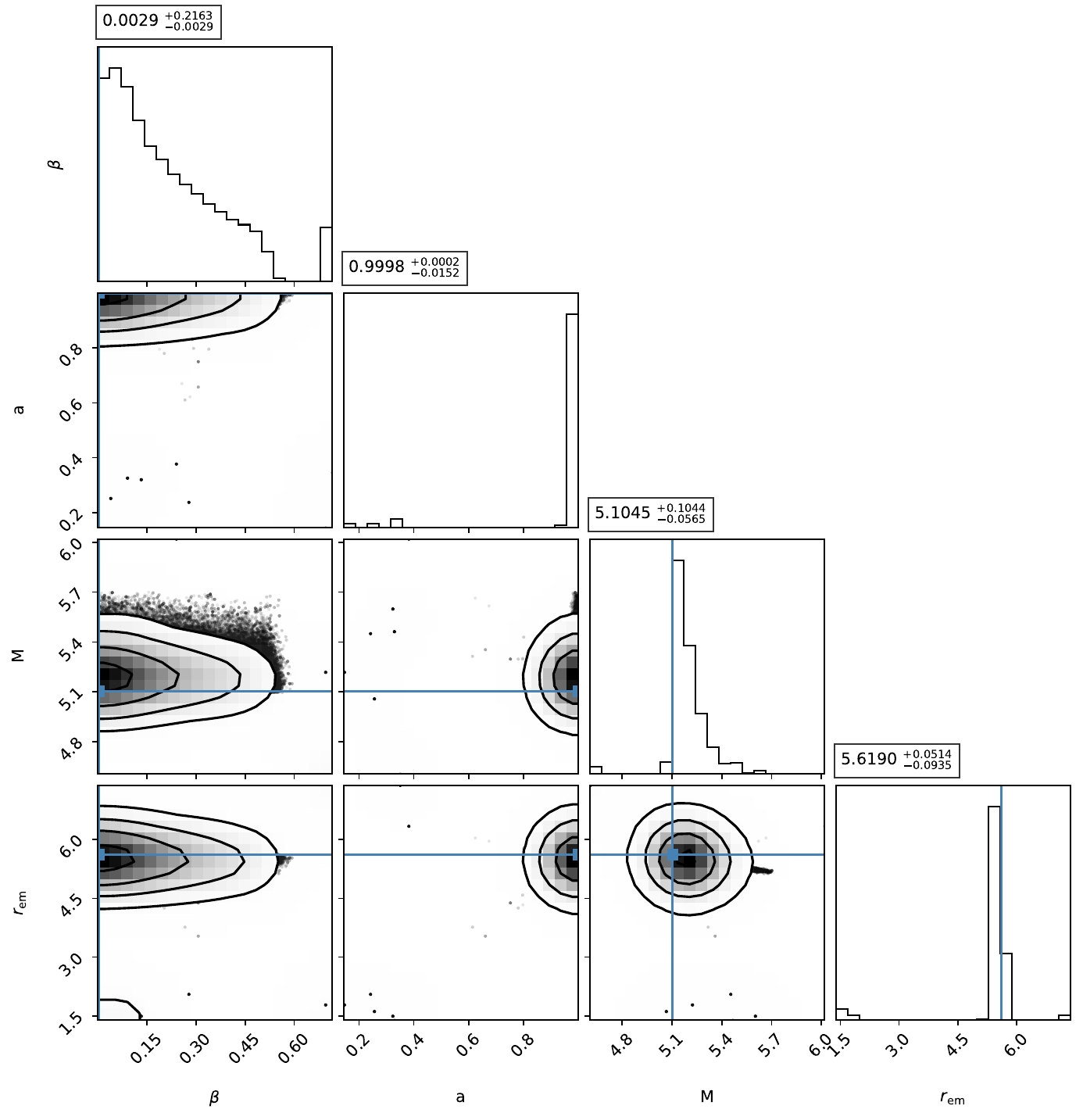}
    \caption*{(c) Keplerian Resonance Model 1}
\end{subfigure}
\hfill
\begin{subfigure}[b]{0.48\textwidth}
    \centering
    \includegraphics[width=\linewidth]{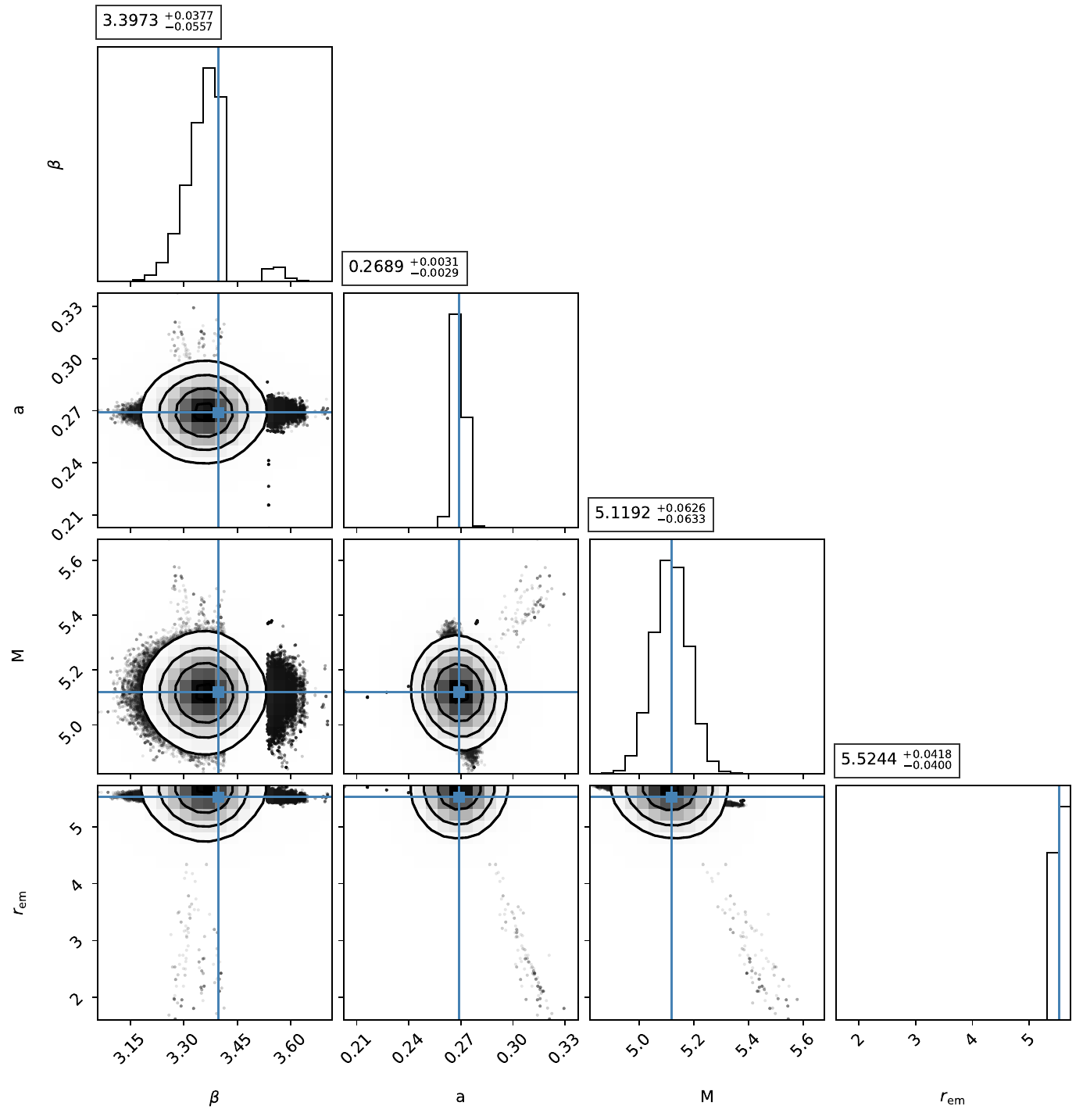}
    \caption*{(c) Relativistic Precesion Model}
\end{subfigure}

\caption{Constraints on the model parameters using the QPO data of GRO J1655-40 considering (a) the Parametric Resonance Model, (b) the Forced Resonance Model 1, (c) the Keplerian Resonance Model 1, and (d) the Relativistic Precession Model.}
\label{GROcorner}
\end{figure}

\begin{figure}[htbp]
%\vspace{-0.1cm}
\centering
\textbf{\underline{XTE J1550-564}}

% First row
\begin{subfigure}[b]{0.48\textwidth}
    \centering
    \includegraphics[width=\linewidth]{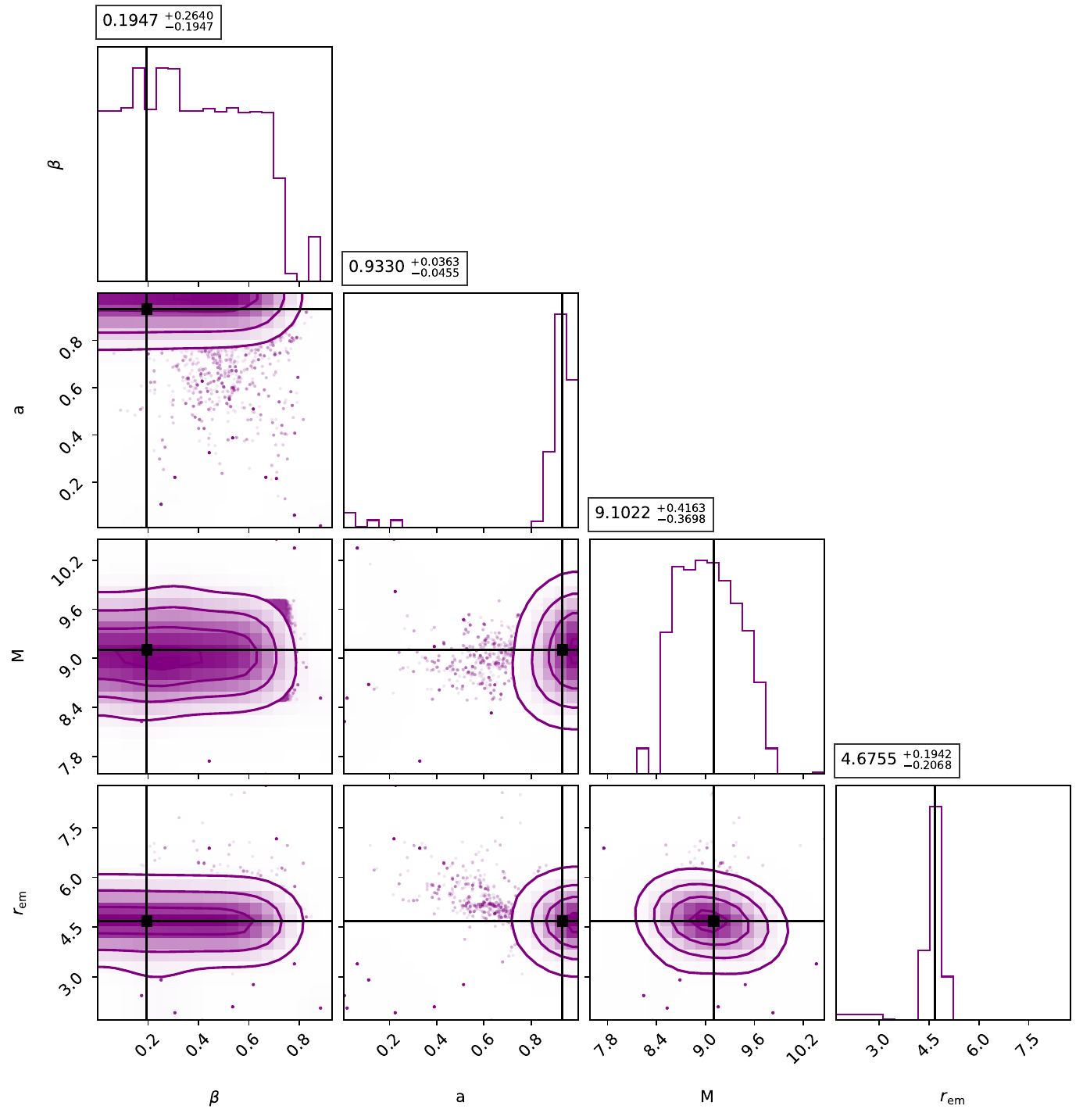}
    \caption*{(a) Parametric Resonance Model}
\end{subfigure}
\hfill
\begin{subfigure}[b]{0.48\textwidth}
    \centering
    \includegraphics[width=\linewidth]{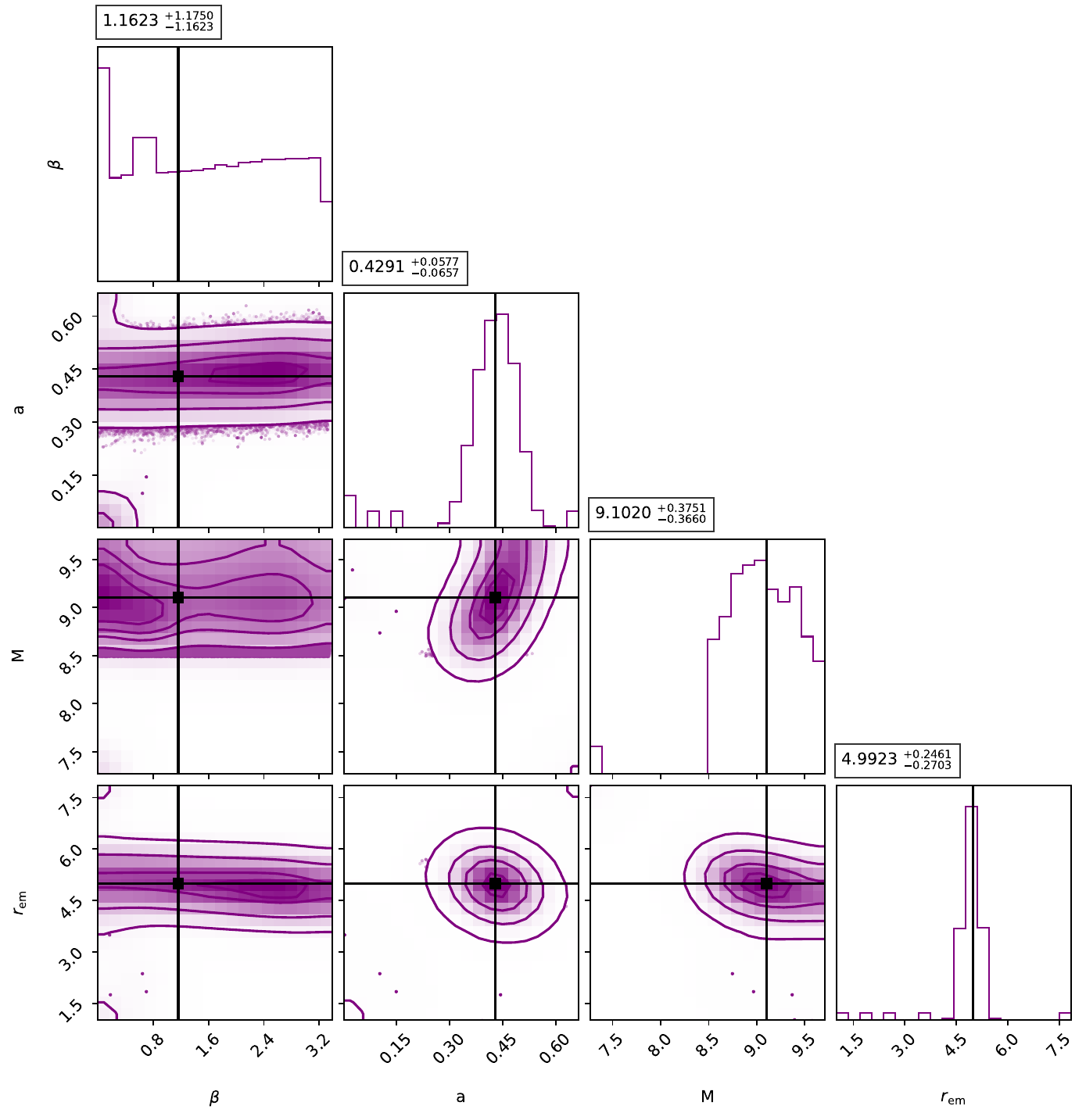}
    \caption*{(b) Forced Resonance Model 1}
\end{subfigure}

\vspace{0.05cm} % space between rows

% Second row
\begin{subfigure}[b]{0.48\textwidth}
    \centering
    \includegraphics[width=\linewidth]{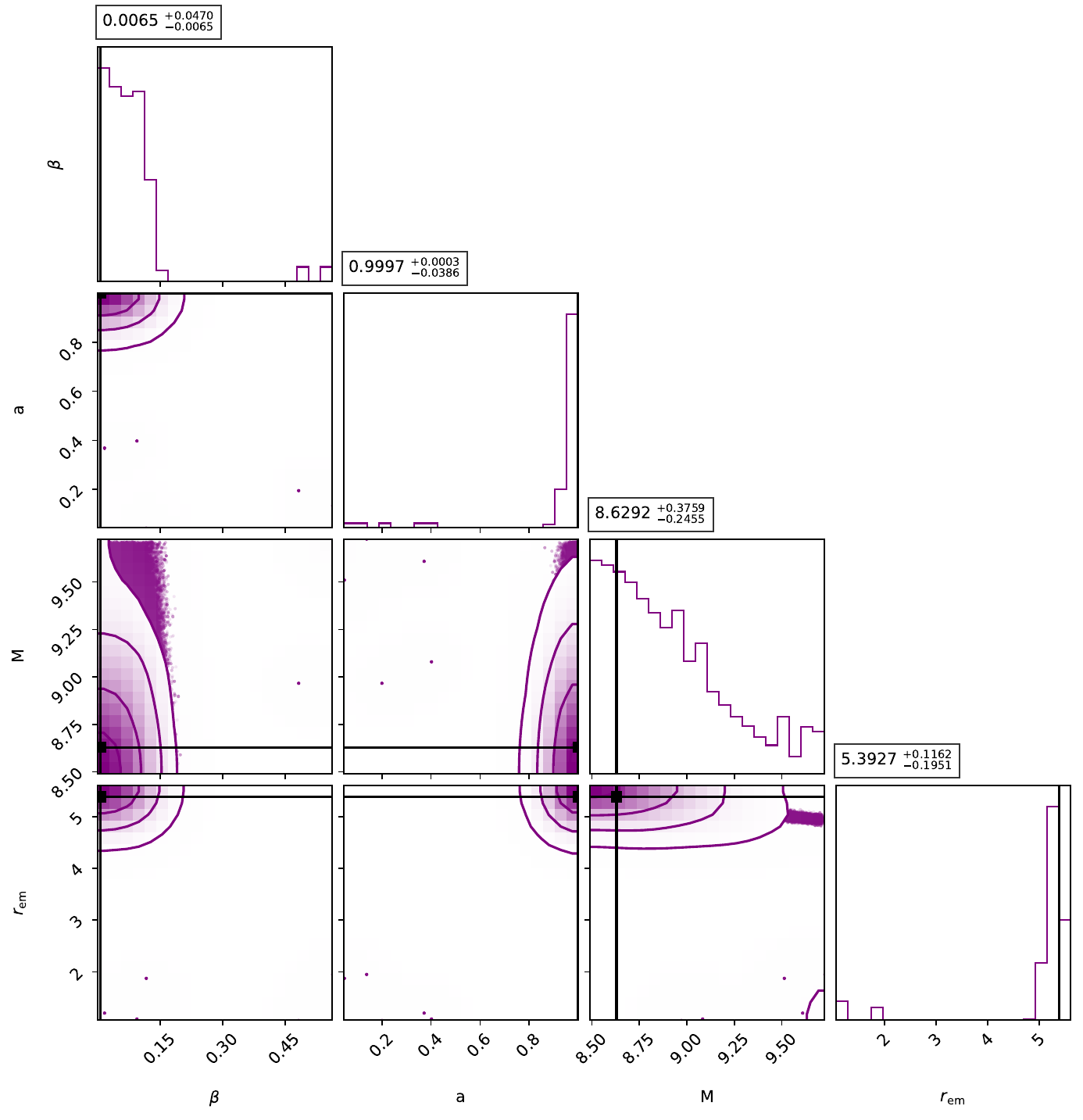}
    \caption*{(c) Keplerian Resonance Model 1}
\end{subfigure}
\hfill
\begin{subfigure}[b]{0.48\textwidth}
    \centering
    \includegraphics[width=\linewidth]{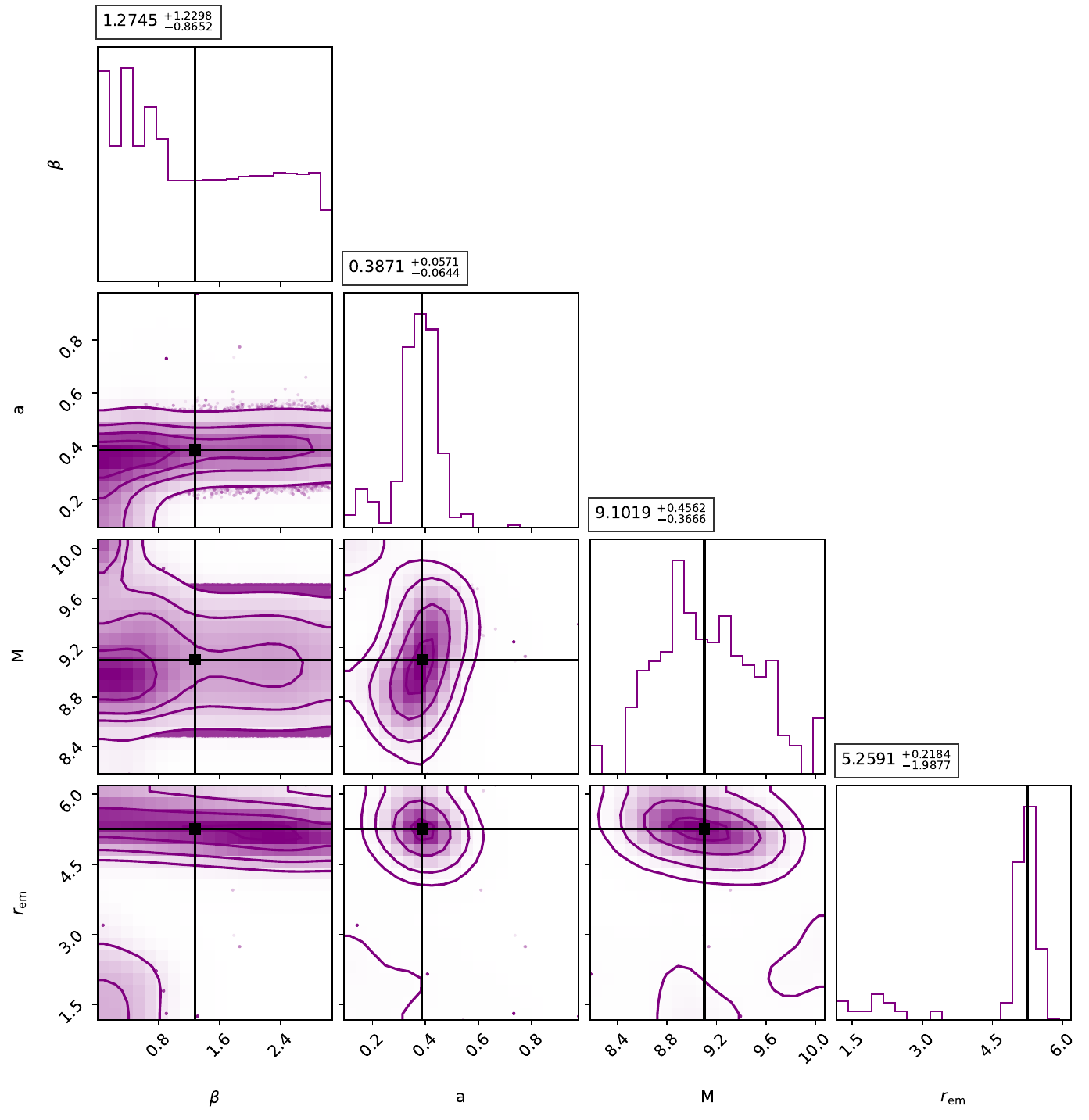}
    \caption*{(d) Relativistic Precession Model }
\end{subfigure}
\caption{Constraints on the model parameters using the HFQPO data of XTE J1550-564 considering (a) the Parametric Resonance Model, (b) the Forced Resonance Model 1, (c) the Keplerian Resonance Model 1, and (d) the Relativistic Precession Model.}
\label{FigXTE1corner}
\end{figure}

\begin{figure}[H]
\vspace{-0.1cm}
\centering
\textbf{\underline{GRS 1915+105}}

% Row with two figures
\begin{subfigure}[b]{0.45\textwidth}
    \centering
    \includegraphics[width=\linewidth]{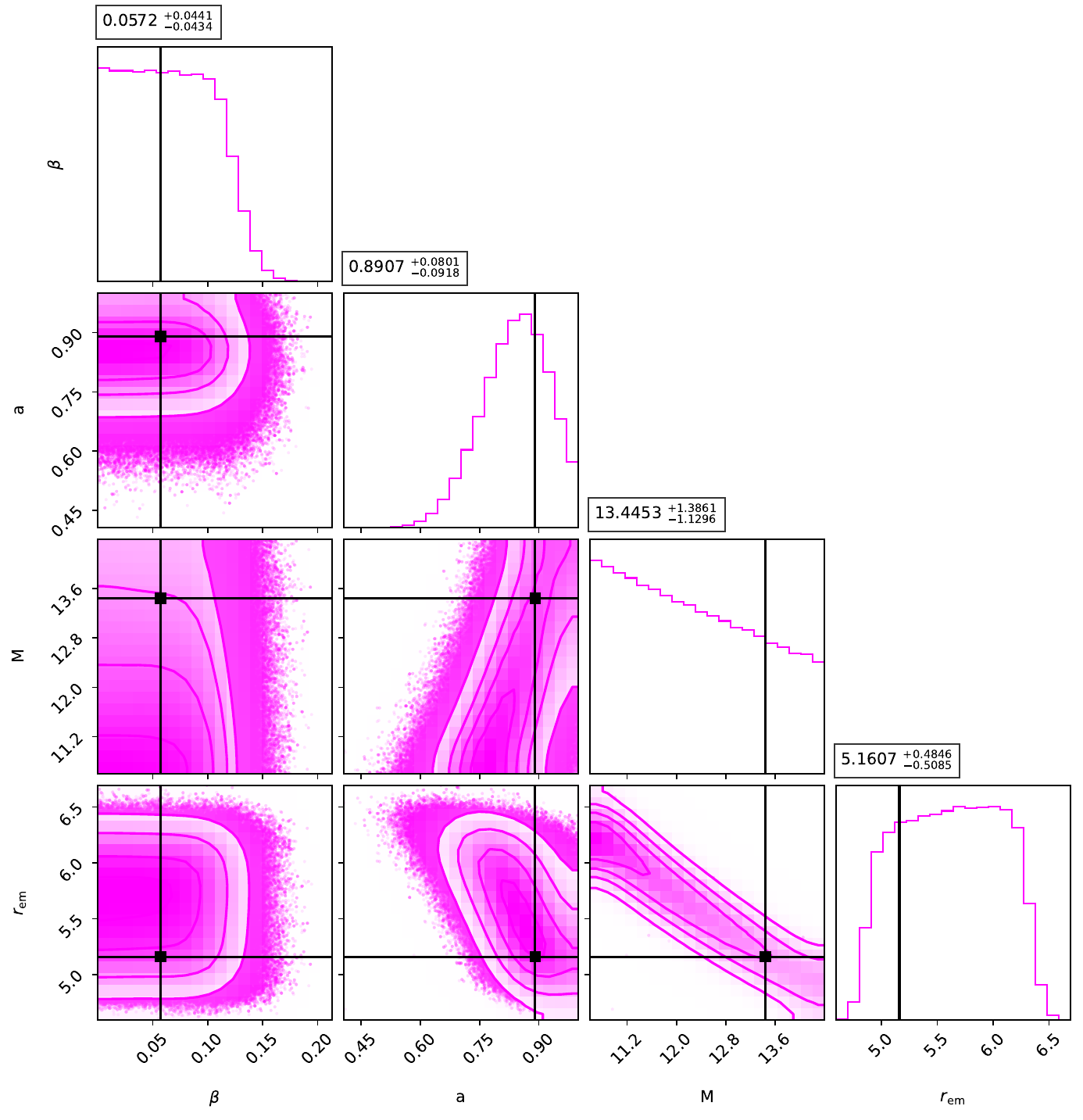}
    \caption*{(a) Parametric Resonance Model}
\end{subfigure}
\hfill
\begin{subfigure}[b]{0.45\textwidth}
    \centering
    \includegraphics[width=\linewidth]{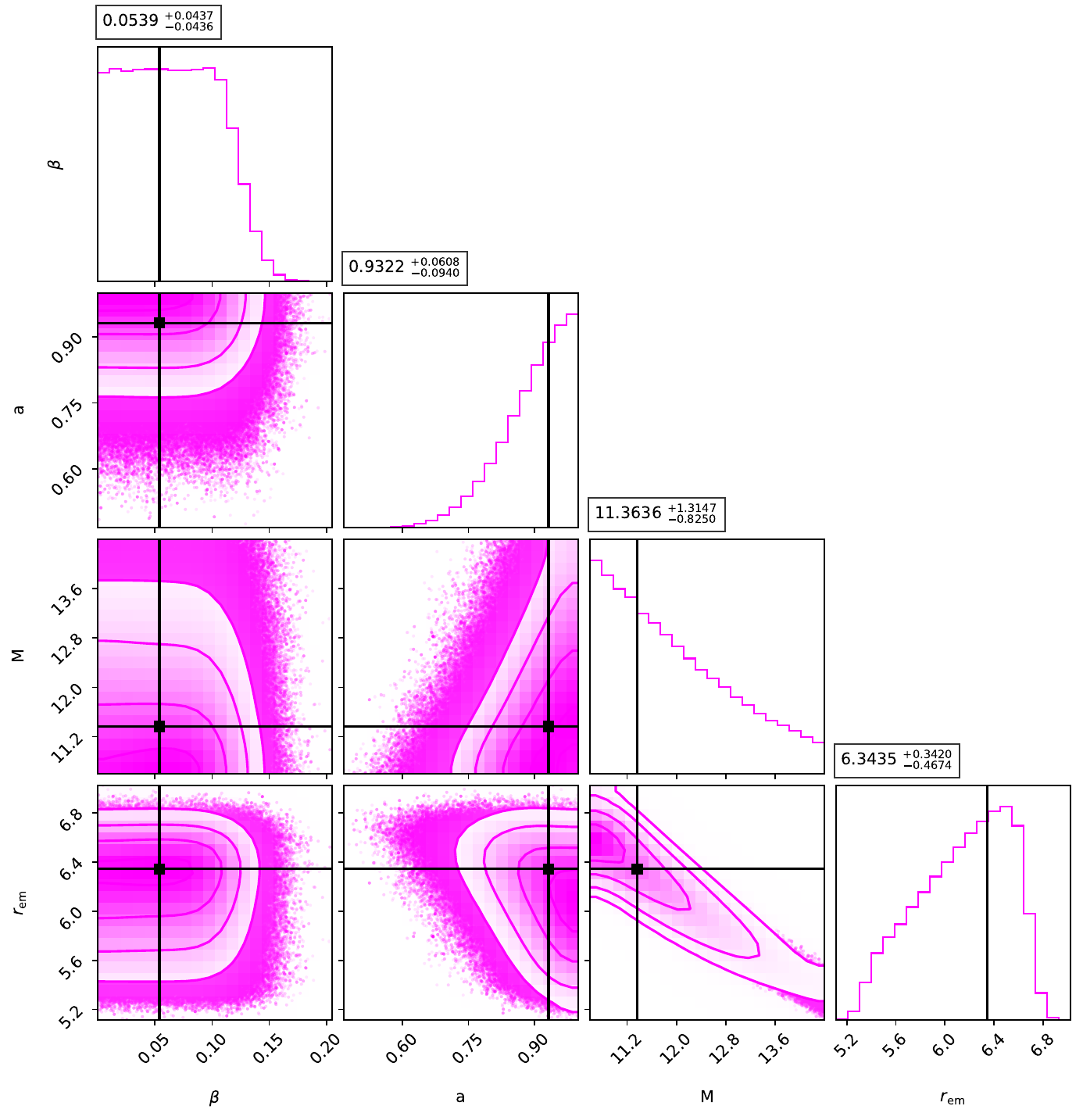}
    \caption*{(b) Keplerian Resonance Model 1}
\end{subfigure}

\caption{Constraints on the model parameters using the HFQPO data of GRS 1915+105 considering (a) the Parametric Resonance Model and (b) the Keplerian Resonance Model 1.}
\label{Fig2-2}
\end{figure}

\begin{figure}[H]
\vspace{-0.1cm}
\centering
\textbf{\underline{H 1743-322}}

% Row with two figures
\begin{subfigure}[b]{0.45\textwidth}
    \centering
    \includegraphics[width=\linewidth]{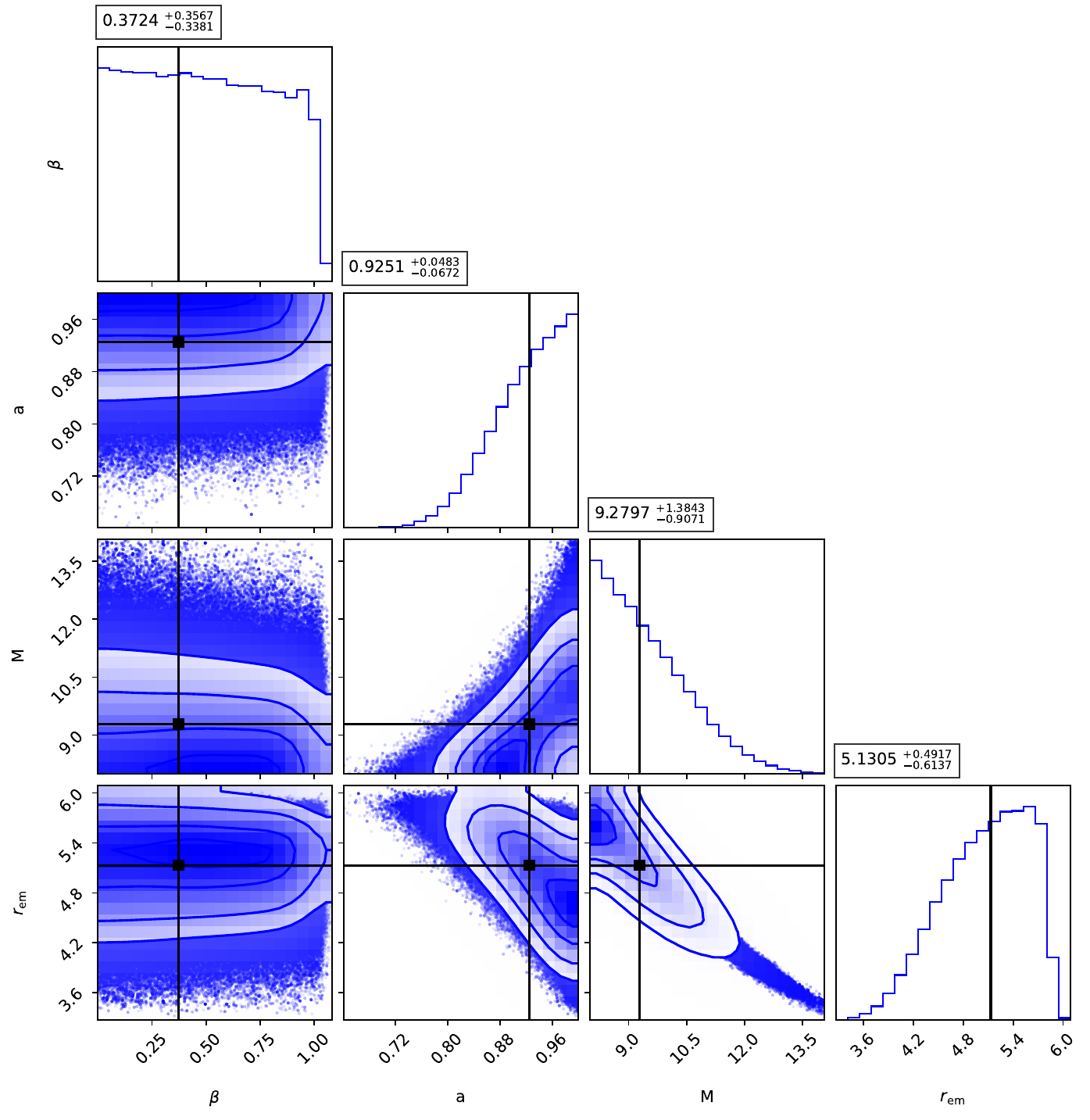}
    \caption*{(a) Parametric Resonance Model}
\end{subfigure}
\hfill
\begin{subfigure}[b]{0.45\textwidth}
    \centering
    \includegraphics[width=\linewidth]{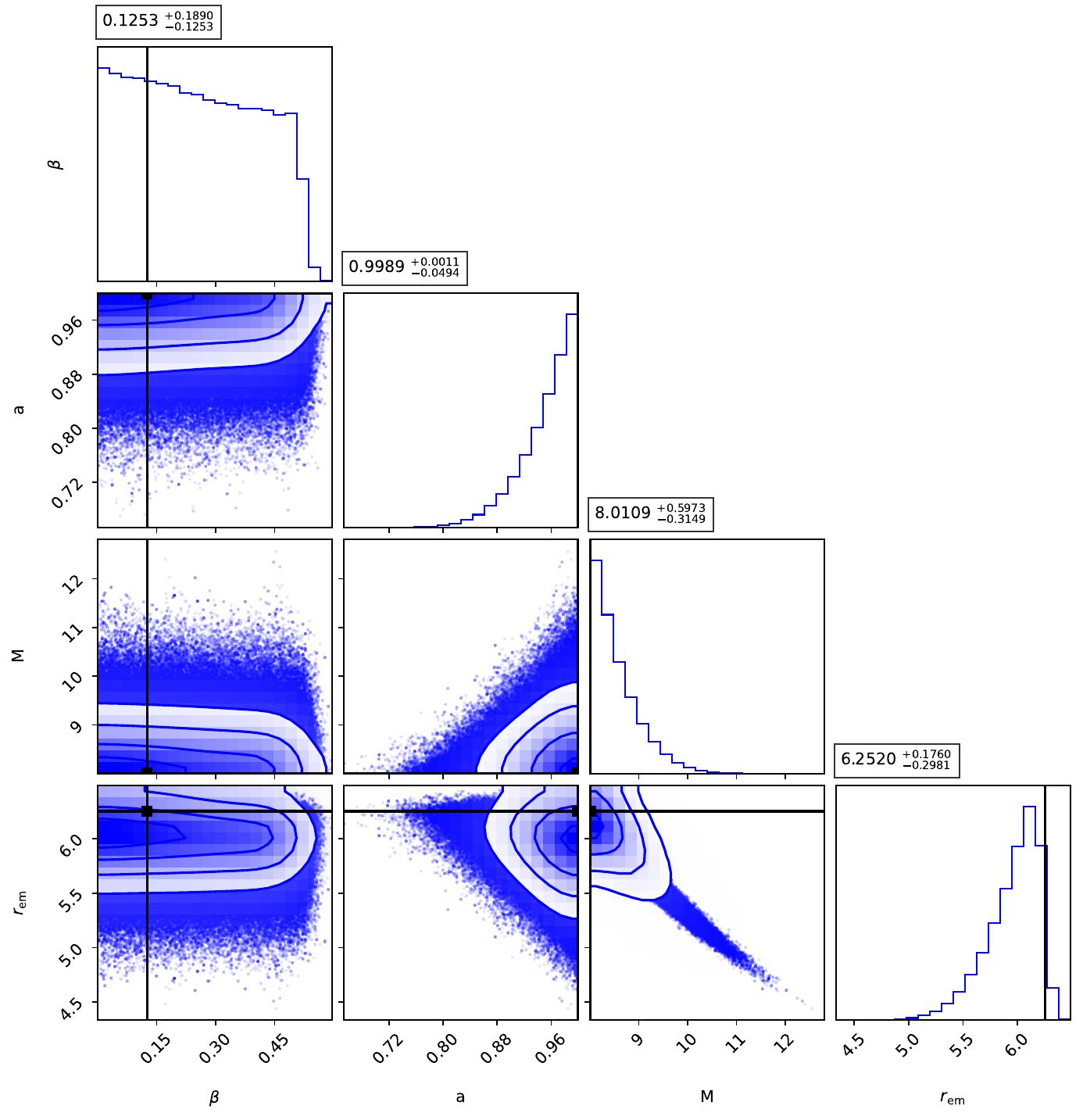}
    \caption*{(b) Keplerian Resonance Model 1}
\end{subfigure}

\caption{Constraints on the model parameters using the HFQPO data of H1743-322 considering (a) the Parametric Resonance Model and (b) the Keplerian Resonance Model 1.}
\label{H-corner}
\end{figure}

\begin{figure}[H]
\vspace{-0.1cm}
\centering
\textbf{\underline{Sgr A*}}

% First row
\begin{subfigure}[b]{0.48\textwidth}
    \centering
    \includegraphics[width=\linewidth]{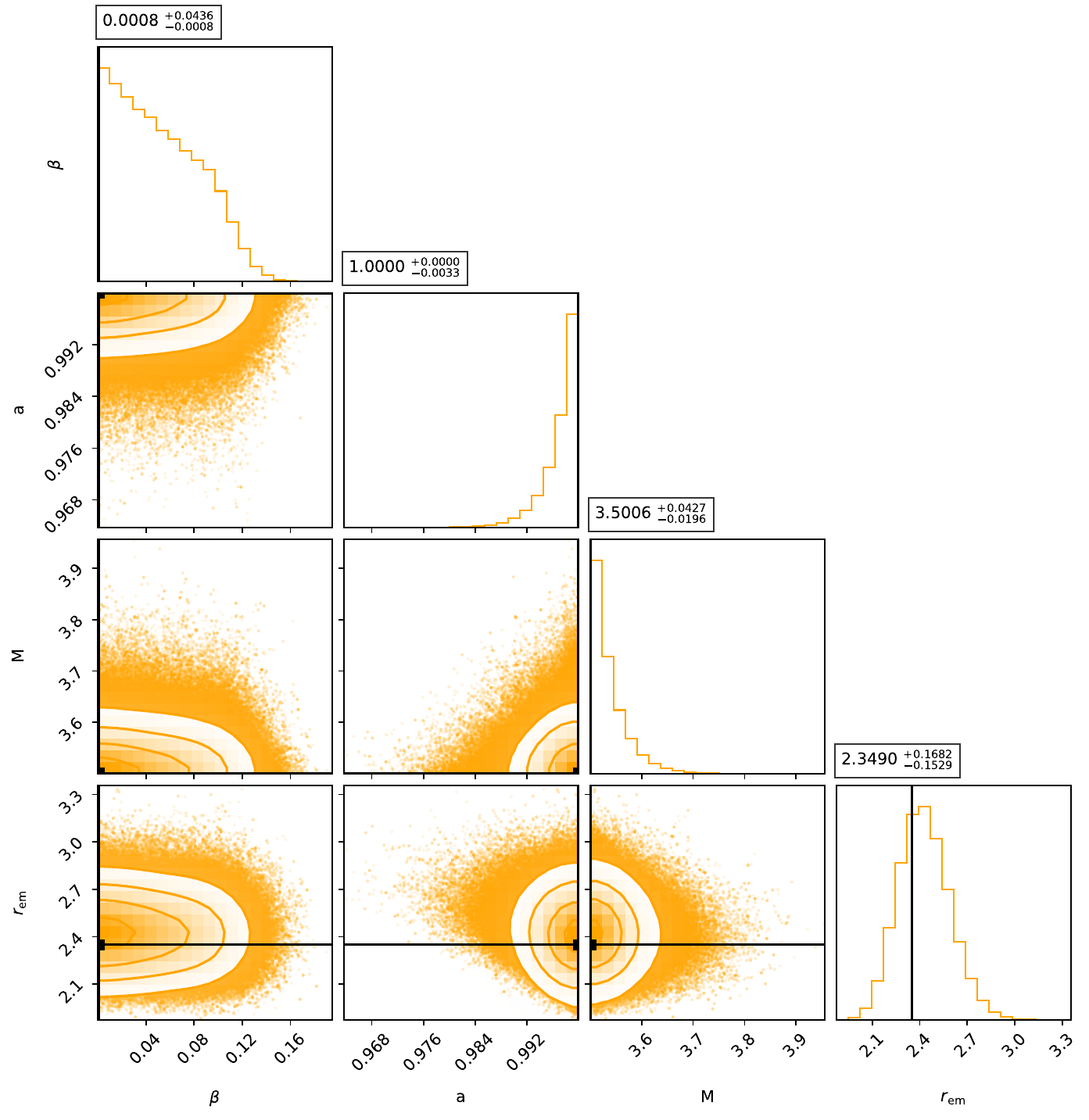}
    \caption*{(a) Parametric Resonance Model}
\end{subfigure}
\hfill
\begin{subfigure}[b]{0.48\textwidth}
    \centering
    \includegraphics[width=\linewidth]{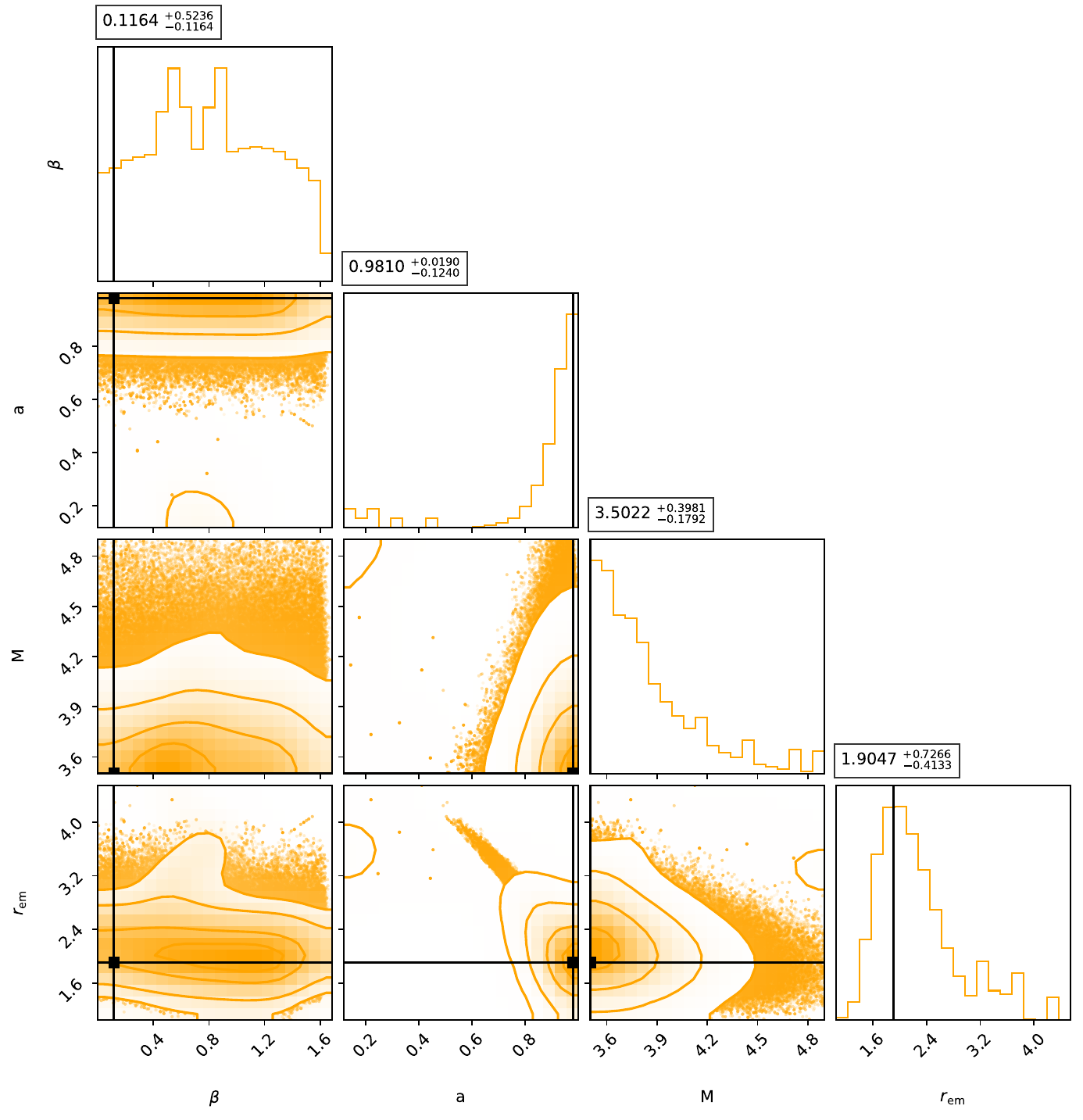}
    \caption*{(b) Forced Resonance Model 1}
\end{subfigure}
\vspace{0.05cm} % space between rows
    \begin{subfigure}[b]{0.48\textwidth}
        \centering
        \includegraphics[width=\textwidth]{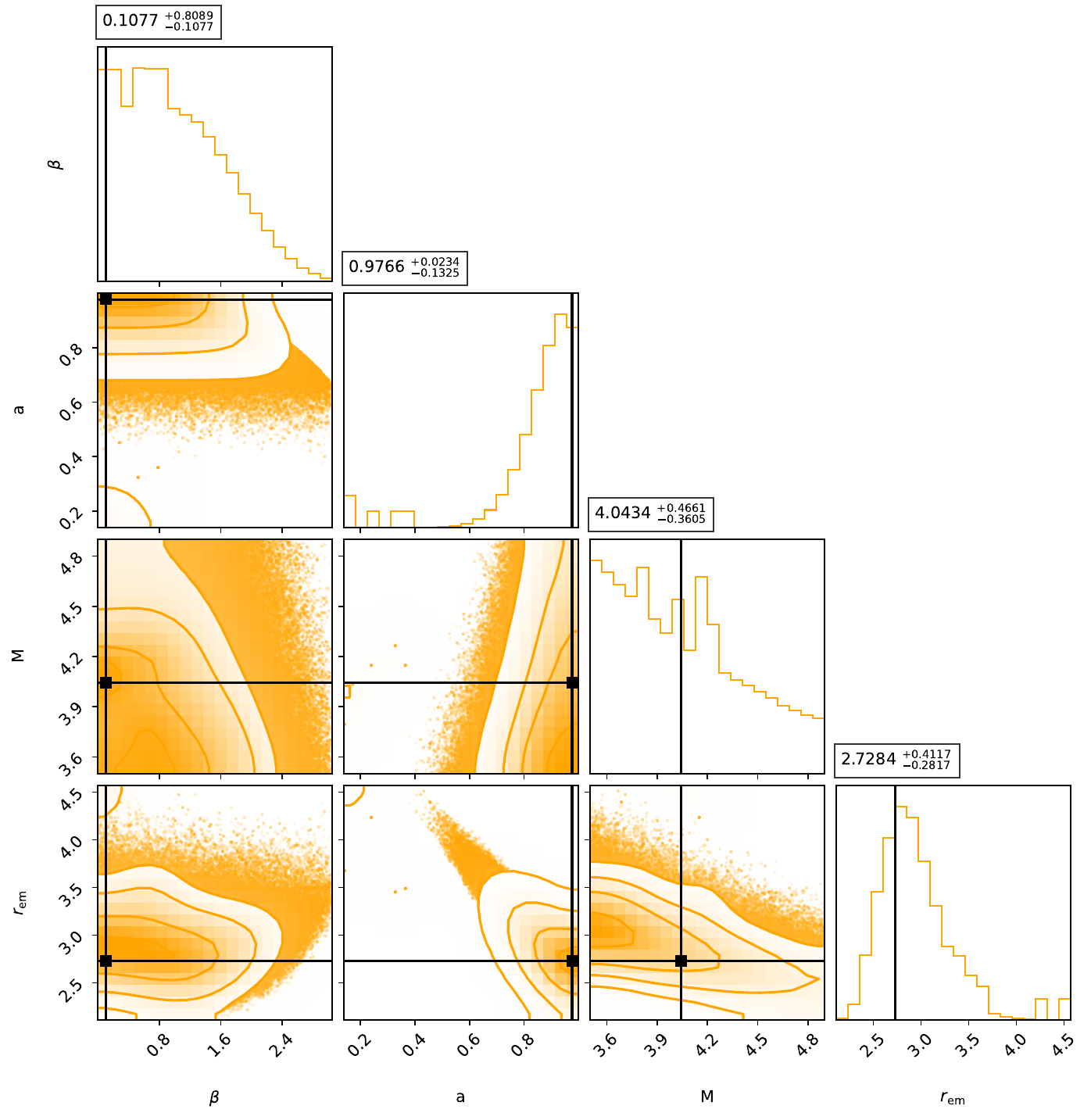}
        \caption{Relativistic Precession Model }
    \end{subfigure}
    \hfill
    % Second row - one figure centered
    \begin{subfigure}[b]{0.48\textwidth}
        \centering
        \includegraphics[width=\textwidth]{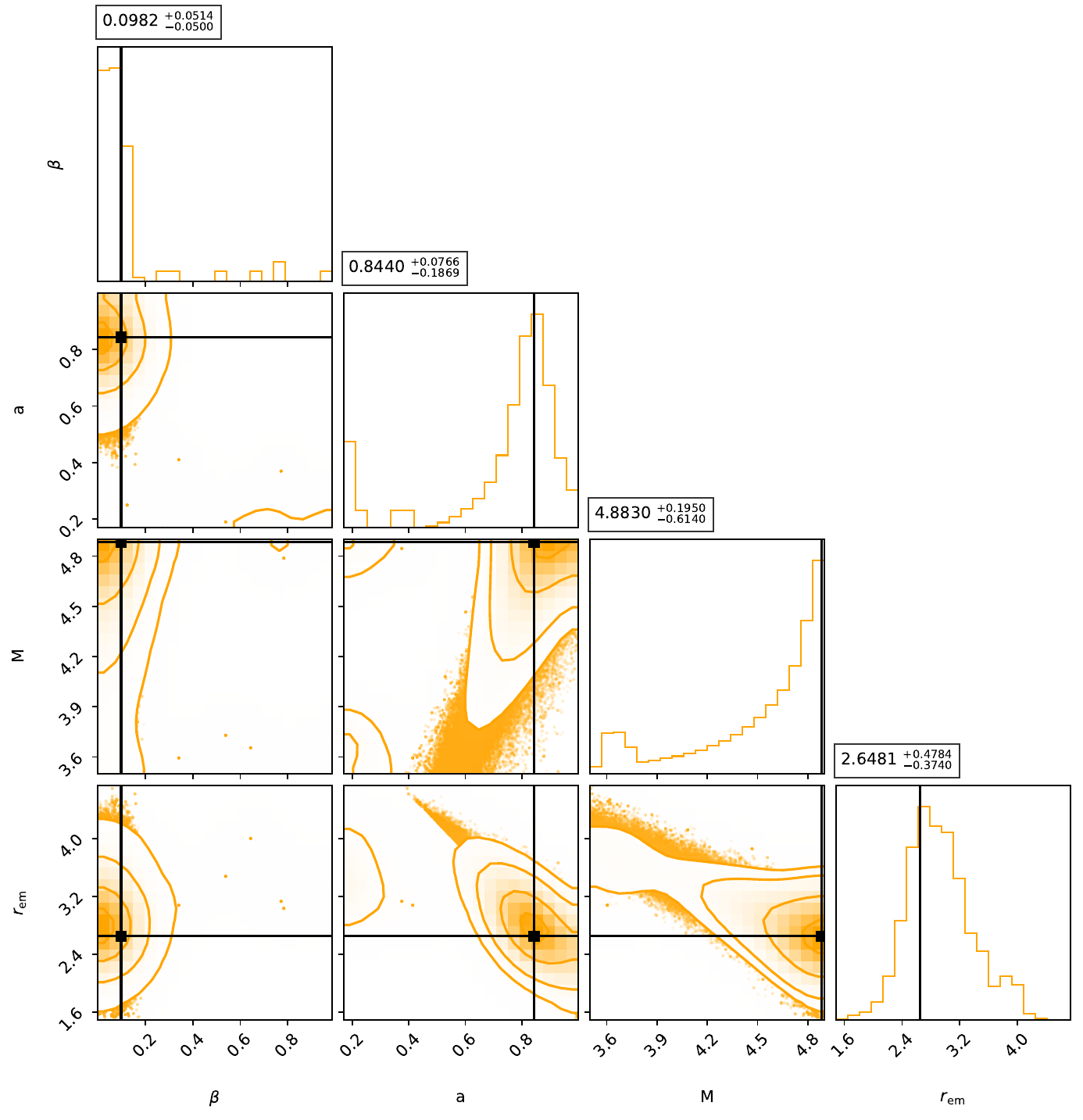}
        \caption{Tidal Disruption Model}
    \end{subfigure}
% Second row

\caption{Constraints on the model parameters using the HFQPO data of Sgr A* considering four different models: (a) Parametric Resonance Model, (b) Forced Resonance Model 1, (c) Relativistic Precession Model, and the (d) Tidal Disruption Model.
}
\label{Sgr-corner}
\end{figure}

\begin{figure}[H]
\vspace{-0.1cm}
\centering
\textbf{\underline{XTE J1859+226}}

% First row
\begin{subfigure}[b]{0.49\textwidth}
    \centering
    \includegraphics[width=\linewidth]{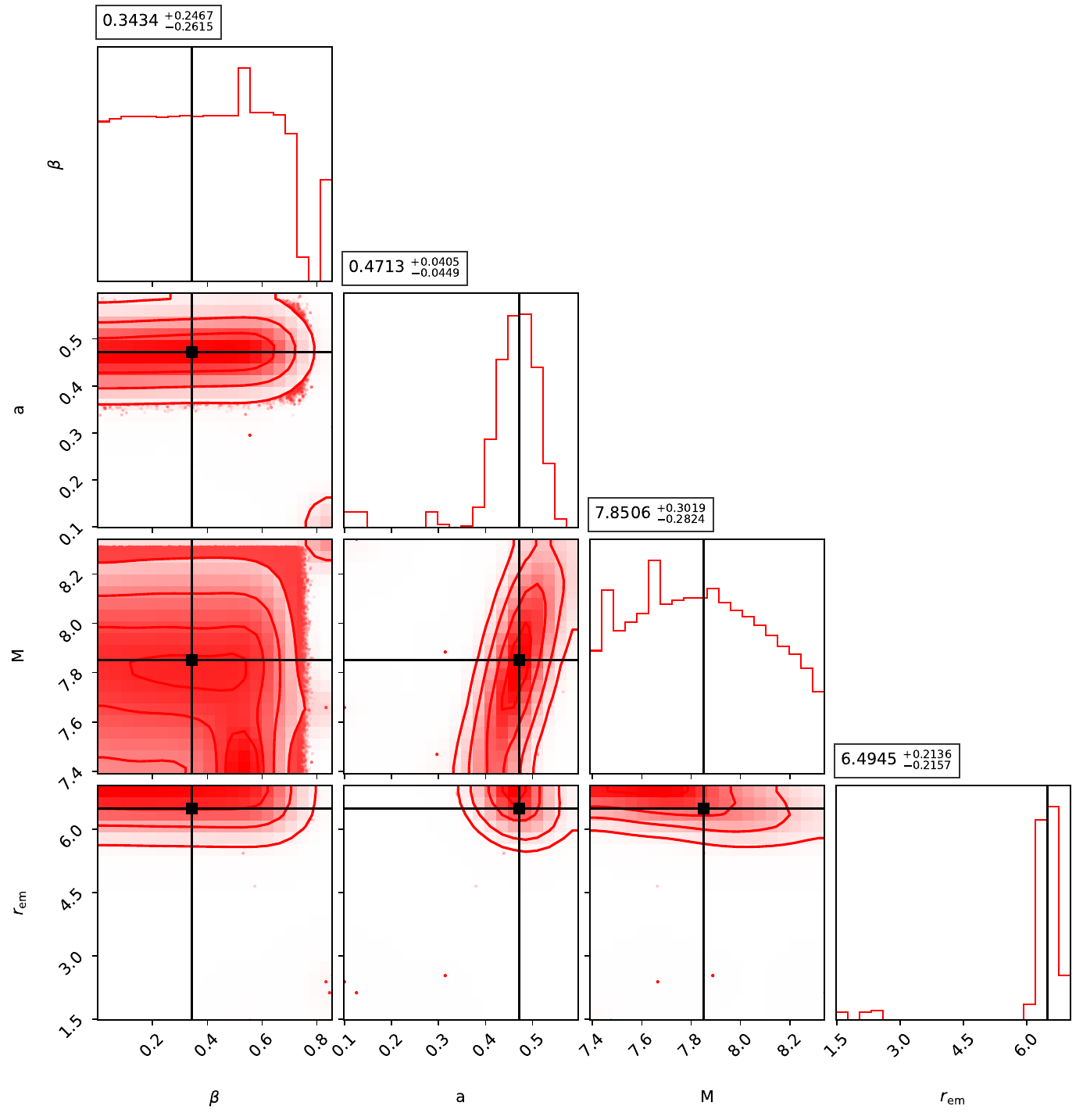}
    \caption*{(a) Parametric Resonance Model}
\end{subfigure}
\hfill
\begin{subfigure}[b]{0.49\textwidth}
    \centering
    \includegraphics[width=\linewidth]{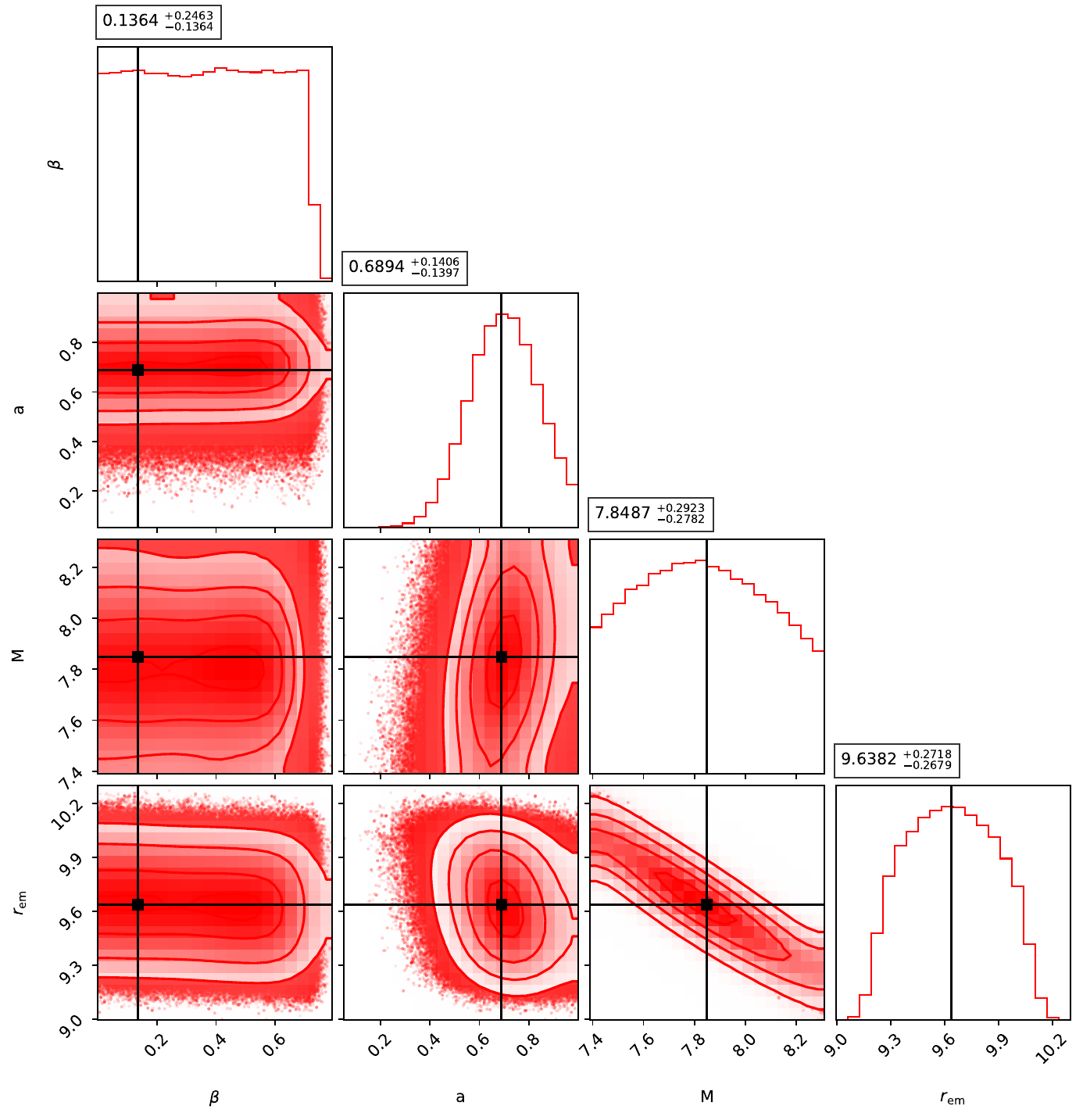}
    \caption*{(b) Forced Resonance Model 2}
\end{subfigure}

\vspace{0.2cm} % space between rows

% Second row
\begin{subfigure}[b]{0.49\textwidth}
    \centering
    \includegraphics[width=\linewidth]{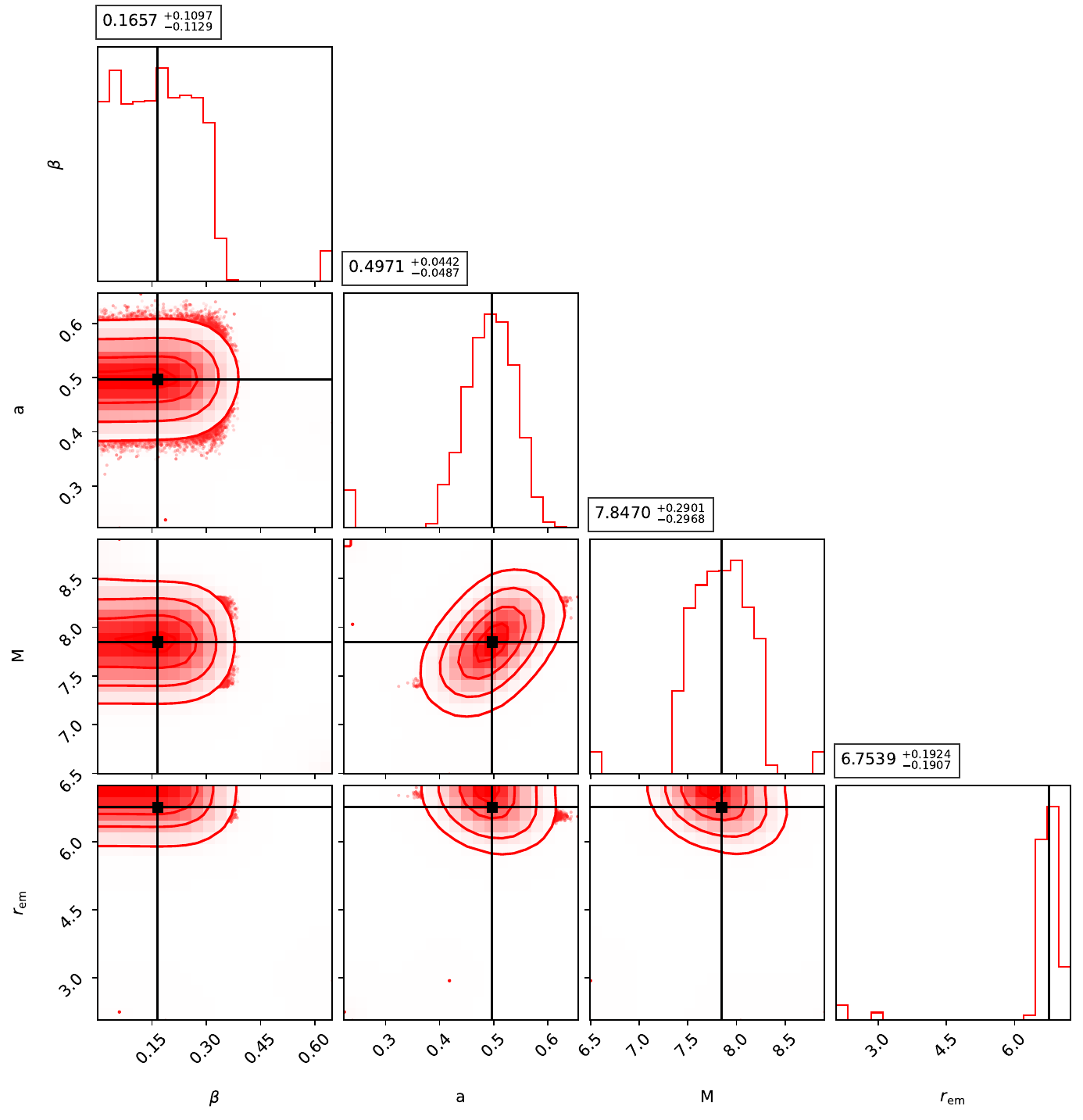}
    \caption*{(c)Keplerian Resonance Model 1}
\end{subfigure}
\hfill
\begin{subfigure}[b]{0.49\textwidth}
    \centering
    \includegraphics[width=\linewidth]{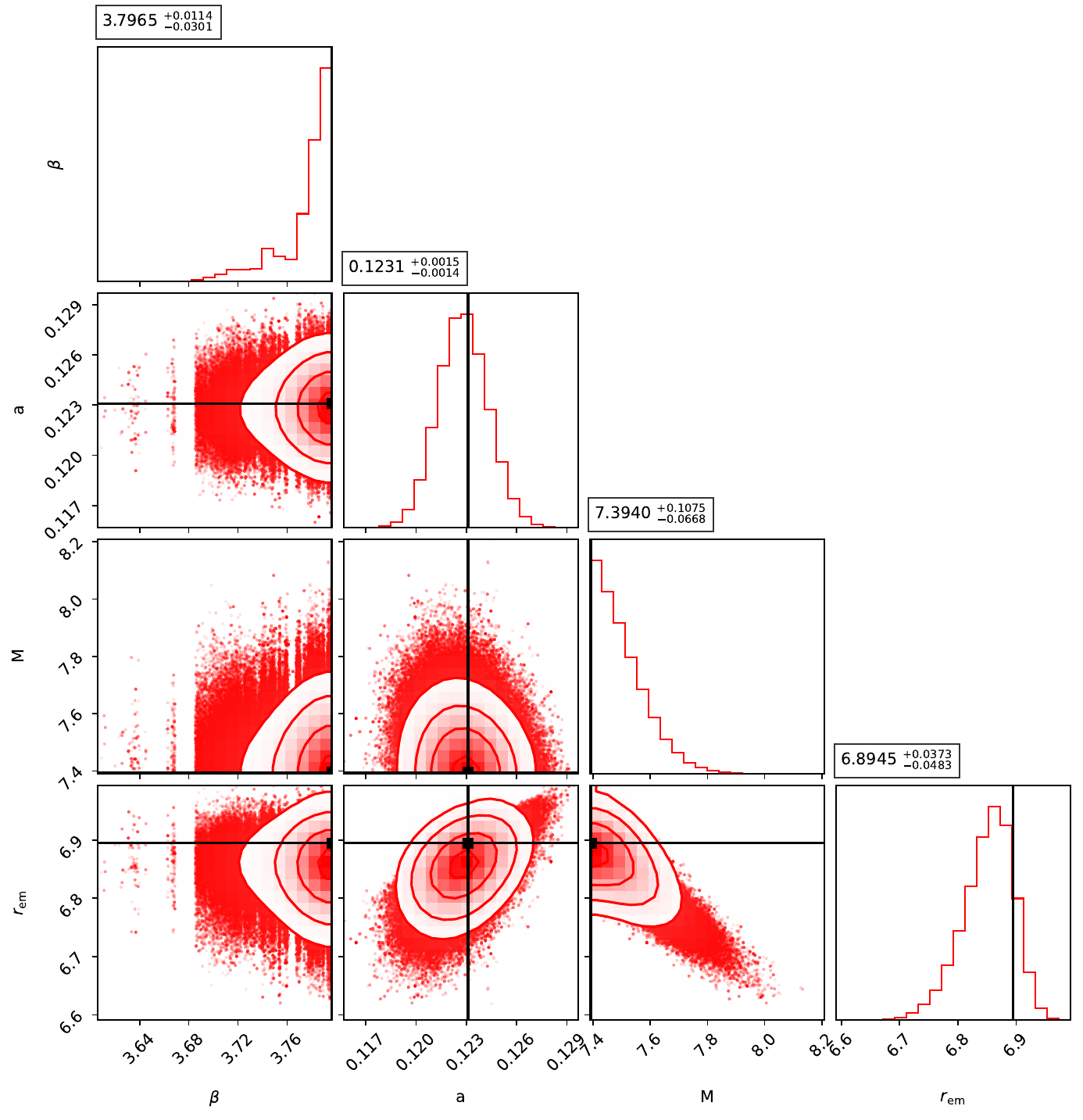}
    \caption*{(d) Relativistic Precession Model  }
\end{subfigure}
\caption{Constraints on the model parameters using the QPO data of XTE J1859+226 considering four different models: (a) Parametric Resonance Model , (b) Forced Resonance Model 2 (c) Keplerian Resonance Model 1, (d) Relativistic Precession Model .}
\label{XTE2-corner}
\end{figure}

In what follows, we present the corner plots (\ref{GROcorner}-\ref{XTE2-corner}) which we have obtained from the MCMC simulation for all the six BH sources. The 1-$\sigma$, 2-$\sigma$ and 3-$\sigma$ confidence intervals for the posterior distributions of
the model parameters are represented by the shaded regions. For a
particular source, we have obtained the corner plots for only those models where 1-$\sigma$ bounds on
$\beta$ could be given, based on our earlier grid-search method (\ref{PRM2}-\ref{TDM2}). {\bf However, for brevity, we provide only a few representative corner plots for each source. After the corner plots are presented for each source, a table summarizing the
results related to the best-fit model parameters both from the MCMC and the grid search method is given.}
For the remaining models, the likelihood function is not sufficiently sensitive to the metric parameters and hence $\beta$ and $a$ cannot be well constrained, in agreement with our earlier grid-search analysis.

From the tables one may note that the results obtained from both the methods are consistent specially when grid search method is able to give a tight bound within 1-$\sigma$ (e.g., GRO J1655-40 with models like RPM, PRM and KRM1). For sources where tight constraints on $\beta$ are not established by the grid-search method, we sometimes get a moderate deviation of result. This is due to the insensitivity of the likelihood function or $\chi^2$ on the metric parameters ($\beta$, a) for some black hole sources with respect to  certain models. We can also see this insensitivity in the $\chi^2$ vs $\beta$ plots presented in the previous section. Overall, the results associated with the grid search and MCMC methods are in agreement.

\begin{table}[t!]
\centering
\setlength{\tabcolsep}{5pt}               % Controls column spacing
\renewcommand{\arraystretch}{1.6}         % Increases row height
\footnotesize
\begin{adjustbox}{max width=\textwidth}
\begin{tabular}{|l|l|l|l|l|l|l|l|l|}
\hline
\multicolumn{9}{|c|}{\textbf{GRO J1655-40}} \\ \hline
\multirow{2}{*}{\textbf{Previous Constraints}} & \multicolumn{2}{c|}{\textbf{$\beta$}} & \multicolumn{3}{c|}{\textbf{Spin}} & \multicolumn{3}{c|}{\textbf{Mass ($M_{\odot}$)}} \\ \cline{2-9}
& \multicolumn{2}{c|}{$0 \lesssim \beta \lesssim 4$\cite{Yang:2024mro}} & $a \sim 0.65-0.75$ \cite{shafee2005estimating} & $a \sim 0.94-0.98$ \cite{Miller:2009cw} & $a = 0.29 \pm 0.003$ \cite{Motta:2013wga} & \multicolumn{3}{c|}{$5.4 \pm 0.3$ \cite{Beer:2001cg}} \\ \hline \hline
\multirow{2}{*}{\textbf{Model}} & \multicolumn{5}{c|}{\textbf{Grid Search}} & \multicolumn{3}{c|}{\textbf{MCMC}} \\ \cline{2-9}
& \textbf{$\beta$} & \textbf{1-$\sigma$} & \textbf{3-$\sigma$} & \textbf{Spin} & \textbf{Mass ($M_{\odot}$)} & \textbf{$\beta$} & \textbf{Spin} & \textbf{Mass ($M_{\odot}$)} \\ \hline
PRM & 0 & $0 \lesssim \beta \lesssim 2.1$ & $0 \lesssim \beta \lesssim 2.3$ & 0.9 ($\beta \sim 0$) & 5.17 ($\beta \sim 0$) & $0.0140^{+0.0472}_{-0.0140}$ & $0.9210^{+0.0310}_{-0.0404}$ & $5.4054^{+0.2864}_{-0.2848}$ \\ \hline
\multirow{2}{*}{FRM1} & 0.4 & $0 \lesssim \beta \lesssim 3.8$ & $0 \lesssim \beta \lesssim 3.9$ & 0.3 ($\beta_{,\text{min}} \sim 0.4$) & 5.64 ($\beta_{,\text{min}} \sim 0.4$) & $0.0288^{+1.2192}_{-0.0288}$ & $0.3460^{+0.0473}_{-0.0488}$ & $5.3988^{+0.2010}_{-0.1809}$ \\ 
& & & & 0.4 ($\beta \sim 0$) & 5.65 ($\beta \sim 0$) & & & \\ \hline
\multirow{1}{*}{KRM1} & 0 & $0 \lesssim \beta \lesssim 0.9$ & $0 \lesssim \beta \lesssim 1.8$ & 0.999 ($\beta \sim 0$) & 8.49 ($\beta \sim 0$) & $0.0065^{+0.0470}_{-0.0065}$ & $0.9997^{+0.0003}_{-0.0386}$ & $8.6292^{+0.3759}_{-0.2455}$ \\ 
\hline
\multirow{2}{*}{KRM2} & 3.3 & $0 \lesssim \beta \lesssim 3.7$ & $0 \lesssim \beta \lesssim 3.8$ & 0.3 ($\beta_{,\text{min}} \sim 3.3$) & 5.14 ($\beta_{,\text{min}} \sim 3.3$) & $2.5023^{+1.0701}_{-1.1340}$ & $0.3460^{+0.0386}_{-0.0414}$ & $5.3976^{+0.1927}_{-0.1835}$ \\ 
& & & & 0.3 ($\beta \sim 0$) & 5.26 ($\beta \sim 0$) & & & \\ \hline
\multirow{2}{*}{NADO1} & 2.6 & $0 \lesssim \beta \lesssim 3.7$ & $0 \lesssim \beta \lesssim 3.8$ & 0.5 ($\beta_{,\text{min}} \sim 2.6$) & 5.49 ($\beta_{,\text{min}} \sim 2.6$) & $0.1.8174^{+1.2772}_{-1.0620}$ & $0.4691^{+0.0662}_{-0.0789}$ & $5.3978^{+0.1933}_{-0.1931}$ \\ 
& & & & 0.4 ($\beta \sim 0$) & 5.28 ($\beta \sim 0$) & & & \\ \hline
\multirow{2}{*}{NADO2} & 0.6 & $0 \lesssim \beta \lesssim 3.7$ & $0 \lesssim \beta \lesssim 3.9$ & 0.2 ($\beta_{,\text{min}} \sim 0.6$) & 5.12 ($\beta_{,\text{min}} \sim 0.6$) & $1.6495^{+1.2987}_{-1.0530}$ & $0.2450^{+0.0359}_{-0.0388}$ & $5.3968^{+0.1913}_{-0.1846}$ \\ 
& & & & 0.2 ($\beta \sim 0$) & 5.14 ($\beta \sim 0$) & & & \\ \hline
\multirow{2}{*}{RPM} & 3.4 & $2.8 \lesssim \beta \lesssim 3.4$ & $1.5 \lesssim \beta \lesssim 3.5$ & 0.3 ($\beta_{,\text{min}} \sim 3.4$) & 5.1 ($\beta_{,\text{min}} \sim 3.4$) & $3.3973^{+0.0377}_{-0.0557}$ & $0.2689^{+0.0031}_{-0.0029}$ & $5.1192^{+0.0626}_{-0.0633}$ \\ 
& & & & 0.3 ($\beta \sim 0$) & 5.14 ($\beta \sim 0$) & & & \\ \hline
\end{tabular}
\end{adjustbox}
\caption{Comparison of the best-fit model parameters for GRO J1655-40 derived using the grid-search and the MCMC methods.}
\label{GROT}
\end{table}

In \ref{GROT}, we present the constraints on the regularisation parameter, spin and mass for the source GRO J1655-40 from different HFQPO models. Only seven models out of eleven are presented in \ref{GROT} as they rule  out certain range of $\beta$ outside the $3-\sigma$ confidence interval.
From the table we note that the tightest constrain on $\beta$ is established by the KRM1, followed by the RPM and the PRM. For these models, the parameter estimates based on the grid-search and the MCMC methods are completely in agreement. For the remaining four models (FRM1, KRM2, NADO1 and NADO2), the $\beta$ estimates from the two methods exhibit variation although they agree within 1-$\sigma$ (this may be attributed to the insensitivity of the likelihood function on $\beta$ evident from the plots in the previous section). However, the spin and mass estimates are consistent. 
This indicates that the results obtained are robust.

Interestingly, the spin estimates of GRO J1655-40 from the four HFQPO models FRM1, KRM2, NADO1 and NADO2 (assuming $\beta\sim 0$) are not in agreement with previous independent spin estimates based on the Continuum Fitting or the Fe-line method \cite{shafee2005estimating,Miller:2009cw} (also reported in \ref{GROT}), which were derived assuming GRO J1655-40 to be a Kerr BH. Both these previous estimates indicate that GRO J1655-40 is a moderately/rapidly spinning BH while these four models indicate that GRO J1655-40 is a slowly spinning BH. It might appear that the spin predicted by these four models is consistent with \cite{Motta:2013wga}, however, this should not be considered, as Motta et al. \cite{Motta:2013wga} predicted the spin of GRO J1655-40 using the RPM model and the Kerr geometry. Thus, FRM1, KRM2, NADO1 and NADO2 models do not seem to be suitable for GRO J1655-40 as they fail to adequately explain the HFQPO data for this source. 

Now, if the model RPM is used to explain the QPO data of GRO J1655-40, then the Kerr scenario is ruled out outside 3-$\sigma$ (\ref{GROT}) and the predicted spin (assuming GR) is also not in agreement with at least one of the earlier estimates \cite{shafee2005estimating,Miller:2009cw}. Hence, RPM also do not seem to be a suitable model for GRO J1655-40. However, it is important to note that the previous spin estimates based on the two independent methods are also not mutually consistent. 
This may suggest that the black hole possesses additional ``hair". Incorporating such a parameter could potentially yield consistent estimates for both the spin and the corresponding hair parameter when analyzed through the Continuum-Fitting and the Fe-line methods. However, this interpretation requires confirmation, contingent upon the availability of more precise X-ray spectral and timing observations of the source.

If we consider PRM and KRM1 to explain the HFQPO data of GRO J1655-40, the predicted spins are in agreement with one of the previous estimates \cite{Miller:2009cw}. Also these models indicate that the Kerr scenario is more favored although $0\lesssim \beta\lesssim 2.1$ is allowed within 1-$\sigma$ (for PRM) and $0\lesssim \beta\lesssim 0.9$ is allowed within 1-$\sigma$ (for KRM1). The fact that the data does not rule out the SV scenario might indicate the presence of an additional hair at play in the strong gravity regime, which is subject to further investigation. 

We now discuss the implications of the four models TDM, FRM2, KRM3 and WDOM, not reported in \ref{GROT}. FRM2, KRM3, WDOM and TDM favor all values of $\beta$ (e.g. \ref{FRM22}) and interestingly, for any $\beta$ the preferred spin is towards the lower side, $\sim 0.1-0.2$, which in turn is inconsistent with previous estimates \cite{Miller:2009cw,shafee2005estimating}. Hence, these models also do not seem to be suitable for GRO J1655-40. \\
Thus, PRM and KRM1 seem to explain the HFQPO data of GRO J1655-40 the best. Although they exhibit a preference towards the Kerr scenario, non-zero $\beta$ is also allowed within 1-$\sigma$.

%%%%%%%%%%%%%%%%%%%%%%%%%%%%%%%%%%%%%%%%%%%%%%%%%%%%%%%%%%%%%%%%%%%%%%%%%%%%%%%%%%%%%%%%%%
%%%%%%%%%%%%%%%%%%%%%%%%%%%%%%%%%%%%%%%%%%%%%%%%%%%%%%%%%%%%%%%%%%%%%%%%%%%%%%%%%%%%%%%%%%

% Defining table
\begin{table}[htbp]
\centering
\setlength{\tabcolsep}{5pt}               % Controls column spacing
\renewcommand{\arraystretch}{1.6}         % Increases row height
\footnotesize
\begin{adjustbox}{max width=\textwidth}
\begin{tabular}{|l|l|l|l|l|l|l|l|l|}
\hline
\multicolumn{9}{|c|}{\textbf{XTE J1550-564}} \\ \hline
\multirow{2}{*}{\textbf{Previous Constraints}} & \multicolumn{2}{c|}{\textbf{$\beta$}} & \multicolumn{3}{c|}{\textbf{Spin}} & \multicolumn{3}{c|}{\textbf{Mass ($M_{\odot}$)}} \\ \cline{2-9}
& \multicolumn{2}{c|}{$0 \lesssim \beta \lesssim 4$\cite{Yang:2024mro}} & \multicolumn{2}{c|}{$-0.11 < a < 0.71$ \cite{steiner2011spin}} & $a = 0.55^{+0.15}_{-0.22}$ \cite{steiner2011spin} & \multicolumn{3}{c|}{$9.1 \pm 0.61$ \cite{Orosz:2011ki}} \\ \hline \hline
\multirow{2}{*}{\textbf{Model}} & \multicolumn{5}{c|}{\textbf{Grid Search}} & \multicolumn{3}{c|}{\textbf{MCMC}} \\ \cline{2-9}
& \textbf{$\beta$} & \textbf{1-$\sigma$} & \textbf{3-$\sigma$} & \textbf{Spin} & \textbf{Mass ($M_{\odot}$)} & \textbf{$\beta$} & \textbf{Spin} & \textbf{Mass ($M_{\odot}$)} \\ \hline
\multirow{2}{*}{PRM} & 0.2 & $0 \lesssim \beta \lesssim 2.3$ & $0 \lesssim \beta \lesssim 2.5$ & 0.95 ($\beta_{,\text{min}} \sim 0.2$) & 9.52 ($\beta_{,\text{min}} \sim 0.2$) & $0.1947^{+0.2640}_{-0.1947}$ & $0.9330^{+0.0363}_{-0.0455}$ & $9.1022^{+0.4163}_{-0.3698}$ \\ 
& & & & 0.95 ($\beta \sim 0$) & 9.63 ($\beta \sim 0$) & & & \\ \hline
\multirow{2}{*}{FRM1} & 1.8 & $0 \lesssim \beta \lesssim 3.7$ & $0 \lesssim \beta \lesssim 3.8$ & 0.5 ($\beta_{,\text{min}} \sim 1.8$) & 9.66 ($\beta_{,\text{min}} \sim 1.8$) & $1.1623^{+1.1750}_{-1.1623}$ & $0.4291^{+0.0577}_{-0.0657}$ & $9.1020^{+0.3751}_{-0.3660}$ \\ 
& & & & 0.4 ($\beta \sim 0$) & 8.95 ($\beta \sim 0$) & & & \\ \hline
KRM1 & 0 & $0 \lesssim \beta \lesssim 0.9$ & $0 \lesssim \beta \lesssim 1.8$ & 0.999 ($\beta \sim 0$) & 8.49 ($\beta \sim 0$) & $0.0065^{+0.0470}_{-0.0065}$ & $0.9997^{+0.0003}_{-0.0386}$ & $8.6292^{+0.3759}_{-0.2455}$ \\ \hline
\multirow{2}{*}{KRM2} & 0.9 & $0 \lesssim \beta \lesssim 3.8$ & $0 \lesssim \beta \lesssim 3.9$ & 0.4 ($\beta_{,\text{min}} \sim 0.9$) & 9.23 ($\beta_{,\text{min}} \sim 0.9$) & $0.2614^{+1.1999}_{-0.2614}$ & $0.3796^{+0.0426}_{-0.0560}$ & $9.1082^{+0.3837}_{-0.4035}$ \\ 
& & & & 0.4 ($\beta \sim 0$) & 9.3 ($\beta \sim 0$) & & & \\ \hline
\multirow{2}{*}{KRM3} & 0.7 & $0 \lesssim \beta \lesssim 3.8$ & $0 \lesssim \beta \lesssim 3.9$ & 0.3 ($\beta_{,\text{min}} \sim 0.7$) & 9.57 ($\beta_{,\text{min}} \sim 0.7$) & $0.2497^{+0.3375}_{-0.2497}$ & $0.2321^{+0.0586}_{-0.0595}$ & $9.1074^{+0.4137}_{-0.3992}$ \\ 
& & & & 0.3 ($\beta \sim 0$) & 9.65 ($\beta \sim 0$) & & & \\ \hline
\multirow{2}{*}{NADO1} & 0.2 & $0 \lesssim \beta \lesssim 3.7$ & $0 \lesssim \beta \lesssim 3.8$ & 0.5 ($\beta_{,\text{min}} \sim 0.2$) & 8.89 ($\beta_{,\text{min}} \sim 0.2$) & $1.2195^{+1.1265}_{-1.0123}$ & $0.5601^{+0.0807}_{-0.0826}$ & $9.0952^{+0.3910}_{-0.3817}$ \\ 
& & & & 0.5 ($\beta \sim 0$) & 8.9 ($\beta \sim 0$) & & & \\ \hline
\multirow{2}{*}{NADO2} & 0.3 & $0 \lesssim \beta \lesssim 3.8$ & $0 \lesssim \beta \lesssim 3.9$ & 0.3 ($\beta_{,\text{min}} \sim 0.3$) & 9.09 ($\beta_{,\text{min}} \sim 0.3$) & $2.4271^{+1.1750}_{-1.1877}$ & $0.3048^{+0.0392}_{-0.0460}$ & $9.0968^{+0.3863}_{-0.3829}$ \\ 
& & & & 0.3 ($\beta \sim 0$) & 9.11 ($\beta \sim 0$) & & & \\ \hline
\multirow{2}{*}{RPM} & 0.9 & $0 \lesssim \beta \lesssim 3.8$ & $0 \lesssim \beta \lesssim 3.9$ & 0.4 ($\beta_{,\text{min}} \sim 0.9$) & 9.23 ($\beta_{,\text{min}} \sim 0.9$) & $1.2745^{+1.2298}_{-0.8652}$ & $0.3871^{+0.0571}_{-0.0644}$ & $9.1019^{+0.4562}_{-0.3666}$ \\ 
& & & & 0.4 ($\beta \sim 0$) & 9.3 ($\beta \sim 0$) & & & \\ \hline
\end{tabular}
\end{adjustbox}
\caption{Comparison of best-fit model parameters for XTE J1550-564 derived using the grid-search and the MCMC methods.}
\label{XTET}
\end{table}

%%%%%%%%%%%%%%%%%%%%%%%%%%%%%%%%%%%%%%%%%%%%%%%%%%%%%%%%%%%%%%%%%%%%%%%%%%%%%%%%%%%%%%%%%%
%%%%%%%%%%%%%%%%%%%%%%%%%%%%%%%%%%%%%%%%%%%%%%%%%%%%%%%%%%%%%%%%%%%%%%%%%%%%%%%%%%%%%%%%%%
In \ref{XTET} we present the constraints on the regularisation parameter $\beta$, spin and mass for the source XTE J1550-564 from the different HFQPO models, using both the grid-search and the MCMC methods. We note that none of the models in \ref{XTET} establish very tight constrains on $\beta$, hence the most probable estimates of $\beta$ obtained from the two methods sometimes vary. However, the most probable estimates of mass and spin and in agreement (\ref{XTET}).

Out of the eleven models, eight models are presented in \ref{XTET}
as they are able to put some constrain on $\beta$ and rule out a certain range of the parameter space. 
From \ref{XTET} we note that PRM and KRM1 predict high/near maximal spin for XTE J1550-564 (when $\beta\sim 0$) which is clearly not in agreement with previous estimates of spin for this source based on independent methods like the Continuum-Fitting or the Fe-line, which predict $a\sim 0.34$ and $a=0.55^{+0.15}_{-0.22}$ respectively, \cite{steiner2011spin}. Hence, these models do not seem to be suitable for XTE J1550-564. It is important to note that when we compare the spin estimates of this source with previous results, we have to consider the $\beta\sim 0$ case as the previous estimates were made assuming the Kerr geometry. 
NADO1, RPM, KRM2 and FRM1 seem to best explain the HFQPO data of XTE J1550-564
as the spins predicted by these models when $\beta\sim 0$ are in agreement with earlier estimates \cite{steiner2011spin}. 

NADO2 and KRM3 predict $a\sim 0.3$ which are not fully consistent with \cite{steiner2011spin} although it falls within the predicted error bar $-0.11<a<0.71$. 
The remaining three models TDM, FRM2 and WDOM (not reported in \ref{XTET}) cannot distinguish between the SV or the Kerr scenario. For any given $\beta$, TDM, FRM2 and WDOM predicts $a\sim 0.2-0.3$, which is not completely in agreement with previous spin estimates. However, if the error-bars in the previous spin estimates are taken into account, then the spin of XTE J1550-564 predicted by the NADO2, KRM3, TDM, FRM2, and WDOM agree with earlier estimates \cite{steiner2011spin}. Hence, based on the current precision of the data, NADO1, RPM, KRM2 and FRM1 seem to be most suitable for the source XTE J1550-564. However, none of these models constrain $\beta$ very strongly since $0\lesssim \beta \lesssim 3.7$ is allowed within 1-$\sigma$. Further, the spin of the source inferred from the Continuum-Fitting and the Fe-line methods are not fully consistent, although they remain statistically compatible within the error bars, although, the uncertainty associated with the Continuum-Fitting method is considerably large \cite{steiner2011spin}.
This may plausibly indicate towards some beyond GR phenomenon at play in the strong gravity regime near black holes (as the data does not strongly rule out non-zero $\beta$).
Alternatively, the procedures employed in estimating BH spins through the two aforementioned methods may themselves warrant further scrutiny.

In \ref{GRST},  we present the constraints on $\beta$, spin and mass for the source GRS 1915+105 from different HFQPO models. Out of the eleven models only two models (PRM and KRM1) are able to put constrain on $\beta$  which are shown in \ref{Fig2-2}. The estimates of $\beta$, spin and mass obtained from these two models using the grid-search and the MCMC methods are consistent (\ref{GRST}). Both these models favor the Kerr scenario, although $0\lesssim \beta \lesssim 2.6$ is allowed within 1-$\sigma$ for KRM1 and $0\lesssim \beta \lesssim 3$ is allowed within 1-$\sigma$ for PRM. The spin predicted by PRM is consistent with previous independent estimates \cite{Mills:2021dxs,Blum:2009ez} while the spin predicted by KRM1 is also in agreement with earlier estimates of spin \cite{Mills:2021dxs,Blum:2009ez,mcclintock2006spin}. However, Keplerian resonance arising from g-mode oscillations has been shown to be suppressed by corotation resonance \cite{Li:2002yi}. Nevertheless, Keplerian resonance driven by the coupling between the orbital angular frequencies of spatially separated vortex pairs with opposite vorticities, oscillating at the radial epicyclic frequency, may still remain viable \cite{torok2005orbital,abramowicz1998theory}. Once again the previous spin estimates of this source by different methods is not very precise. Further, PRM and KRM1 do not rule out non-zero $\beta$. These may signal presence of some beyond GR phenomenon being instrumental in the vicinity of BHs. 
The remaining nine models (not listed in \ref{GRST}) fail to establish constrains on $\beta$ as they equally favor the Kerr and the SV scenario. Interestingly, the spins predicted by these models (when $\beta\sim 0$) are on the lower side $a<0.3$ and sometimes retrograde, thus differing from all the previous spin estimates \cite{Mills:2021dxs,Blum:2009ez,mcclintock2006spin,Middleton:2006kj}. Hence, these models do not seem to be suitable for GRS 1915+105, rather PRM and KRM1 seem to be more appropriate.

\begin{table}[htbp]
\centering
\setlength{\tabcolsep}{5pt}               % Controls column spacing
\renewcommand{\arraystretch}{1.6}         % Increases row height
\footnotesize
\begin{adjustbox}{max width=\textwidth}
\begin{tabular}{|l|l|l|l|l|l|l|l|l|}
\hline
\multicolumn{9}{|c|}{\textbf{GRS 1915+105}} \\ \hline
\multirow{2}{*}{\textbf{Previous Constraints}} & \multicolumn{2}{c|}{\textbf{$\beta$}} & \multicolumn{4}{c|}{\textbf{Spin}} & \multicolumn{2}{c|}{\textbf{Mass ($M_{\odot}$)}} \\ \cline{2-9}
& \multicolumn{2}{c|}{$0 \lesssim \beta \lesssim 4$\cite{Yang:2024mro}} & $a \sim 0.98$ \cite{mcclintock2006spin} & $a \sim 0.7$ \cite{Middleton:2006kj} & $a \sim 0.6-0.98$ \cite{Blum:2009ez} & $a \sim 0.4-0.98$ \cite{Mills:2021dxs} & \multicolumn{2}{c|}{$12.4^{+2.0}_{-1.8}$ \cite{Reid:2014ywa}} \\ \hline \hline
\multirow{2}{*}{\textbf{Model}} & \multicolumn{5}{c|}{\textbf{Grid Search}} & \multicolumn{3}{c|}{\textbf{MCMC}} \\ \cline{2-9}
& \textbf{Best $\beta$} & \textbf{1-$\sigma$} & \textbf{3-$\sigma$} & \textbf{Spin} & \textbf{Mass ($M_{\odot}$)} & \textbf{$\beta$} & \textbf{Spin} & \textbf{Mass ($M_{\odot}$)} \\ \hline
\multirow{1}{*}{PRM} & 0 & $0 \lesssim \beta \lesssim 3$ & $0 \lesssim \beta \lesssim 3.3$ & 0.9 ($\beta \sim 0$) & 13.75 ($\beta \sim 0$) & $0.0572^{+0.0441}_{-0.0434}$ & $0.8907^{+0.0801}_{-0.0918}$ & $13.4453^{+1.3861}_{-1.1296}$ \\ \hline
\multirow{1}{*}{KRM1} & 0 & $0 \lesssim \beta \lesssim 2.6$ & $0 \lesssim \beta \lesssim 3.1$ & 0.999 ($\beta \sim 0$) & 12.64 ($\beta \sim 0$) & $0.0539^{+0.0437}_{-0.0436}$ & $0.9322^{+0.0608}_{-0.0940}$ & $11.3636^{+1.3147}_{-0.8250}$ \\ \hline
\end{tabular}
\end{adjustbox}
\caption{Comparison of the best-fit model parameters for GRS 1915+105 derived using the grid-search and the MCMC methods.}
\label{GRST}
\end{table}

In \ref{HT}, we present the constraints on $\beta$, spin and mass for the source H 17433+322 from different QPO models using both the grid-search and the MCMC methods. For this black hole also only PRM and KRM1 can put some constrain on the values of $\beta$. The best-fit values of model parameters from both these methods are more or less consistent. However, the spin of this source estimated using PRM and KRM1 (assuming GR) is not consistent with previous estimates obtained from the Continuum-Fitting method \cite{steiner2011distance}. The spin predicted by these two models are also inconsistent with the spin estimates of this source from the observed jet power $0.25\lesssim a \lesssim 0.5$ \cite{Banerjee:2020ubc}. 
In fact Steiner et al. 
\cite{steiner2011distance} ruled out large spin values for this source outside 99.7\% confidence interval, contrary to our predictions based on PRM and KRM1. 
Hence, these two models do not seem to be the correct description of the HFQPO data of H1743-322.  

For the remaining nine models, all values of $\beta$ from 0 to 4 are favored.  Assuming GR, the models FRM1, FRM2, KRM2, KRM3, NADO1, NADO2, RPM, TDM and WDOM respectively predict $a \sim 0.6$, $a\sim 0.4$, $a\sim 0.1$, $a\sim 0.2$, $a\sim 0.6$, $a\sim 0.3$, $a\sim 0.6$, $a\sim 0.1$, and $a\sim 0.4$ for this source. The best agreement with the previous spin estimate $a=0.2\pm0.3$ \cite{steiner2011distance} (68\% confidence) seems to be with respect to the models KRM3, KRM2 and NADO2 although the predictions from FRM2, TDM and WDOM agree within 1-$\sigma$. The above discussion thus elucidates that PRM and KRM1 are not appropriate models to explain the HFQPO data of H1743-322 although from the present data it is difficult to establish the most favored HFQPO model for this source. It is however worth noting that all the nine HFQPO models which predict a low/moderate spin of H1743-322 favor the Kerr and the SV scenario equally.

\begin{table}[htbp]
\centering
\setlength{\tabcolsep}{5pt}               % Controls column spacing
\renewcommand{\arraystretch}{1.4}         % Increases row height
\footnotesize
\begin{adjustbox}{max width=\textwidth}
\begin{tabular}{|l|l|l|l|l|l|l|l|l|}
\hline
\multicolumn{9}{|c|}{\textbf{H1743-322}} \\ \hline
\multirow{2}{*}{\textbf{Previous Constraints}} & \multicolumn{2}{c|}{\textbf{$\beta$}} & \multicolumn{3}{c|}{\textbf{Spin}} & \multicolumn{3}{c|}{\textbf{Mass ($M_{\odot}$)}} \\ \cline{2-9}
& \multicolumn{2}{c|}{$0 \lesssim \beta \lesssim 4$\cite{Yang:2024mro}} & \multicolumn{3}{c|}{$a = 0.2 \pm 0.3$ \cite{steiner2011distance}, $0.25\lesssim a \lesssim 0.5$ \cite{Banerjee:2020ubc}} & \multicolumn{3}{c|}{$8.0-14.07$ \cite{Pei:2016kka, Bhattacharjee:2017rbl,Nathan:2024eme}} \\ \hline \hline
\multirow{2}{*}{\textbf{Model}} & \multicolumn{5}{c|}{\textbf{Grid Search}} & \multicolumn{3}{c|}{\textbf{MCMC}} \\ \cline{2-9}
& \textbf{$\beta$} & \textbf{1-$\sigma$} & \textbf{3-$\sigma$} & \textbf{Spin} & \textbf{Mass ($M_{\odot}$)} & \textbf{$\beta$} & \textbf{Spin} & \textbf{Mass ($M_{\odot}$)} \\ \hline
\multirow{2}{*}{PRM} & 0.2 & $0 \lesssim \beta \lesssim 2.5$ & $0 \lesssim \beta \lesssim 2.9$ & 0.95 ($\beta_{,\text{min}} \sim 0.2$) & 9.96 ($\beta_{,\text{min}} \sim 0.2$) & $0.3724^{+0.3567}_{-0.3381}$ & $0.9251^{+0.0483}_{-0.0672}$ & $9.2797^{+1.3843}_{-0.9071}$ \\ 
& & & & 0.95 ($\beta \sim 0$) & 10.07 ($\beta \sim 0$) & & & \\ \hline
\multirow{1}{*}{KRM1} & 0 & $0 \lesssim \beta \lesssim 1.6$ & $0 \lesssim \beta \lesssim 2.3$ & 0.999 ($\beta \sim 0$) & 8.14 ($\beta \sim 0$) & $0.1253^{+0.1890}_{-0.1253}$ & $0.9989^{+0.0011}_{-0.0494}$ & $8.0109^{+0.5973}_{-0.3149}$ \\ \hline
\end{tabular}
\end{adjustbox}
\caption{Comparison of the best-fit model parameters for H1743-322 derived using the grid-search and the MCMC methods.}
\label{HT}
\end{table}

%%%%%%%%%%%%%%%%%%%%%%%%%%%%%%%%%%%%%%%%%%%%%%%%%%%%%%%%%%%%%%%%%%%%%%%%%%%%%%%%%%%%%%%%%%
%%%%%%%%%%%%%%%%%%%%%%%%%%%%%%%%%%%%%%%%%%%%%%%%%%%%%%%%%%%%%%%%%%%%%%%%%%%%%%%%%%%%%%%%%%

\begin{table}[t!]
\centering
\setlength{\tabcolsep}{5pt}               % Controls column spacing
\renewcommand{\arraystretch}{1.6}         % Increases row height
\footnotesize
\begin{adjustbox}{max width=\textwidth}
\begin{tabular}{|l|l|l|l|l|l|l|l|l|l|}
\hline
\multicolumn{10}{|c|}{\textbf{Sgr A*}} \\ \hline
\multirow{2}{*}{\textbf{Previous Constraints}} & \multicolumn{1}{c|}{\textbf{$\beta$}} & \multicolumn{7}{c|}{\textbf{Spin}} & \multicolumn{1}{c|}{\textbf{Mass ($10^6 M_{\odot}$)}} \\ \cline{2-10}
& $0 \lesssim \beta \lesssim 4$\cite{Yang:2024mro} & $a \sim 0.92$ \cite{Moscibrodzka:2009gw} & $a \sim 0.5$ \cite{Shcherbakov:2010ki} & $a \lesssim 0.1$ \cite{fragione2020upper} & $a = 0.9 \pm 0.06$ \cite{Daly:2023axh} & $a > 0.4$ \cite{Meyer:2006fd} & $a \sim 0.52$ \cite{Genzel:2003as} & $a \sim 0.22$ \cite{Belanger:2006gm} & $3.5-4.9$ \cite{Ghez:2008ms, Gillessen:2008qv} \\ \hline \hline
\multirow{2}{*}{\textbf{Model}} & \multicolumn{5}{c|}{\textbf{Grid Search}} & \multicolumn{4}{c|}{\textbf{MCMC}} \\ \cline{2-10}
& \textbf{$\beta$} & \textbf{1-$\sigma$} & \textbf{3-$\sigma$} & \textbf{Spin} & \textbf{Mass ($10^6 M_{\odot}$)} & \textbf{$\beta$} & \textbf{Spin} & \multicolumn{2}{c|}{\textbf{Mass ($10^6 M_{\odot}$)}} \\ \hline
\multirow{1}{*}{PRM} & 0 & $0 \lesssim \beta \lesssim 0.3$ & $0 \lesssim \beta \lesssim 0.5$ & 0.999 ($\beta \sim 0$) & 3.5 ($\beta \sim 0$) & $0.0008^{+0.0436}_{-0.0008}$ & $1.0000^{+0.0000}_{-0.0033}$ & \multicolumn{2}{c|}{$3.5006^{+0.0427}_{-0.0196}$} \\ \hline
\multirow{2}{*}{FRM1} & 0.9 & $0 \lesssim \beta \lesssim 2$ & $0 \lesssim \beta \lesssim 3$ & 0.999 ($\beta_{,\text{min}} \sim 0.9$) & 3.5 ($\beta_{,\text{min}} \sim 0.9$) & $0.1164^{+0.5236}_{-0.1164}$ & $0.9810^{+0.0190}_{-0.1240}$ & \multicolumn{2}{c|}{$3.5022^{+0.3981}_{-0.1792}$} \\ 
& & & & 0.95 ($\beta \sim 0$) & 3.5 ($\beta \sim 0$) & & & \multicolumn{2}{c|}{} \\ \hline
\multirow{2}{*}{FRM2} & 0.6 & $0 \lesssim \beta \lesssim 2.7$ & $0 \lesssim \beta \lesssim 3.4$ & 0.99 ($\beta_{,\text{min}} \sim 0.6$) & 3.7 ($\beta_{,\text{min}} \sim 0.6$) & $0.1281^{+0.2143}_{-0.1281}$ & $0.9882^{+0.0118}_{-0.1072}$ & \multicolumn{2}{c|}{$3.8136^{+0.5351}_{-0.3606}$} \\ 
& & & & 0.999 ($\beta \sim 0$) & 4.0 ($\beta \sim 0$) & & & \multicolumn{2}{c|}{} \\ \hline
\multirow{1}{*}{KRM1} & 0 & $0 \lesssim \beta \lesssim 0.2$ & $0 \lesssim \beta \lesssim 0.6$ & 0.999 ($\beta \sim 0$) & 3.5 ($\beta \sim 0$) & $0.0004^{+0.0472}_{-0.0004}$ & $1.0000^{+0.0000}_{-0.0046}$ & \multicolumn{2}{c|}{$3.5015^{+0.0520}_{-0.0239}$} \\ \hline
\multirow{2}{*}{KRM2} & 0.6 & $0 \lesssim \beta \lesssim 2.5$ & $0 \lesssim \beta \lesssim 2.7$ & 0.9 ($\beta_{,\text{min}} \sim 0.6$) & 3.8 ($\beta_{,\text{min}} \sim 0.6$) & $0.3968^{+0.4529}_{-0.3968}$ & $0.9200^{+0.0436}_{-0.0558}$ & \multicolumn{2}{c|}{$4.0390^{+0.4937}_{-0.4472}$} \\ 
& & & & 0.95 ($\beta \sim 0$) & 4.5 ($\beta \sim 0$) & & & \multicolumn{2}{c|}{} \\ \hline
\multirow{2}{*}{KRM3} & 0.7 & $0 \lesssim \beta \lesssim 2.8$ & $0 \lesssim \beta \lesssim 3.3$ & 0.99 ($\beta_{,\text{min}} \sim 0.6$) & 4.0 ($\beta_{,\text{min}} \sim 0.7$) & $0.3952^{+0.3247}_{-0.3007}$ & $0.9652^{+0.0348}_{-0.0850}$ & \multicolumn{2}{c|}{$3.8800^{+0.5093}_{-0.3075}$} \\ 
& & & & 0.999 ($\beta \sim 0$) & 4.3 ($\beta \sim 0$) & & & \multicolumn{2}{c|}{} \\ \hline
\multirow{1}{*}{NADO1} & 0 & $0 \lesssim \beta \lesssim 1.5$ & $0 \lesssim \beta \lesssim 3.2$ & 0.999 ($\beta \sim 0$) & 3.5 ($\beta \sim 0$) & $0.0547^{+0.2812}_{-0.0547}$ & $0.9978^{+0.0022}_{-0.1484}$ & \multicolumn{2}{c|}{$3.5018^{+0.3322}_{-0.1723}$} \\ \hline
\multirow{2}{*}{NADO2} & 0.7 & $0 \lesssim \beta \lesssim 3.2$ & $0 \lesssim \beta \lesssim 3.5$ & 0.8 ($\beta_{,\text{min}} \sim 0.7$) & 3.9 ($\beta_{,\text{min}} \sim 0.7$) & $0.7488^{+0.3530}_{-0.3536}$ & $0.8319^{+0.0847}_{-0.1174}$ & \multicolumn{2}{c|}{$4.3454^{+0.5370}_{-0.4431}$} \\ 
& & & & 0.8 ($\beta \sim 0$) & 4.0 ($\beta \sim 0$) & & & \multicolumn{2}{c|}{} \\ \hline
\multirow{1}{*}{WDOM} & 0 & $0 \lesssim \beta \lesssim 3$ & -- & 0.999 ($\beta \sim 0$) & 3.6 ($\beta \sim 0$) & $0.0455^{+0.0461}_{-0.0438}$ & $0.9986^{+0.0014}_{-0.2867}$ & \multicolumn{2}{c|}{$3.5508^{+0.5426}_{-0.3788}$} \\ \hline
\multirow{1}{*}{RPM} & 0 & $0 \lesssim \beta \lesssim 2.5$ & $0 \lesssim \beta \lesssim 3.3$ & 0.999 ($\beta \sim 0$) & 4.4 ($\beta \sim 0$) & $0.1077^{+0.8089}_{-0.1077}$ & $0.9766^{+0.0234}_{-0.1325}$ & \multicolumn{2}{c|}{$4.0434^{+0.4661}_{-0.3605}$} \\ \hline
\multirow{1}{*}{TDM} & 0 & $0 \lesssim \beta \lesssim 2.6$ & $0 \lesssim \beta \lesssim 3.5$ & 0.999 ($\beta \sim 0$) & 3.5 ($\beta \sim 0$) & $0.0982^{+0.0514}_{-0.0500}$ & $0.8440^{+0.0766}_{-0.1869}$ & \multicolumn{2}{c|}{$4.8830^{+0.1950}_{-0.6140}$} \\ \hline
\end{tabular}
\end{adjustbox}
\caption{Comparison of the best-fit model parameters for Sgr A* derived using the grid-search and the MCMC methods.}
\label{SgrT}
\end{table}

%%%%%%%%%%%%%%%%%%%%%%%%%%%%%%%%%%%%%%%%%%%%%%%%%%%%%%%%%%%%%%%%%%%%%%%%%%%%%%%%%%%%%%%%%%
%%%%%%%%%%%%%%%%%%%%%%%%%%%%%%%%%%%%%%%%%%%%%%%%%%%%%%%%%%%%%%%%%%%%%%%%%%%%%%%%%%%%%%%%%%
\newpage
In \ref{SgrT}, we present constraints on the regularisation parameter $\beta$, spin and mass for Sgr A* from different HFQPO models using both the grid-search and the MCMC methods. In this case, all the eleven models can constrain $\beta$ as all of them rule out some values of $\beta$ outside $1-\sigma$ interval and except WDOM, all of them also rule out a certain range of the parameter space of $\beta$ outside $3-\sigma$ interval. The spin constraints of this source obtained using the MCMC and the grid-search methods are very much in agreement (\ref{SgrT}). The constraints established on $\beta$ and M using both the aforesaid methods show mild variation although they agree within 1-$\sigma$.  
While PRM, KRM1, NADO1, WDOM, RPM and TDM favor the Kerr scenario (since the best-fit $\beta\sim0$), the remaining five models favor the Simpson-Visser case. PRM and KRM1 best constrain $\beta$ as they respectively rule out $\beta>0.5$ and $\beta>0.6$ outside the 3$-\sigma$ confidence interval. 

From \ref{SgrT}, we can see that there is a huge discrepancy in the previous spin estimates done by different methods. While some mention low/very low spin \cite{fragione2020upper,Belanger:2006gm}, some advocate intermediate spin \cite{Shcherbakov:2010ki,Meyer:2006fd,Genzel:2003as} and some high spin  \cite{Moscibrodzka:2009gw,Daly:2023axh,Meyer:2006fd}. However, most of them rule out near extremal spin for Sgr A*, except \cite{Meyer:2006fd}. Hence, we may state that plausibly PRM, FRM1, FRM2, KRM1, NADO1, WDOM and RPM are not suitable models to address the HFQPO data of Sgr A* as they predict near extremal spin for this source (\ref{SgrT}). Interestingly, none of the models predict intermediate/low spin for Sgr A*. The spin predicted by KRM2 and NADO2 are in agreement with \cite{Moscibrodzka:2009gw,Daly:2023axh,Meyer:2006fd} within the error bars. However, they favor the SV scenario (with $\beta\sim 0.6-0.7$) compared to GR although the Kerr scenario is allowed within 1-$\sigma$.
This, along with the wide disparity in spin estimates for this source may indicate the presence of beyond-GR effects in its near-horizon regime, warranting further investigation.

In \ref{XTETT}, we present constraints on the regularisation parameter $\beta$, spin and mass for XTE J1859+226 black hole from different HFQPO models using both the grid-search and the MCMC methods. The constraints on the model parameters obtained using both the methods more or less agree. 
Among the eleven models, only five models are able to constrain $\beta$ and those models are enlisted in \ref{XTETT}. 

The spin of this object has been determined using the RPM which turns out to be low $0.149 \pm 0.005$ \cite{Motta:2022rku}. However, this is based on the QPO data like the present work and is in agreemeent with the spin estimated by us using the RPM \ref{XTETT}. However, if we consider spin estimated by other independent methods then it turns out to be $0.987 \pm 0.003$ \cite{Mall:2023nab}. However, none of the above five models predict a spin in agreement with \cite{Mall:2023nab}. 

The remaining six models favor the Kerr scenario as much as the SV scenario.  
When we take $\beta\sim 0$, the most favored spins from FRM1, KRM2, KRM3, NADO1, NADO2 and WDOM respectively correspond to $a\sim 0.2$, $a\sim -0.1$, $a\sim -0.2$, $a\sim 0.3$, $a\sim 0.1$ and $a\sim 0.3$. Thus, none of the eleven models predict a near extremal spin for this source, as mentioned in  \cite{Mall:2023nab}. It was reported in \cite{Mall:2023nab} that the reflection features of this source is mostly blurred, out of which they selected the best 23 spectra to derive the spin using the Fe-line method.  
But this spin is inconsistent with the spin estimates based on any of the eleven HFQPO models. Hence, one may need to revisit the spin estimate of this source. 
Also, one may consider some non-Kerr geometry and attempt to simultaneously determine the spin and the deviation parameter of this object, which might yield consistent results with one of the HFQPO models. Therefore, for this BH we cannot conclusively mention the most favored HFQPO model. But the five models which constrain $\beta$ (\ref{XTETT}) favour the SV scenario compared to the Kerr scenario (in fact, RPM rules out $\beta\sim 0$ outside 3-$\sigma$) and the remaining six models fail to distinguish between the Kerr and the SV scenario. This may signal some beyond GR phenomena at play in the near-horizon regime of this BH, which requires further investigation.

\begin{table}[htbp]
\centering
\setlength{\tabcolsep}{5pt}               % Controls column spacing
\renewcommand{\arraystretch}{1.6}         % Increases row height
\footnotesize
\begin{adjustbox}{max width=\textwidth}
\begin{tabular}{|l|l|l|l|l|l|l|l|l|}
\hline
\multicolumn{9}{|c|}{\textbf{XTE J1859+226}} \\ \hline
\multirow{2}{*}{\textbf{Previous Constraints}} & \multicolumn{2}{c|}{\textbf{$\beta$}} & \multicolumn{3}{c|}{\textbf{Spin}} & \multicolumn{3}{c|}{\textbf{Mass ($M_{\odot}$)}} \\ \cline{2-9}
& \multicolumn{2}{c|}{$0 \lesssim \beta \lesssim 4$\cite{Yang:2024mro}} & $0.149 \pm 0.005$ \cite{Motta:2022rku} & $0.987 \pm 0.003$ \cite{Mall:2023nab} & & \multicolumn{3}{c|}{$7.85 \pm 0.46$ \cite{Orosz:2011ki}} \\ \hline \hline
\multirow{2}{*}{\textbf{Model}} & \multicolumn{5}{c|}{\textbf{Grid Search}} & \multicolumn{3}{c|}{\textbf{MCMC}} \\ \cline{2-9}
& \textbf{$\beta$} & \textbf{1-$\sigma$} & \textbf{3-$\sigma$} & \textbf{Spin} & \textbf{Mass ($M_{\odot}$)} & \textbf{$\beta$} & \textbf{Spin} & \textbf{Mass ($M_{\odot}$)} \\ \hline
\multirow{2}{*}{PRM} & 0.6 & $0 \lesssim \beta \lesssim 3.6$ & $0 \lesssim \beta \lesssim 3.7$ & 0.5 ($\beta_{,\text{min}} \sim 0.6$) & 8.04 ($\beta_{,\text{min}} \sim 0.6$) & $0.3434^{+0.2467}_{-0.2615}$ & $0.4713^{+0.0405}_{-0.0449}$ & $7.8506^{+0.3019}_{-0.2824}$ \\ 
& & & & 0.5 ($\beta \sim 0$) & 8.13 ($\beta \sim 0$) & & & \\ \hline
\multirow{2}{*}{FRM2} & 0.6 & $0 \lesssim \beta \lesssim 3.4$ & $0 \lesssim \beta \lesssim 3.8$ & 0.7 ($\beta_{,\text{min}} \sim 0.6$) & 7.86 ($\beta_{,\text{min}} \sim 0.6$) & $0.1364^{+0.2463}_{-0.1364}$ & $0.6894^{+0.1406}_{-0.1397}$ & $7.8487^{+0.2923}_{-0.2782}$ \\ 
& & & & 0.7 ($\beta \sim 0$) & 8.02 ($\beta \sim 0$) & & & \\ \hline
\multirow{2}{*}{KRM1} & 0.2 & $0 \lesssim \beta \lesssim 3.6$ & $0 \lesssim \beta \lesssim 3.7$ & 0.5 ($\beta_{,\text{min}} \sim 0.2$) & 7.89 ($\beta_{,\text{min}} \sim 0.2$) & $0.1657^{+0.1097}_{-0.1129}$ & $0.4971^{+0.0442}_{-0.0487}$ & $7.8470^{+0.2901}_{-0.2968}$ \\ 
& & & & 0.5 ($\beta \sim 0$) & 7.91 ($\beta \sim 0$) & & & \\ \hline
\multirow{2}{*}{RPM} & 3.99 & $3.97 \lesssim \beta \lesssim 3.99$ & $3.96 \lesssim \beta \lesssim 3.99$ & 0.14($\beta_{,\text{min}} \sim 3.99$) & 8.13 ($\beta_{,\text{min}} \sim 3.99$) & $3.7965^{+0.0114}_{-0.0301}$ & $0.1231^{+0.0015}_{-0.0014}$ & $7.3940^{+0.1075}_{-0.0668}$ \\ 
& & & & 0.2 ($\beta \sim 0$) & 7.39 ($\beta \sim 0$) & & & \\ \hline
\multirow{2}{*}{TDM} & 1.7 & $0 \lesssim \beta \lesssim 3.3$ & $0 \lesssim \beta \lesssim 3.7$ & 0.9 ($\beta_{,\text{min}} \sim 1.7$) & 7.72 ($\beta_{,\text{min}} \sim 1.7$) & $0.3266^{+0.2509}_{-0.2511}$ & $0.5069^{+0.1117}_{-0.1082}$ & $7.8459^{+0.2982}_{-0.2841}$ \\ 
& & & & 0.9 ($\beta \sim 0$) & 8.15 ($\beta \sim 0$) & & & \\ \hline
\end{tabular}
\end{adjustbox}
\caption{Comparison of best-fit model parameters for XTE J1859+226 derived using the grid-search and the MCMC methods.}
\label{XTETT}
\end{table}

%%%%%%%%%%%%%%%%%%%%%%%%%%%%%%%%%%%%%%%%%%%%%%%%%%%%%%%%%%%%%%%%%%%%%%%%%%%%%%%%%%%%%%%%%%
%%%%%%%%%%%%%%%%%%%%%%%%%%%%%%%%%%%%%%%%%%%%%%%%%%%%%%%%%%%%%%%%%%%%%%%%%%%%%%%%%%%%%%%%%%
\newpage
\section{Conclusion}
\label{S6}
In this work, we investigate the role of high-frequency quasi-periodic oscillations (HFQPOs) in black holes (BHs) as probes of strong gravity. We consider the regular BH scenario described by the Simpson-Visser (SV) spacetime, which departs from standard general relativity through the presence of a regularizing parameter $\beta$. Regular BHs are worth exploring as they provide a plausible framework to evade the singularities arising in GR, particularly in the absence of a well-established quantum theory of gravity.
It is therefore natural to explore potential astrophysical signatures of this theory using the available observational data.
To this end, we compare HFQPO observations of BHs with various kinematic and resonant QPO models proposed in the literature. HFQPOs are generally attributed to the local or collective motion of accreting plasma near the innermost stable circular orbit (ISCO). Since they are determined primarily by the spacetime geometry and are only weakly influenced by the accretion flow, HFQPOs provide a cleaner probe of the background metric than continuum spectra or Fe-line diagnostics. 

In this work, we restrict our analysis to six black hole sources$-$GRO J1655-40, XTE J1550+564, GRS 1915+105, H 1743-322, XTE J1859+226 and Sgr A*-as these BHs exhibit HFQPOs in their power spectrum. We also consider eleven HFQPO models available in the literature and interestingly the model dependent QPO frequencies are linear functions of the three fundamental frequencies associated with the motion of test particles around the BH. The model dependent HFQPO frequencies are compared with the observed frequencies for each BH and the model parameters are estimated using both the grid-search and the MCMC techniques. The convergence of MCMC has been tested using the Gelman-Rubin convergence criterion such that $0.99\lesssim R \lesssim 1.1$ is obtained for each model parameter, corresponding to all the six BHs.
The results obtained from both the grid-search and the MCMC methods are consistent and below we discuss the implications of our results, sourcewise.

For GRO J1655-40, we compare the observed HFQPOs with predictions from the eleven QPO models considered in this study. Among these, the Parametric Resonance Model (PRM) and the KeplerianResonance Model 1 (KRM1) provide the most consistent 
explanation, as the spin values inferred from these models (assuming a Kerr background) agree with at least one of the earlier estimates (the Fe-line method \cite{Miller:2009cw}). In fact, the spin inferred using PRM shows a better agreement with Miller et al. \cite{Miller:2009cw}.
Both these models exhibit a preference towards the Kerr scenario (or very low $\beta$) although $0\lesssim \beta\lesssim2.1$ is allowed within 1-$\sigma$ (for PRM) and $0\lesssim \beta \lesssim 0.9$ is allowed within 1-$\sigma$ (for KRM1). Both these models rule out large values of $\beta$ outside 3-$\sigma$. Interestingly, a similar result was also obtained when the Kerr-Sen scenario was tested using the HFQPO data of this source \cite{Dasgupta:2025fuh}.
The fact that the previous spin estimates based on the Fe-line and the Continuum-Fitting methods are inconsistent and also the HFQPO data does not completely rule out the present SV scenario, might indicate the possibility of some deviations from GR being instrumental in the strong gravity
regime near the BH. Incorporating the suitable parameter could potentially lead to consistent estimates of both the spin and the associated hair parameter when analyzed using the Continuum-Fitting and the
Fe-line methods. However, this interpretation remains tentative and requires confirmation, contingent upon the availability of more precise X-ray spectral and timing observations of the source.

When the eleven QPO models are compared with the HFQPO data of XTE J1550-564, we note that PRM and KRM1 predict a high or near-maximal spin for XTE J1550--564 (when $\beta \sim 0$), which is clearly inconsistent with previous spin estimates for this source obtained from independent methods such as the Continuum-Fitting and the Fe-line. The former predicts $a \sim 0.34$ while the latter reports $a = 0.55^{+0.15}_{-0.22}$, respectively~\cite{steiner2011spin}. 
Therefore, these models do not seem to be suitable for XTE J1550-564.
The NADO2 and KRM3 models yield $a \sim 0.3$, which does not fully agree with the 
spin estimates of~\cite{steiner2011spin}, although it remains within the reported 
uncertainty range $-0.11 < a < 0.71$. The three models, TDM, FRM2, and WDOM 
(not listed in \ref{XTET}), are unable to discriminate between the SV and Kerr 
scenarios. For any chosen value of $\beta$, these models predict 
$a \sim 0.2$--$0.3$, which shows some tension with earlier spin measurements. 
Nevertheless, once the observational uncertainties are taken into account, the 
spin values of XTE J1550--564 inferred from NADO2, KRM3, TDM, FRM2, and WDOM 
remain consistent with those reported in~\cite{steiner2011spin}.\\
NADO1, RPM, KRM2 and FRM1 seem to best explain the HFQPO data of XTE J1550-564
as the spins predicted by these models when $\beta\sim 0$ are in agreement with earlier estimates \cite{steiner2011spin}. However, these models fail to constrain $\beta$ very strongly since $0\lesssim \beta \lesssim 3.7$ is allowed within 1-$\sigma$. Once again, the spin estimates of the source obtained from the Continuum-Fitting and the Fe-line methods show a degree of inconsistency; nevertheless, they remain statistically 
compatible within the error-bars. It is worth noting that the 
uncertainty associated with the Continuum-Fitting method is substantially larger, which contributes to this apparent agreement \cite{steiner2011spin}.
This may plausibly suggest the presence of a beyond-GR effect operating in the strong 
gravity regime near black holes, since the data does not strongly rule out a non-zero 
value of $\beta$. Alternatively, the methodologies employed for estimating black hole 
spins through the two aforementioned techniques may themselves require further scrutiny.

When the eleven QPO models are compared with the HFQPO data of GRS 1915+105, we note that only PRM and KRM1 provide meaningful constraints on $\beta$ (\ref{GRST}). Both models favor the 
Kerr scenario, though values up to $0\lesssim \beta \lesssim 2.6$ (KRM1) and 
$0\lesssim \beta \lesssim 3$ (PRM) are allowed within 1-$\sigma$. The spin inferred from PRM 
matches previous estimates \cite{Mills:2021dxs,Blum:2009ez}, while KRM1 also yields values consistent 
with earlier results \cite{Mills:2021dxs,Blum:2009ez,mcclintock2006spin}. Although Keplerian resonance from $g$-mode 
oscillations is known to be suppressed by corotation resonance \cite{Li:2002yi}, a resonance mechanism involving vortex pairs oscillating at the radial epicyclic frequency may remain viable \cite{torok2005orbital,abramowicz1998theory}. Once again the previous 
spin measurements of this source are not fully consistent and also neither PRM nor KRM1 rules out a non-zero $\beta$, which might indicate towards some deviation from the Kerr scenario.
The remaining nine models (not listed in \ref{GRST}) predict spins on the lower side $a<0.3$ and sometimes retrograde, thus differing from all the previous spin estimates \cite{Mills:2021dxs,Blum:2009ez,mcclintock2006spin,Middleton:2006kj}. Therefore, these models do not appear to be suitable for GRS 1915+105, whereas PRM and KRM1 seems to provide a more consistent description of the source.

For the source H17433+322, only PRM and KRM1 are capable of constraining the 
values of $\beta$. However, the spin estimates predicted by these models 
(assuming GR) are in clear disagreement with previous measurements from the 
Continuum-Fitting method~\cite{steiner2011distance}. They also contradict the 
spin range inferred from the observed jet power, 
$0.25 \lesssim a \lesssim 0.5$~\cite{Banerjee:2020ubc}. Moreover, 
Steiner et al.~\cite{steiner2011distance} ruled out high spin values for this source at the $99.7\%$ confidence level, directly opposing the 
predictions from PRM and KRM1. 
For the remaining nine models, all values of $\beta$ from 0 to 4 are favored.
The models KRM3, KRM2, and NADO2 show the best agreement with the previous spin 
estimate, $a = 0.2 \pm 0.3$~\cite{steiner2011distance} (68\% confidence level), 
although the predictions from FRM2, TDM, and WDOM also remain consistent within 
1-$\sigma$. The above discussion indicates that PRM and KRM1 are not suitable models for explaining the HFQPO data of H1743-322. However, based on the current data, it remains difficult to identify a single most favored HFQPO model for this source. It is noteworthy that all nine HFQPO models predicting a low to moderate spin for H1743-322 are equally compatible with both the Kerr and SV scenarios.

For Sgr A*, all the eleven models can place constraints on $\beta$, as each of them excludes 
certain values of $\beta$ outside the 1-$\sigma$ interval. 
Interestingly, there exists a significant discrepancy in previous spin estimates for Sgr A* derived 
from different methods. While some studies suggest a low or very low spin \cite{fragione2020upper,Belanger:2006gm}, 
others advocate an intermediate spin \cite{Shcherbakov:2010ki,Meyer:2006fd,Genzel:2003as}, and yet others 
propose a high spin \cite{Moscibrodzka:2009gw,Daly:2023axh,Meyer:2006fd}. Most of these works, however, 
rule out a near-extremal spin for Sgr A*, with the exception of \cite{Meyer:2006fd}. 
Consequently, PRM, FRM1, FRM2, KRM1, NADO1, WDOM, and RPM are likely unsuitable for explaining the HFQPO data of Sgr A*, as they predict near-extremal spins for this source (see \ref{SgrT}). Interestingly, none of the HFQPO models 
predict an intermediate or low spin. The spin values predicted by KRM2 and NADO2, however, are consistent with 
previous estimates \cite{Moscibrodzka:2009gw,Daly:2023axh,Meyer:2006fd} within the error bars. These models favor 
the SV scenario (with $\beta \sim 0.6$--$0.7$) over GR, although the Kerr scenario remains allowed within 1$-\sigma$. 
Taken together, the wide variation in spin estimates, along with these model preferences, may point to the presence 
of beyond-GR effects in the near-horizon regime of Sgr A*, warranting further investigation.

Finally, for the source  XTE J1859+226, among the eleven HFQPO models, only five models are able to constrain $\beta$ and those models are enlisted in \ref{XTETT}. Independent (not from HFQPO data) spin estimates of this source indicate a rapidly spinning object with $a\simeq 0.987 \pm 0.003$ \cite{Mall:2023nab}. However, neither the five aforesaid models, nor the remaining six models predict a near extremal spin for this source, as mentioned in  \cite{Mall:2023nab}. It may be worthwhile to note that Mall et al. \cite{Mall:2023nab} reported that the reflection features of this source are largely blurred, 
from which the authors selected the best 23 spectra to estimate the spin using the Fe-line method. 
However, this spin estimate is inconsistent with those derived from any of the eleven HFQPO models, 
suggesting that the spin of this source may need to be revisited. An alternative approach would be 
to consider a non-Kerr geometry and attempt a simultaneous determination of both the spin and the 
deviation parameter, which could potentially yield results consistent with one of the HFQPO models. 
Consequently, for this black hole, it is not possible to identify a single most favored HFQPO model. Interestingly, the five models that constrain $\beta$ (\ref{XTETT}) favor the SV scenario over 
the Kerr scenario, while the remaining 
six models are unable to distinguish between the Kerr and the SV scenarios.

The above analysis indicates that no single HFQPO model can consistently explain 
the HFQPO data across all black hole sources, which likely explains the ongoing 
lack of consensus in the literature. Models that provide spin estimates consistent 
with previous independent measurements generally fail to constrain the regularization parameter. Hence, the Simpson-Visser scenario with a certain range of $\beta$ is equally favored as the Kerr scenario for most of the BHs considered here. This, when combined with the wide variation in spin measurements from different 
methods for a particular source, may suggest the presence of additional black hole hairs \cite{Bambi:2013fea,Bambi:2012pa,Dasgupta:2025fuh}, which possibly requires further investigation.
The present study is limited by the small number of sources exhibiting HFQPOs and by the lack of precise data. Future missions, such as ESA's LOFT, are expected to improve 
measurement precision significantly, enabling stronger constraints on both spin and 
potential additional hair parameters from HFQPO observations.

\section*{Acknowledgements}
Research of I.B. is funded by the Start-Up
Research Grant from SERB, DST, Government of India
(Reg. No. SRG/2021/000418).\\

\bibliographystyle{unsrt}
\bibliography{Re,QPO,SV}

%\bibliography{SV}

\end{document}